\documentclass[twocolumn,australian,english,aps, physrev, eqsecnum]{revtex4-1}
\usepackage[T1]{fontenc}
\usepackage[latin9]{inputenc}
\usepackage{color}
\usepackage{verbatim}
\usepackage{bm}
\usepackage{amsmath}
\usepackage{amssymb}
\usepackage{graphicx}
\usepackage{esint}
\usepackage{babel}
\begin{document}
\title{Q-based, objective-field model for wave-function collapse: Analyzing
measurement on a macroscopic superposition state}
\author{Channa Hatharasinghe, Ashleigh Willis, Run Yan Teh, P. D. Drummond
and M. D. Reid}
\affiliation{Centre for Quantum Science and Technology Theory, Swinburne University
of Technology, Melbourne 3122, Australia}
\begin{abstract}
The measurement problem remains unaddressed in modern physics, with
an array of proposed solutions but as of yet no agreed resolution.
In this paper, we examine measurement using the Q-based, objective-field
model for quantum mechanics. Schrodinger considered a microscopic
system prepared in a superposition of states which is then coupled
to a macroscopic meter.  In this paper, we analyze the entangled
meter and system, and measurements on it, by solving coupled forward-backward
stochastic differential equations for real amplitudes $x(t)$ and
$p(t)$ that correspond to the phase-space variables of the Q function
of the system at a time $t$. We model the system and meter as single-mode
fields, and measurement of $\hat{x}$ by amplification of the amplitude
$x(t)$. Our conclusion is that the outcome for the measurement is
determined at (or by) the time $t_{m}$, when the coupling to the
meter is complete, the meter states being macroscopically distinguishable.
There is consistency with macroscopic realism. By evaluating the distribution
of the amplitudes $x$ and $p$ postselected on a given outcome of
the meter, we show how the $Q$-based model represents a more complete
description of quantum mechanics: The variances associated with amplitudes
$x$ and $p$ are too narrow to comply with the uncertainty principle,
ruling out that the distribution represents a quantum state. We analyze
further, and conclude that (in this model) the collapse of the wavefunction
occurs as a two-stage process: First there is an amplification that
creates branches of amplitudes $x(t)$ of the meter, associated with
distinct eigenvalues. The final outcome of measurement is determined
by $x(t)$ once amplified, explaining Born's rule. Second, the distribution
that determines the final collapse (when the system is in an eigenstate)
is the state inferred for the system conditioned on the outcome of
the meter: information is lost about the meter, in particular, about
the complementary variable $p$.
\end{abstract}
\maketitle

\section{Introduction}

The measurement problem was summarized by Bell \citep{bell-against,beables}
and Born \citep{born-reality} as being a significant challenge in
quantum physics. According to the measurement postulates in the Copenhagen
interpretation of quantum mechanics, if a system is in a superposition
\begin{equation}
|\psi\rangle=\sum_{j}c_{j}|\lambda_{j}\rangle\label{eq:sup-3}
\end{equation}
of eigenstates $|\lambda_{j}\rangle$ of an observable $\hat{O}$,
where $c_{j}$ are complex amplitudes satisfying $\sum_{j}|c_{j}|^{2}=1$,
then the possible outcomes are the eigenvalues $\lambda_{j}$, with
relative probabilities $|c_{j}|^{2}$. The system after measurement
collapses to the eigenstate $|\lambda_{j}\rangle$ associated with
the outcome $\lambda_{j}$.

The measurement problem is to address the following:
\begin{enumerate}
\item How does the collapse from the superposition to the eigenstate arise.
\item At what stage in the measurement process is the outcome of the measurement
determined?
\end{enumerate}
In the Copenhagen interpretation, the measurement is the crucial process
for creating the outcome. The underlying question is what, if any,
real properties could be assigned to the system prior to the measurement.

As early as 1935, Schrodinger questioned the completeness of the formalism
of quantum mechanics in relation to measurement \citep{s-cat-1}.
He analyzed the measurement on a system in a superposition (\ref{eq:sup-3}).
A system $A$ is coupled to a second macroscopic system $B$, a meter.
The measurement on the meter gives a readout, which is then used to
infer a result for the measurement on the first system $A$. The coupled
system is a superposition state, of type 
\begin{equation}
\sum_{j}c_{j}|\lambda_{j}\rangle|S_{j}\rangle\label{eq:ent-sup-1}
\end{equation}
where $|S_{j}\rangle$ are macroscopic distinguishable states of the
meter, the meter being entangled with the system. The outcome of a
measurement $\hat{S}$ on the meter that distinguishes the different
$|S_{j}\rangle$ gives the final outcome $\lambda_{j}$ of the measurement
$\hat{O}$ of system $A$.

Schrodinger proposed it would be ridiculous to suggest that the outcome
of the measurement on the meter was not predetermined prior to the
measurement made on the meter. Ultimately, considering a series of
couplings, he likened the meter to a cat, where the states $|S_{j}\rangle$
are the cat ``dead'' or ``alive''. The measurement on the meter
is analogous to an observer opening a box at a time $t_{m}$, to view
whether a cat is dead or alive. The open question is that of (2),
when is the outcome of $\hat{O}$, or of $\hat{S}$, actually determined?
Surely, for macroscopically-distinct states, there is a real property
associated with the macroscopic meter, prior to its measurement at
a time $t_{m}$, when an observer makes a record. That real property
distinguishes the states of the cat: It would seem that the cat is
either dead or alive, prior to the final act at time $t_{m}$. Yet,
this real property is not evident in the Copenhagen interpretation
of quantum mechanics \citep{Einstein-letters-sch}.

There have been various approaches to addressing the measurement problem.
These include many-worlds interpretations and the de Broglie-Bohm
pilot wave theory \citep{bohm-hv,struyve-ward-pilot-wave}. It is
known that decoherence due to interaction with the environment rapidly
destroys a macroscopic superposition of type (\ref{eq:ent-sup-1}),
creating a mixture of the states $|\lambda_{j}\rangle|S_{j}\rangle$
\citep{decoherence-lg-wmil,yurke-stoler-1,decoherence-wm,milburn-holmes-2-1}.
Decoherence can explain why macroscopic superposition states do not
(normally) exist. However, these states have been generated in the
laboratory \citep{vlastakis-cat,cat-states-haroche-wine,kirchmair-cat,wang-two-mode-cat,cat-states-review-1-1,brune-cat,monroe-cat,grangier-cat,wrigth-walls-gar-1-1-2,collapse-revival-bec-3-1},
and decoherence does not address what reality existed prior to the
transition to the mixture.

In this paper, we analyse Schrodinger's question from the point of
view of a model for quantum mechanics based on phase-space solutions
for quantum fields \citep{q-frederic,q-measurement,q-retrocausal-model-measurement,q-objective-fields-entropy,drummond-time,dewitt}.
The model is motivated by an equivalence, where the phase-space amplitudes
$x$ and $p$ of the Q function $Q(x,p)$ in quantum optics \citep{q-husimi}
satisfy forward-backward stochastic differential equations with both
past and future boundary conditions \citep{q-retrocausal-model-measurement,drummond-time,q-measurement}.
 In this Q-based objective-field model, it is intended that the
amplitudes $x$ and $p$ provide a more complete, hidden-variable
description of quantum mechanics. There is no conflict with Bell's
theorem, which negates the possibility of local hidden variable theories
(LHV) completing quantum mechanics \citep{Bell-2,bell-contextual,bell-cs-review,bell-1971}.

Here, we develop previous work \citep{q-measurement}, which analyzed
the measurement problem in terms of the $Q$-based model by considering
a measurement $\hat{x}$ on a superposition of eigenstates $|x_{j}\rangle$
of $\hat{x}$, where $\hat{x}$ is a quadrature phase amplitude of
the field \citep{yuen,wu-squeezing-exp,schnabel-1-1-1,schnabel-2-1-1}.
The results of Ref. \citep{q-measurement} implied consistency with
macroscopic realism (MR) and gave insight into the timing and mechanism
of the collapse of the wave function, showing how real properties
defined by the $Q$-based model are consistent with three premises
(referred to as \emph{weak local realism} \citep{ghz-cat-1,weak-versus-det-1,manushan-bell-cat-lg-1,macro-delayed-choice,wigner-friend-macro-1,threebox-channa})
weaker than those assumed by Bell \citep{Bell-2,bell-contextual,bell-cs-review,bell-1971}
and Einstein, Podolsky and Rosen \citep{epr-1}.  In Ref. \citep{q-measurement},
a postselected state associated with the amplitudes $x$ and $p$
of the $Q$-based model is defined and calculated, and shown to specify
$\hat{x}$ and $\hat{p}$ more sharply than can possibly be given
by a quantum state. However, although general methods were explained
in \citep{q-measurement}, full solutions were limited to only certain
types of superpositions and typical Schrodinger-cat states such as
those generated in experiments were not examined.

In this paper, we present full details of the $Q$-based objective-field
model of measurement, confirming how the model predicts Born's rule
for complex superposition states. In particular, we consider a Schrodinger-cat
system, and solve for the dynamics of amplitudes $x$ and $p$ that
model measurement on the meter, where the measurement is a direct
amplification of the observable $\hat{x}$ \citep{glauber-amp,bohr-amp-measurement,q-retrocausal-model-measurement}.
Particularly important is that we treat the meter as a superposition
of two states that are \emph{classical in nature}, the two directions
of the pointer being modeled as two macroscopically-distinct \emph{coherent}
states, which have a uniform uncertainty with respect to $\hat{x}$
and $\hat{p}$, as in the experiment of Ref. \citep{brune-cat}. We
confirm the consistency of weak local realism in this limit, giving
insight into the nature of the quantum-to-classical transition.

We examine two types of Schrodinger-cat systems. In both cases, the
meter is modeled as a single-mode field, and $\hat{S}$ is the sign
of the outcome of $\hat{x}$. In the $Q$-based model of the measurement
$\hat{x}$, there are two real amplitudes $x(t)$ and $p(t)$ which
represent the field at a given time $t$, with $x(t)$ being amplified
to a macroscopic level, becoming directly detectable (here, $p(t)$
is de-amplified to an undetectable level). The outcome inferred for
$\hat{x}$ is 
\begin{equation}
x_{0}=x(t)/G\label{eq:inferred-outcome}
\end{equation}
 where $G$ is the amplification induced by the measurement.

First, we examine a single-mode cat-state, which is a superposition
of two coherent states \citep{yurke-stoler-1}
\begin{equation}
|\psi_{cat}\rangle=N(|\alpha_{0}\rangle+e^{i\varphi}|-\alpha_{0}\rangle)/\sqrt{2}\label{eq:cat}
\end{equation}
($N$ is a normalization constant), similar to that prepared in experiments
\citep{vlastakis-cat,kirchmair-cat,grangier-cat}. The ``state of
the cat'' is determined by a direct measurement on the cat state,
by amplification of $\hat{x}$. The system becomes its own meter,
once $x(t)$ are amplified to a macroscopic level. Here, we extend
the work presented in \citep{q-measurement}, where a superposition
of two eigenstates of $\hat{x}$ is treated. The eigenstate is modeled
as a highly squeezed state in $\hat{x}$. In particular, we analyze
a variety of superposition states, with a variable phase factor $\varphi$,
and present forward-backward simulations where boundary conditions
are deduced from the Wigner function \citep{wigner,wigner-dn}. Extending
the earlier work, this enables a simple causal model for measurement
to be developed from the simulation.

Second, we examine a Schrodinger-cat state based on two correlated
field-modes e.g., the two-mode cat state
\begin{equation}
|\psi_{Cat}\rangle=N_{2}(|\alpha_{0}\rangle|\beta_{0}\rangle+e^{i\varphi}|-\alpha_{0}\rangle|-\beta_{0}\rangle)/\sqrt{2}\label{eq:tm-cat}
\end{equation}
($N_{2}$ is a normalization constant, $|\pm\alpha_{0}\rangle$ and
$|\pm\beta_{0}\rangle$ are coherent states). One mode is macroscopic
($\beta_{0}\rightarrow\infty$) as in a meter, in analogy with the
state (\ref{eq:ent-sup-1}) of Schrodinger's original proposal. This
type of cat state has been generated in experiments \citep{wang-two-mode-cat}.
There are two systems, $A$ and $B$, for which the respective amplitudes
$\hat{x}_{A}$ and $\hat{x}_{B}$ are defined. We solve for the measurement
of $\hat{x}_{A}$ and $\hat{x}_{B}$, by considering independent amplification
of each field. There are four stochastic amplitudes: $x_{A}(t)$,
$p_{A}(t)$, $x_{B}(t)$ and $p_{B}(t)$. The outcomes in the Q-based
model for quantum measurement are determined by $x_{A}(t)$ and $x_{B}(t)$
in the large amplification limit \citep{q-measurement}. We confirm
consistency within the Q-based model, that the meter outcome $\hat{S}$,
as inferred from the sign of $x_{B}(t)$, agrees with the outcome
inferred from $x_{A}(t)$ for $A$, if the system $A$ is directly
measured. This confirms consistency with the causal model for measurement
presented in \citep{q-measurement,threebox-channa,wigner-friend-macro-1,ghz-cat-1,delayed-choice-causal-model-chaves-2},
which is based on real properties for the system as it exists after
measurement settings are fixed.

\textbf{\emph{Main conclusions of paper:}} The main conclusions of
this paper are threefold, and are the same for each type of cat-system.
\textbf{(1)} \emph{First, there is consistency with macroscopic realism
(MR).}  As with earlier results for the superpositions of eigenstates
\citep{q-measurement}, we conclude that in the Q-based model, the
outcome of $\hat{S}$ of the meter is determined at the time $t_{m}$
when the system is prepared in the coupled state (\ref{eq:tm-cat}),
\emph{provided the meter is macroscopic}. This is in agreement with
Schrodinger's argument that MR must hold: The system has a predetermined
value for the outcome of $\hat{S}$ once coupled to the meter, at
the time $t_{m}$: The cat is either dead or alive, prior to the observer
opening the box. 

\textbf{(2) }\emph{Second, there is insight into how the collapse
of the wave-function occurs in the Q-based model and how MR is consistent
with this collapse.} Here, we probe the Schrodinger-cat paradox, by
evaluating the distribution {[}denoted $Q(x,p|S)${]} for the amplitudes
$x$ and $p$ of the cat state, conditioned on a final readout of
the meter that gives the outcome $S$ of $\hat{S}$, indicating whether
the cat is ``dead'' or ``alive''. This \emph{postselected} state
has been calculated for the system in a superposition of eigenstates
$|x\rangle$ of $\hat{x}$, and in a model for the meter, and shown
to correspond to the projected state given by quantum mechanics. We
extend the analysis, and give further details for the two-mode cat
state. Following \citep{q-measurement}, the postselected state $Q(x_{A},p_{A}|S)$
inferred for the system $A$ conditioned on a final readout of $S$
of the meter $B$ is evaluated by integrating over the complementary
observables $p_{B}$ of the meter. For a \emph{macroscopic} meter,
we confirm for a variety of superposition states by solving the forward-backward
equations numerically that the postselected state for $A$ \emph{is}
the projected state predicted by quantum mechanics e.g. eigenstate
$|x_{j}\rangle$ in (\ref{eq:ent-sup-1}) corresponding to the outcome
$x_{j}$ of $\hat{x}$. In particular, we demonstrate that for the
coherent-state meter given by system $B$ in (\ref{eq:tm-cat}), the
postselected state for system $A$ is the coherent state, either $|\alpha_{0}\rangle$
or $|-\alpha_{0}\rangle$ in (\ref{eq:tm-cat}), corresponding to
the outcome ($+$ or $-$) of $\hat{S}$. This supports and generalizes
conclusions in \citep{q-measurement}.

In summary, the collapse of the wave-function is a two-stage process:
Amplification is the key to the measurement process, and the real
property that determines the outcome of the meter (whether the cat
is found dead or alive) manifests in the simulations as $\hat{x}$
is amplified. The collapse to the precise distribution corresponding
to the eigenstate (or coherent state) arises as a result of the inference
\emph{about the system} based on the value of the amplified $x(t)$,
There is a loss of information about the complementary variable $p_{B}$
of the meter system. Hence, the cat (\ref{eq:cat}) (or (\ref{eq:tm-cat}))
cannot be regarded as having been in one or other of the coherent
states, $|\alpha_{0}\rangle$ or $|-\alpha_{0}\rangle$ prior to the
observer opening the box.

\textbf{(3)} \emph{Third, there is insight into how MR can be consistent
with the superposition state.} Schrodinger's argument is that if the
cat is indeed dead or alive, prior to the observer opening the box,
then what state could the cat be in \citep{s-cat-1}? By the nature
of the superposition, this state cannot be a quantum state (there
is no mixture of quantum states with definite outcomes for $\hat{S}$
that can be equivalent to the superposition $|\psi\rangle$). Schrodinger's
argument hence suggests that quantum mechanics is incomplete \citep{Einstein-letters-sch}.
In this paper, we follow Ref. \citep{q-measurement} and examine the
postselected state further, by evaluating the variances associated
with the postselected state, showing incompatibility with the Heisenberg
uncertainty relation. The amplitudes $x$ and $p$ defining the state
of the cat, inferred from a given outcome, cannot be equivalent to
a quantum state. The compatibility with MR is obtained at the expense
of ``states'' for the cat that are defined more sharply than can
be given by a quantum state.

\textbf{\emph{Layout of paper: }}In Section II, we summarize the two
models for a Schrodinger cat state. In Section III, we present the
model for measurement by amplification of $\hat{x}$, and in Section
IV, we solve the forward-backward equations for $x$ and $p$, presenting
solutions showing individual trajectories for $x(t)$ and $p(t)$
amplitudes. We consider in Sections IV-VI measurement on single-mode
states, including superpositions of squeezed states, and superpositions
of coherent states (the cat state). We confirm from the simulations
Born's rule for the probability densities of $x(t)$ arising after
large amplification in Section IV.C. In Section V, we explain the
Q-based model of reality, and in Section VI, calculate the postselected
state and associated variances for the amplitudes $x$ and $p$ of
the system in a superposition state, conditioned on the outcome of
the measurement $\hat{x}$.

The two-mode entangled Schrodinger-cat state is studied in Section
VII, where we present solutions of the forward-backward equations
for measurements $\hat{x}_{A}$ and $\hat{x}_{B}$. In Section VIII,
we evaluate the bipartite postselected distribution for the two modes,
and also the distribution of $x_{A}$ and $p_{A}$ for system $A$
alone, conditioned on the outcome of the meter $B$, thus modeling
the collapse of the wave function. By examining the variances of the
postselected state, we show incompatibility of the postselected state
with the uncertainty relation. A conclusion is given in Section IX.

\section{Two models of a schrodinger-cat}

\subsection{Single-mode superposition}

As we wish to analyze the measurement problem, we consider the measurement
of a specific observable $\hat{x}$, and assume that a system $S$
is prepared in a superposition 
\begin{equation}
|\psi_{S}\rangle=c_{1}|x_{1}\rangle+c_{2}|x_{2}\rangle\label{eq:sup-4}
\end{equation}
of eigenstates $|x_{j}\rangle$ of $\hat{x}$. Here, $x_{j}$ are
the eigenvalues of $\hat{x}$ and $c_{j}$ are probability amplitudes.
For simplicity, we take only two eigenstates and may consider that
$x_{2}=-x_{1}$. The outcome of the measurement will be one or other
of the eigenvalues, $x_{1}$ and $x_{2}$. After the measurement,
the system collapses to the eigenstate $|x_{j}\rangle$ corresponding
to the outcome $x_{j}$. In the first part of this paper, we study
the system prepared in a superposition of type (\ref{eq:sup-4}),
and model the measurement of $\hat{x}$ as a direct amplification,
induced by an interaction Hamiltonian $H_{amp}$, which produces a
final state 
\begin{equation}
|\psi_{M,S}\rangle=c_{1}e^{-iH_{amp}t/\hbar}|x_{1}\rangle+c_{2}e^{-iH_{amp}t/\hbar}|x_{2}\rangle\label{eq:M-supcat}
\end{equation}
which is a superposition of two amplified states, where $G$ is the
amplification factor. The superposition $|\psi_{M,S}\rangle$ is a
Schrodinger-cat state, since the states $e^{-iH_{amp}t\hbar}|x_{1}\rangle$
and $e^{-iH_{amp}t/\hbar}|x_{2}\rangle$ become macroscopically distinct
for $G$ sufficiently large.

\subsection{Correlated state for the system and meter}

An alternative strategy for measurement involves the coupling of the
system prepared in (\ref{eq:sup-4}) to a second system, a meter $M$.
The first stage of measurement involves an interaction, modeled by
the Hamiltonian $H_{int}$, which couples the system to a meter. The
final state after the coupling is an entangled state of type 
\begin{equation}
|\psi_{ent}\rangle=c_{1}|x_{1}\rangle|+\rangle_{M}+c_{2}|x_{2}\rangle|-\rangle_{M}\label{eq:ent-sup}
\end{equation}
where the $|x_{j}\rangle$ are states for the system, and $|+\rangle_{M}$
and $|-\rangle_{M}$ are macroscopically-distinguishable states of
the meter $M$. The measurement on the meter indicates a value $+$
or $-$, which indicates the outcome of the measurement $\hat{x}$
of the system being either $x_{1}$ or $x_{2}$ respectively. The
entangled meter-system is an example of the Schrodinger-cat state
\citep{s-cat-1}.

In the second part of this paper, we examine a simple realization
of the entangled meter-system state, involving two fields. We will
consider the entangled state
\begin{equation}
|\psi_{ent}\rangle=c_{1}|x_{1}\rangle|\beta_{0}\rangle_{M}+c_{2}|x_{2}\rangle|-\beta_{0}\rangle_{M}\label{eq:ent-two-coherent}
\end{equation}
where both the system $S$ and meter $M$ are single-field modes.
Here, the $|x_{j}\rangle$ are eigenstates of $\hat{x}$ for the system
$S$, and $|\beta_{0}\rangle_{M}$, $|-\beta_{0}\rangle_{M}$ are
coherent states for the meter mode, modeled as an intense field, the
$\beta_{0}$ being a macroscopic amplitude. To examine the transition
as the meter becomes macroscopic, we will allow that both fields can
have an arbitrary intensity, and we denote the system and meter fields
by $A$ and $B$ respectively.

\subsection{Model for the system and meter}

We consider that the system A is a single field mode, denoted by $A$,
with boson operators $\hat{a}$ and $\hat{a}^{\dagger}$. In the model
(\ref{eq:ent-sup}), the meter $M$ is a second single-mode field,
denoted by $B$, with boson operators $\hat{b}$ and $\hat{b}^{\dagger}$.
The observables of interest for each system $A$ and $B$ are the
quadrature phase amplitudes defined as \citep{yurke-stoler-1}
\begin{eqnarray}
\hat{x}_{A} & = & \hat{a}+\hat{a}^{\dagger}\nonumber \\
\hat{x}_{B} & = & \hat{b}+\hat{b}^{\dagger}\label{eq:x}
\end{eqnarray}
The complementary observables are $\hat{p}_{A}=(\hat{a}-\hat{a}^{\dagger})/i$
and $\hat{p}_{B}=(\hat{b}-\hat{b}^{\dagger})/i$. Where we consider
the first model (\ref{eq:sup-4}), we denote $\hat{x}_{A}$ by $\hat{x}$,
and $\hat{p}_{A}$ by $\hat{p}$.

The eigenstates of $\hat{x}_{A}$ are approximated as highly squeezed
states in $\hat{x}_{A}$ \citep{q-retrocausal-model-measurement}.
Generally, we consider the squeezed state for the field $A$, defined
as
\begin{equation}
|\alpha_{0},r\rangle=D(\alpha_{0})S(r)|0\rangle\label{eq:sq}
\end{equation}
where $r$ is the squeeze parameter. Here, $D(\alpha_{0})=e^{\alpha_{0}a^{\dagger}-\alpha_{0}^{*}a}$,
$S(r)=e^{\frac{1}{2}(ra^{2}-ra^{\dagger2})}$ and $|0\rangle$ is
the vacuum state of the field mode $A$. We take both $r$ and $\alpha_{0}$
to be real. This implies variances in $\hat{x}_{A}$ and $\hat{p}_{A}$
of
\begin{eqnarray}
(\Delta\hat{x})^{2} & \equiv\langle\hat{x}^{2}\rangle-\langle\hat{x}\rangle^{2}= & e^{-2r}\nonumber \\
(\Delta\hat{p})^{2} & \equiv\langle\hat{p}^{2}\rangle-\langle\hat{p}\rangle^{2}= & e^{2r}\label{eq:var-1}
\end{eqnarray}
(here, we use the notation $\Delta\hat{x}\equiv\sqrt{\langle\hat{x}^{2}\rangle-\langle\hat{x}\rangle^{2}}$).
With $r\rightarrow\infty$, the variance in $\hat{x}$ becomes zero,
and the state of the field becomes the eigenstate $|x_{j}\rangle$,
where $x_{j}=2\alpha_{0}$. For $r=0$, the state of the field is
the coherent state $|\alpha_{0}\rangle$ where $\alpha_{0}=x_{j}/2$,
which has variances $(\Delta\hat{x}_{A})^{2}=(\Delta\hat{p}_{A})^{2}=1$,
at the standard quantum limit.

It is convenient to rewrite the general state (\ref{eq:sq}) as 
\begin{equation}
|x_{j}/2,r\rangle\equiv|\psi(\alpha,r)\rangle=D(x_{j}/2)S(r)|0\rangle\label{eq:sq-state-x}
\end{equation}
where we recognize that $r$ is the squeeze parameter, $\langle\hat{x}_{A}\rangle=x_{j}$,
$\langle\hat{p}_{A}\rangle=0$, $(\Delta\hat{x}_{A})^{2}=e^{-2r}$
and $(\Delta\hat{p}_{A})^{2}=e^{2r}$. Hence, we write the single-mode
superposition state (\ref{eq:sup-4}) of system $A$ as 
\begin{equation}
|\psi_{S}\rangle=N(c_{1}|x_{1}/2,r\rangle+c_{2}|-x_{1}/2,r\rangle)\label{eq:sq-sup}
\end{equation}
where for $r$ large, the state reduces to (\ref{eq:sup-4}): $N\rightarrow1$
as the states $|x_{1}/2,r\rangle$ and $|-x_{1}/2,r\rangle$ become
orthogonal. Our interest is to also examine $r=0$, where $|\psi_{S}\rangle$
reduces to the cat state (\ref{eq:cat}).

Defining similar squeezed states for system $B$, we write the entangled
two-mode state as 
\begin{eqnarray}
|\psi_{ent}\rangle & = & N_{2}(c_{1}|\frac{x_{1}}{2},r\rangle|\frac{x_{1B}}{2},r_{2}\rangle\nonumber \\
 &  & +c_{2}|-\frac{x_{1}}{2},r\rangle|\frac{x_{1B}}{2},r_{2}\rangle)\label{eq:ent-two-coherent-1}
\end{eqnarray}
where $r_{2}$ and $x_{1B}$ denote the squeeze parameter and mean
amplitude for system $B$, which will model a meter $M$. Here, $N_{2}$
is a normalization constant necessary when $r$ or $r_{2}$ are finite.
For $r$ large and $r_{2}=0$, the state (\ref{eq:ent-two-coherent-1})
reduces to (\ref{eq:ent-two-coherent}), where system $A$ is measured
by a coherent-state meter. For $r=r_{2}=0$, the state reduces (\ref{eq:ent-two-coherent-1})
reduces to the two-mode cat state (\ref{eq:tm-cat}). 

\section{Measurement by amplification: Forward-backward stochastic equations}

In this paper, we address the questions relating to the measurement
problem within the framework of the objective-field Q-based interpretation
for quantum mechanics \citep{q-retrocausal-model-measurement}. Following
previous treatments \citep{q-retrocausal-model-measurement,glauber-amp},
we model the direct measurement of $\hat{x}_{A}$ and $\hat{x}_{B}$
as an amplification of $\hat{x}_{A}$ and $\hat{x}_{B}$, respectively.
This can be achieved for a single-mode field using parametric down
conversion \citep{yuen,wu-squeezing-exp,schnabel-1-1-1,schnabel-2-1-1},
as given by the Hamiltonian (for a field $A$) 
\begin{equation}
H_{amp}=\frac{i\hbar g}{2}\left[\hat{a}^{\dagger2}-\hat{a}^{2}\right]\label{eq:hamb-1}
\end{equation}
where $g$ is real. The solution is
\begin{eqnarray}
\hat{x}(t) & = & \hat{x}(0)e^{gt}\nonumber \\
\hat{p}(t) & = & \hat{p}(0)e^{-gt}\label{eq:amp}
\end{eqnarray}
where $\hat{x}=\hat{a}+\hat{a}^{\dagger}$ and $\hat{p}=(\hat{a}-\hat{a}^{\dagger})/i$
, which amplifies $\hat{x}$ when $g>0$ (and $\hat{p}$ when $g<0$).

The single-mode Husimi $Q$ function for a state of the system at
the time $t$ is defined \citep{q-husimi}
\begin{equation}
Q(x,p,t)=\frac{1}{\pi}\langle\alpha|\rho(t)|\alpha\rangle\label{eq:Q}
\end{equation}
where $\alpha=(x+ip)/2$ and $\rho(t)$ is the density operator of
the quantum system $A$ at the time $t$. The Q function defines the
quantum state $\rho(t)$ uniquely, and is always positive, hence representing
a probability distribution for variables $x$ and $p$.

By solving for the equation of motion of $Q$, a generalized Fokker-Planck
equation is obtained, for which positive and negative diffusion terms
can be identified. This transforms to a set of forward-backward stochastic
equations in the amplitudes $x$ and $p$. These are written, for
the amplified variable $x$, as \citep{q-retrocausal-model-measurement,drummond-time}
\begin{equation}
\frac{dx}{dt_{-}}=-gx+\sqrt{2g}\xi_{1},\label{eq:forwardSDE-2-1-1-1}
\end{equation}
where $t_{-}=-t$. The equation is solved in the negative time direction,
with a boundary condition at the final time $t_{f}$, when the interaction
$H_{amp}$ is completed. Hence, the initial condition for the differential
equation in $x_{A}$ is referred to as a future boundary condition
\citep{dirac-fbc}. The equation for the de-amplified or complementary
variable is
\begin{align}
\frac{dp}{dt} & =-gp+\sqrt{2g}\xi_{2},\label{eq:backwardSDE-2-1-1-1}
\end{align}
which is a stochastic differential equation with a boundary condition
at the initial time $t_{0}$. We refer to this as a past boundary
condition. The equations are stochastic, the Gaussian random noises
$\xi_{\mu}\left(t\right)$ satisfying $\left\langle \xi_{\mu}\left(t\right)\xi_{\nu}\left(t'\right)\right\rangle =\delta_{\mu\nu}\delta\left(t-t'\right)$.

\section{Measurement by direct amplification: a single-mode cat state}

\subsection{Measurement of $\hat{x}$ on squeezed and coherent states}

We first examine measurement of $\hat{x}$ on a single-mode system
prepared in the state $|x_{1}/2,r\rangle$ defined as (\ref{eq:sq-state-x}).
The Q function of the squeezed state $|x_{1}/2,r\rangle$ is\textcolor{green}{}
\begin{equation}
Q_{sq}(x,p)=\frac{1}{2\pi\sigma_{x}\sigma_{p}}e^{-(x-x_{1})^{2}/2\sigma_{x}^{2}}e^{-p^{2}/2\sigma_{p}^{2}}\label{eq:q-sq}
\end{equation}
with variances in $x$ and $p$ given by $\sigma_{x}^{2}=1+e^{-2r}$
and $\sigma_{p}^{2}=1+e^{2r}$.

As $r\rightarrow\infty$, the squeezed state has variance $\sigma_{x}^{2}=1$
and $\sigma_{p}^{2}\rightarrow\infty$, which models the eigenstate
$|x_{1}\rangle$ of $\hat{x}$. The coherent state $|\alpha_{0}\rangle$
corresponds to $\alpha_{0}=x_{1}/2$ and $r=0$. The eigenstate $|x_{1}\rangle$
has a non-zero variance in $x$, but the measured variance in $\hat{x}$
is zero. Hence, the distribution $Q_{sq}(x,p)$ has an unobservable
noise at the level of the vacuum: For the eigenstate $|x_{1}\rangle$,
the variance in $x$, given by $\sigma_{x}^{2}$, is $1$. Since it
is not measured as $\hat{x}$ is amplified, we refer to this noise
as ``hidden noise'' \citep{q-measurement}. 

\begin{figure}
\begin{centering}
\includegraphics[width=0.8\columnwidth]{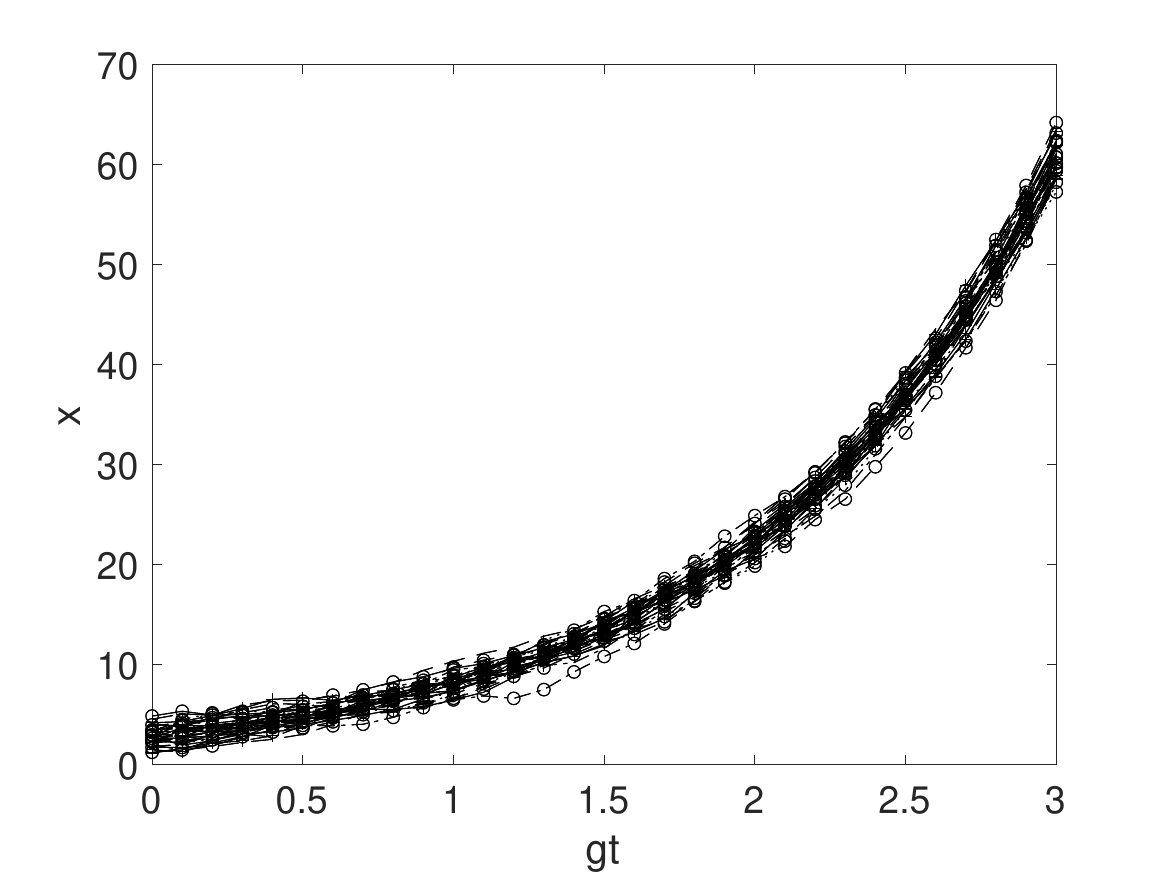}
\par\end{centering}
\begin{centering}
\negthickspace{}\includegraphics[width=0.8\columnwidth]{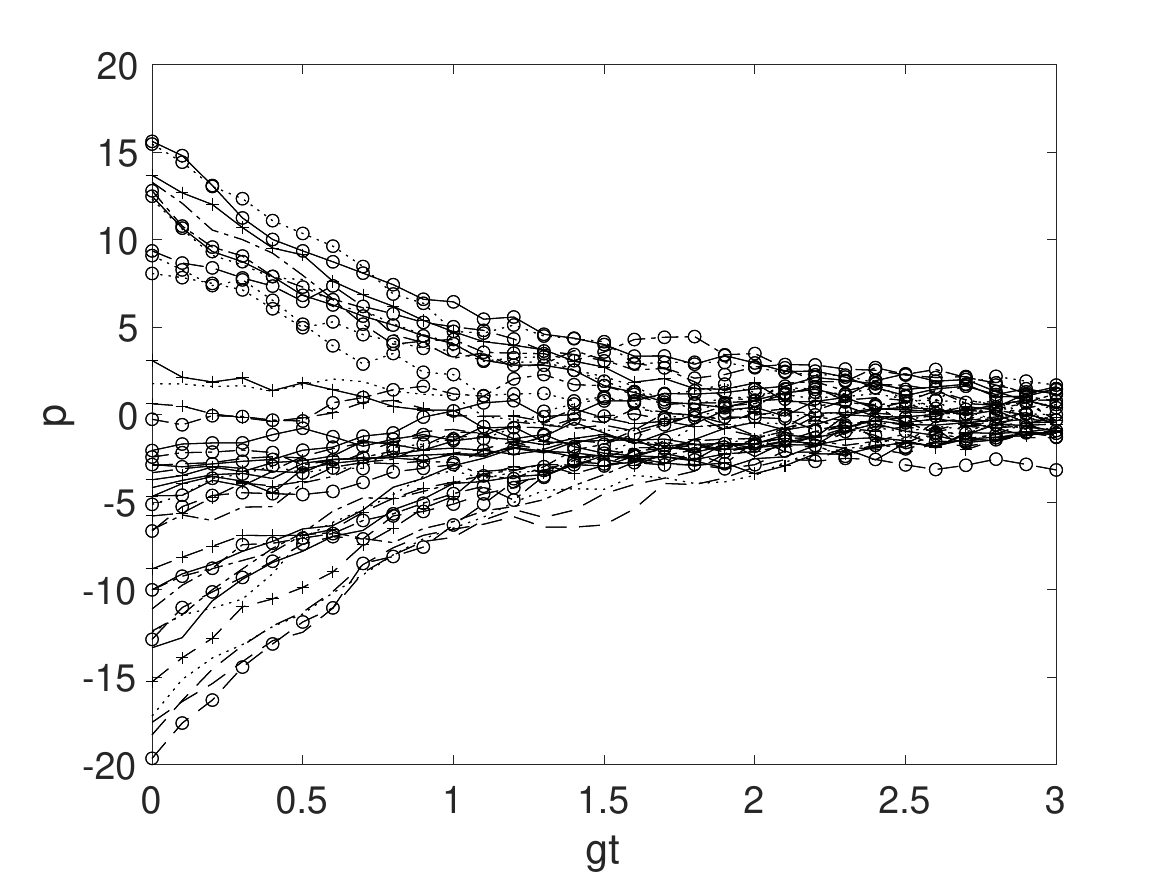}
\par\end{centering}
\caption{Solutions of the forward-backward equations (\ref{eq:forwardSDE-2-1-1-1})
(top) and (\ref{eq:backwardSDE-2-1-1-1}) (lower) , modeling the measurement
of $\hat{x}$ on a system prepared in the state (Eq. (\ref{eq:sq-state-x})),
where $x_{1}=3$. The figures are for the measurement on the highly
squeezed state modeling the eigenstate $|x_{1}\rangle$ of $\hat{x}$,
where $r=3$. The plots are generated with $10^{6}$ trajectories.\textcolor{red}{\label{fig:fb-1}}\textcolor{green}{}}
\end{figure}

\begin{figure}
\begin{centering}
\negthickspace{}
\par\end{centering}
\begin{centering}
\includegraphics[width=0.8\columnwidth]{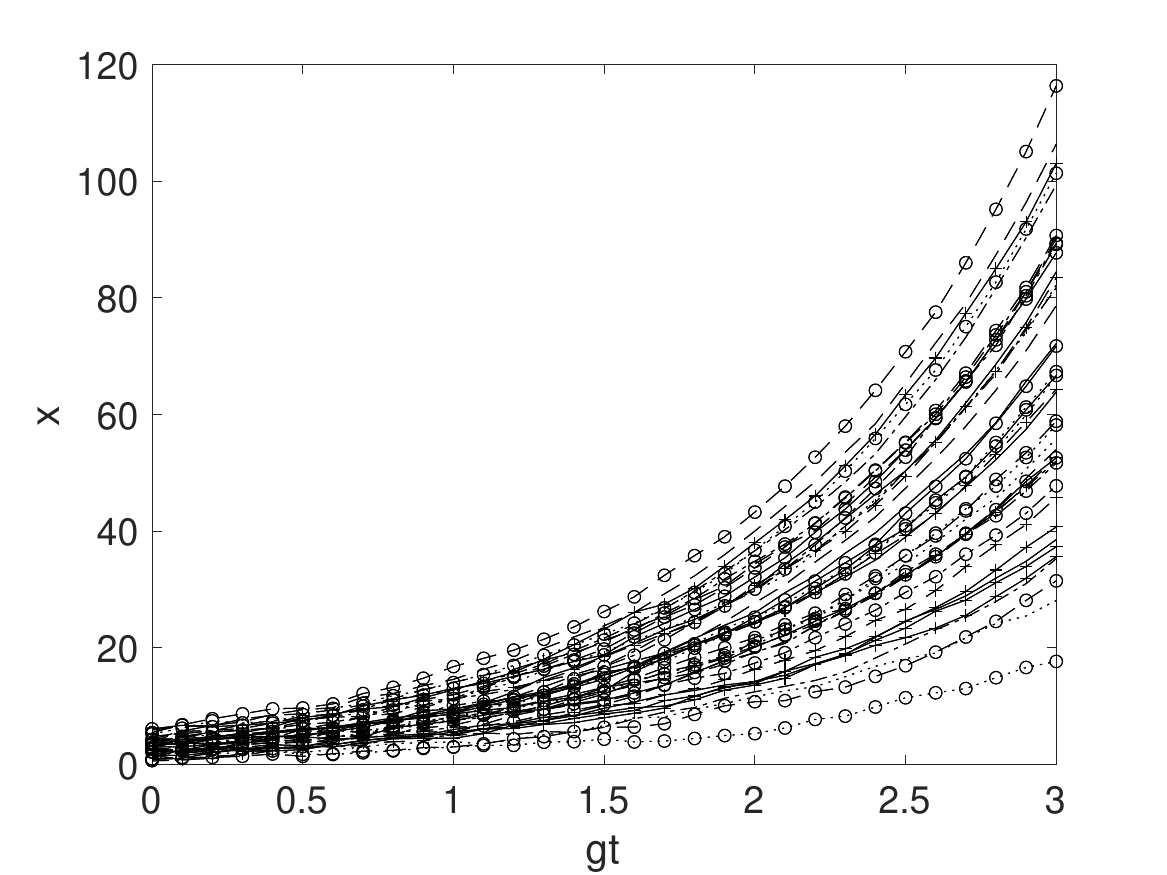}
\par\end{centering}
\begin{centering}
\negthickspace{}\includegraphics[width=0.8\columnwidth]{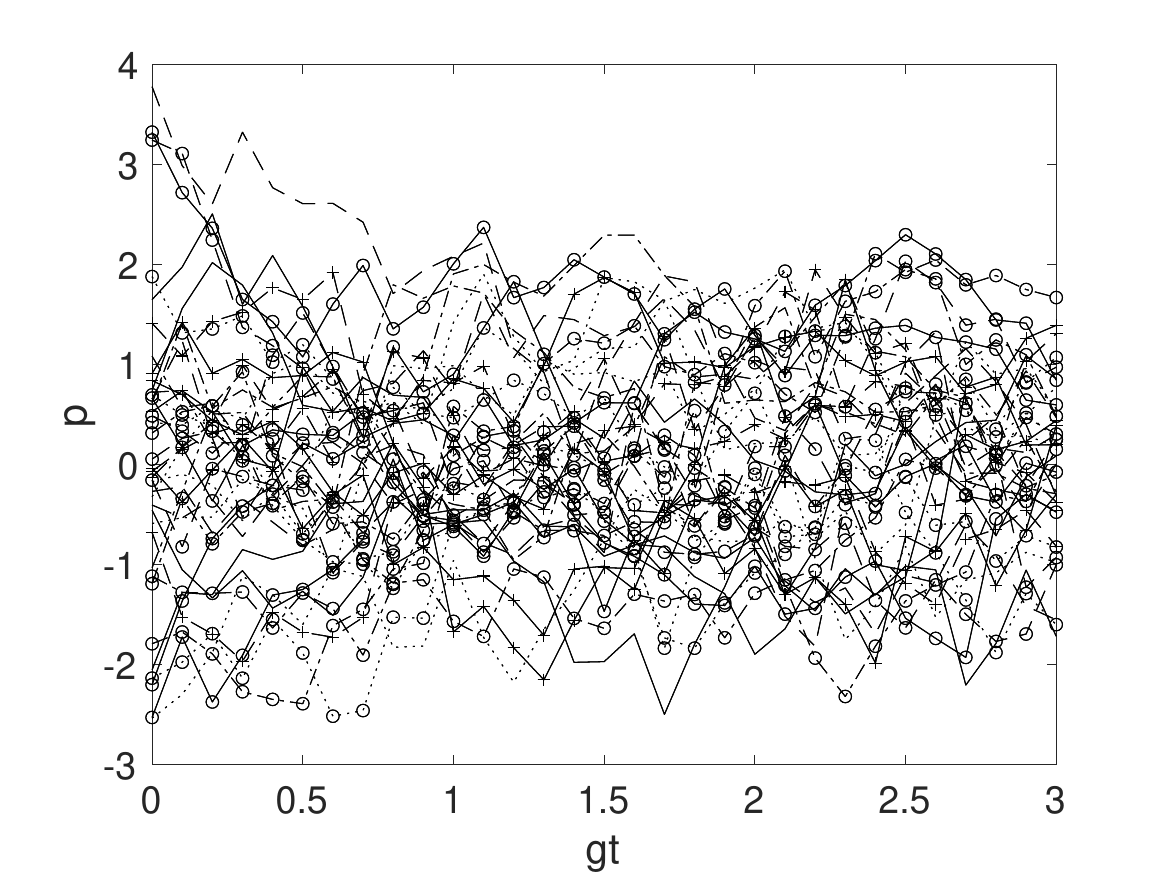}
\par\end{centering}
\caption{As for \ref{fig:fb-1}, where $x_{1}=3$. Here, the two figures solve
for the measurement on a coherent state $|\alpha_{0}\rangle$, where
$\alpha_{0}=1.5$ and $r=0$. The plots are generated with $10^{6}$
trajectories.\textcolor{red}{\label{fig:fb-1-1}}\textcolor{green}{}}
\end{figure}
We follow the techniques of Ref\textcolor{black}{s. \citep{q-measurement,q-retrocausal-model-measurement,drummond-time}},
extending the results to solve for a measurement on the state $|x_{1}/2,r\rangle$,
where $r$ is general and can represent a coherent state. The amplified
state after evolution for a time $t$ is 
\begin{equation}
e^{-iH_{amp}t/\hbar}|x_{1}/2,r\rangle=|Gx_{1}/2,r'\rangle\label{eq:amp-state}
\end{equation}
where $r'=-gt+r$ and $G(t)=e^{gt}$. The Q function for this state
is
\begin{equation}
Q(x,p,t)=\frac{1}{2\pi\sigma_{x}(t)\sigma_{p}(t)}e^{-(x-G(t)x_{1})^{2}/2\sigma_{x}^{2}(t)}e^{-p^{2}/2\sigma_{p}^{2}(t)}\label{eq:q-eigen-2}
\end{equation}
where\textcolor{black}{
\begin{eqnarray}
\sigma_{x}^{2}(t) & = & 1+e^{-2r+2gt}\nonumber \\
\sigma_{p}^{2}(t) & = & 1+e^{2r-2gt}\label{eq:amp-var}
\end{eqnarray}
The variances can be rewritten as $\sigma_{x}^{2}\left(t\right)=1+G(t)^{2}\left[\sigma_{x}^{2}\left(0\right)-1\right]$}
and $\sigma_{p}^{2}(t)=1+[\sigma_{p}^{2}(0)-1]/G(t)^{2}$. The hidden
noise ($\sigma_{x}^{2}(0)=1$) associated with the eigenstate is not
amplified but the noise in $p$ decays, according to $\sigma_{p}^{2}(t)\rightarrow1$
as $G(t)\rightarrow\infty$ (refer to Figure \ref{fig:fb-1}). The
marginal 
\begin{equation}
Q(x,t_{f})=\int Q(x,p,t_{f})dp\label{eq:marg-x}
\end{equation}
given by $Q(x,t_{f})=e^{-(x-Gx_{j})^{2}/2\sigma_{x}^{2}(t_{f})}/\sigma_{x}(t_{f})\sqrt{2\pi}$
determines the future boundary condition for the backward equation
(\ref{eq:backwardSDE-2-1-1-1}). The marginal 
\begin{equation}
Q(p,t_{0})=\int Q(x,p,t_{0})dx\label{eq:marg-p}
\end{equation}
given by $Q(p,t_{0})=e^{-p^{2}/2\sigma_{p}^{2}(0)}/\sigma_{p}(t_{0})\sqrt{2\pi}$
determines the boundary condition of the forward equation (\ref{eq:forwardSDE-2-1-1-1}).

Plots of the solutions for the eigenstate $|x_{1}\rangle$ where $r$
is large, and the coherent state $|\alpha\rangle$ where $r=0$ and
$\alpha=x_{1}/2$, are given in Figure \ref{fig:fb-1}. The mean value
$x_{j}$ is amplified to $Gx_{j}$ in each case. The noise level $\sigma_{x}(t)$
for the eigenstate is constant (at $\sigma_{x}(t)=1$) along the trajectory
for $x$. The noise level for the coherent state begins at $\sigma_{x}(t_{0})=2$
but amplifies to $\sigma_{x}(t_{f})\rightarrow1+G(t_{f})^{2}$.

\subsection{Measurement on a superposition and cat-state}

We next examine the measurement on a simple Schrodinger-cat state,
a superposition of two single-mode squeezed or coherent states. We
first follow\textcolor{red}{{} }\citep{q-measurement} and consider
measurement on a superposition of squeezed states. We assume the system
is prepared in the state
\begin{eqnarray}
|\psi_{S}\rangle & = & N(c_{1}|\frac{x_{1}}{2},r\rangle+c_{2}|\frac{x_{2}}{2},r\rangle)\label{eq:sup-sq}
\end{eqnarray}
We will take $x_{2}=-x_{1}$ and $|c_{1}|=|c_{2}|=1/\sqrt{2}$ and
$N$ is a normalization constant, necessary when $|x_{1}/2,r\rangle$
and $|x_{2}/2,r\rangle$ are not orthogonal. We will take $c_{1}$
to be real and $c_{2}=|c_{2}|e^{i\varphi}$, so that $\varphi$ is
the phase factor associated with the superposition. The Q function
of (\ref{eq:sq-sup}) is \textcolor{green}{ }%
\begin{eqnarray}
Q(x,p,t_{0}) & = & N\frac{e^{-p^{2}/2\sigma_{p}^{2}}}{2\pi\sigma_{x}\sigma_{p}}\Bigr(|c_{1}|^{2}e^{-(x-x_{1})^{2}/2\sigma_{x}^{2}}\nonumber \\
 &  & +|c_{2}|^{2}e^{-(x-x_{2})^{2}/2\sigma_{x}^{2}}+2|c_{1}c_{2}|I\Bigl)\nonumber \\
\label{eq:Q-sup-1}
\end{eqnarray}
with $\sigma_{x}^{2}=1+e^{-2r}$ and $\sigma_{p}^{2}=1+e^{2r}$.
Here
\begin{eqnarray}
I & = & e^{-[(x-x_{1})^{2}+(x-x_{2})^{2}]/4\sigma_{x}^{2}}F\label{eq:I}
\end{eqnarray}
where
\begin{equation}
F=\cos\left(\varphi+\frac{p}{2\sigma_{x}^{2}}(x_{1}-x_{2})\right)\label{eq:F}
\end{equation}
 and 
\begin{equation}
N=(1+2c_{1}|c_{2}|[\cos\varphi]e^{-\frac{G^{2}(x_{1}-x_{2})^{2}}{8\sigma_{x}^{2}}\{1+\frac{\sigma_{p}^{2}}{\sigma_{x}^{2}}\}})^{-1}\label{eq:norm2-3}
\end{equation}
which becomes $1$ for the superposition of eigenstates of $\hat{x}$,
where $r\rightarrow\infty$. The Q function is the sum of the two
Gaussian distributions associated with each squeezed state, as well
as a sinusoidal term arising due to the system being in a superposition.
This term vanishes for the $Q$ function of the mixed state
\begin{equation}
\rho_{mix}=|c_{1}|^{2}|x_{1}\rangle\langle x_{1}|+|c_{2}|^{2}|x_{2}\rangle\langle x_{2}|\label{eq:Qmix}
\end{equation}

\begin{figure}
\begin{centering}
\includegraphics[width=0.8\columnwidth]{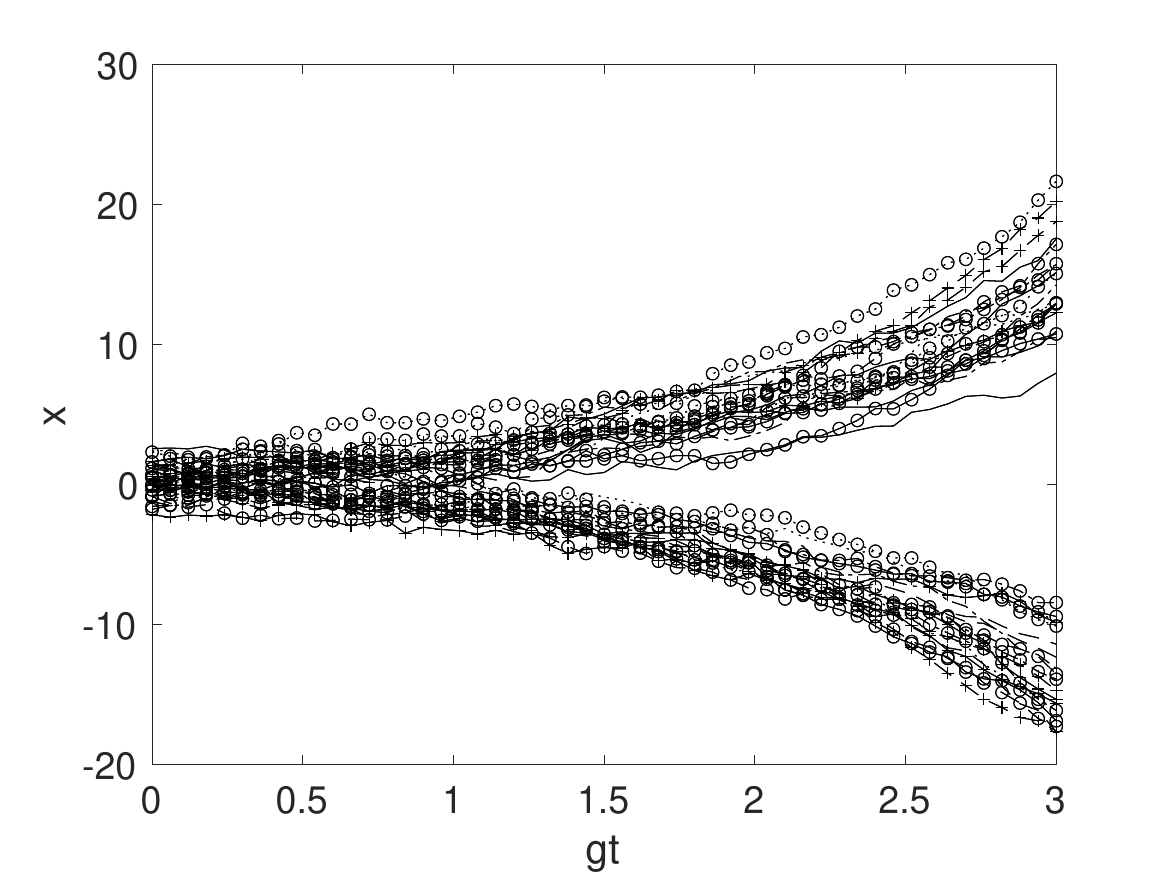}
\par\end{centering}
\begin{centering}
\negthickspace{}\includegraphics[width=0.8\columnwidth]{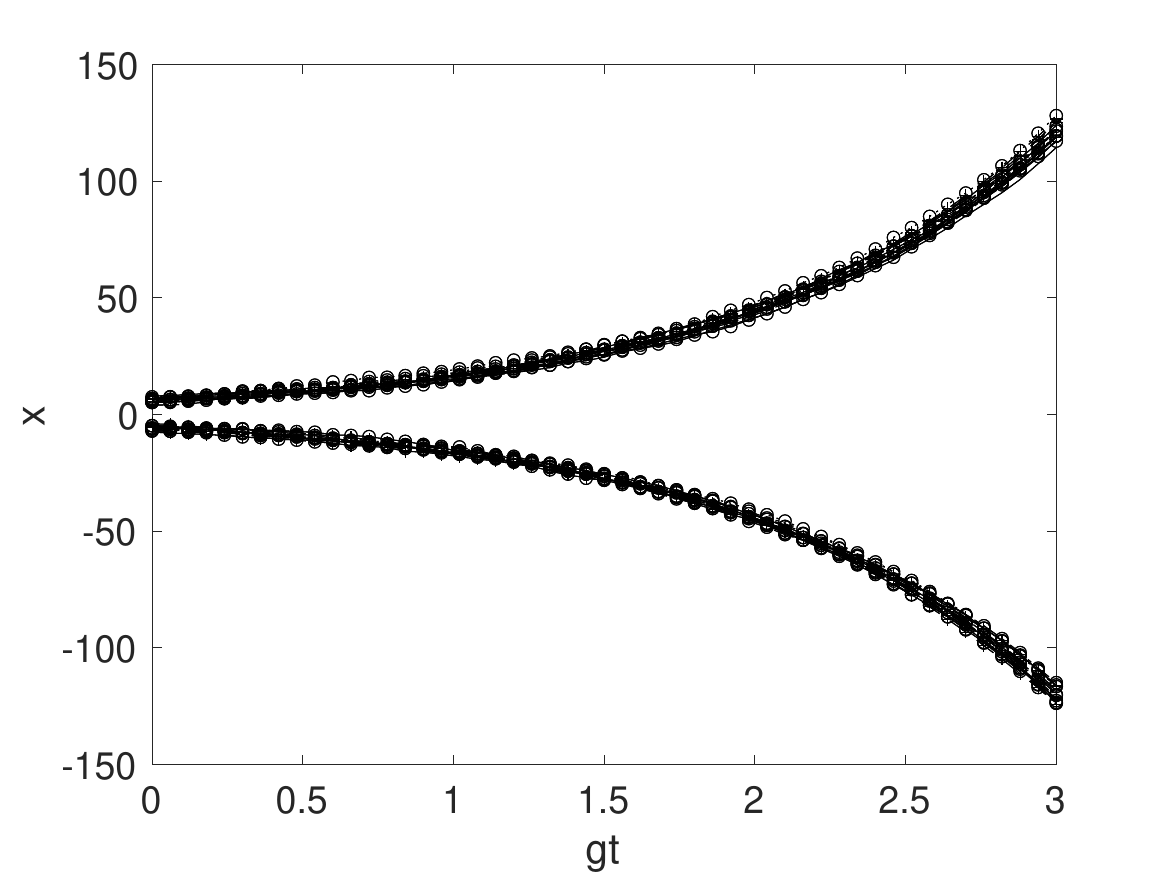}
\par\end{centering}
\caption{Forward-backward stochastic solutions modeling the measurement of
$\hat{x}$ on a system prepared in a superposition given by $|\psi_{S}\rangle$
(Eq. (\ref{eq:sup-sq})) with $c_{1}=-ic_{2}=1/\sqrt{2}$. we choose
$r=2$ which models measurement on a superposition of eigenstates
of $\hat{x}$ (Eq. (\ref{eq:sup-eigen})). The top plot shows $x_{1}=0.7$
and $r=2$, which models measurement on a microscopic superposition
of eigenstates of $\hat{x}$. The lower plot shows $x_{1}=6$ and
$r=2$ which models measurement on a macroscopic superposition of
eigenstates. \textcolor{red}{}\textcolor{green}{{} }Plots show $10^{6}$
trajectories. $t_{f}=3/g$. \label{fig:sup}\textcolor{magenta}{}\textcolor{green}{}}
\end{figure}
We suppose $\hat{x}$ is measured by the amplification process modeled
by the interaction $H_{amp}$ {[}(\ref{eq:hamb-1}){]}. Following
\citep{q-measurement}, the state of the system after amplification
\begin{eqnarray}
|\psi_{S}(t)\rangle & = & N_{t}(c_{1}|Gx_{1},r'\rangle+c_{2}|Gx_{2},r'\rangle)\nonumber \\
\label{eq:amp-sup-3}
\end{eqnarray}
where $G=e^{gt}$ is the amplification factor, $r'=-gt+r$ and $t$
is the time of evolution. The Q function $Q(x,p,t)$ at the time $t$
is readily evaluated. We find
\begin{eqnarray}
Q(x,p,t) & = & N_{t}\frac{e^{-p^{2}/2\sigma_{p}^{2}(t)}}{2\pi\sigma_{x}(t)\sigma_{p}(t)}\Bigr(|c_{1}|^{2}e^{-(x-G(t)x_{1})^{2}/2\sigma_{x}^{2}(t)}\nonumber \\
 &  & +|c_{2}|^{2}e^{-(x-G(t)x_{2})^{2}/2\sigma_{x}^{2}(t)}+2|c_{1}c_{2}|I(t)\Bigl)\nonumber \\
\label{eq:Q-sup-1-1}
\end{eqnarray}
where $G(t)=e^{gt}$, and
\begin{eqnarray}
I(t) & = & e^{-[(x-G(t)x_{1})^{2}+(x-G(t)x_{2})^{2}]/4\sigma_{x}^{2}(t)}F\label{eq:It}
\end{eqnarray}
wher\textcolor{black}{e
\begin{equation}
F(t)=\cos\left(\varphi+\frac{G(t)p}{2\sigma_{x}^{2}(t)}(x_{1}-x_{2})\right)\label{eq:Ft}
\end{equation}
}$N_{t}$ is the normalization constant given by $N$ on replacing
$G$ with $G(t)$, $\sigma_{x}$ with $\sigma_{x}(t)$ and $\sigma_{p}$
with $\sigma_{p}(t)$.\textcolor{black}{{} The variances $\sigma_{x}(t)$
and $\sigma_{p}(t)$ of the amplified state are as for Eq. (}\ref{eq:amp-var}):
\textcolor{black}{
\begin{eqnarray}
\sigma_{x}^{2}\left(t\right) & = & 1+G(t)^{2}\left[\sigma_{x}^{2}\left(0\right)-1\right]\nonumber \\
\sigma_{p}^{2}(t) & = & 1+[\sigma_{p}^{2}(0)-1]/G(t)^{2}\label{eq:var-gen-cat}
\end{eqnarray}
}

The boundary condition for the backward stochastic equation (\ref{eq:backwardSDE-2-1-1-1})
is determined by the marginal $Q(x,t_{f})$ {[}(\ref{eq:marg-x}){]}.
We find
\begin{eqnarray}
Q(x,t) & = & N_{t}\frac{1}{\sqrt{2\pi}\sigma_{x}(t)}\Bigr(|c_{1}|^{2}e^{-(x-G(t)x_{1})^{2}/2\sigma_{x}^{2}(t)}\nonumber \\
 &  & +|c_{2}|^{2}e^{-(x-G(t)x_{2})^{2}/2\sigma_{x}^{2}(t)}+2|c_{1}c_{2}|I(x,t)\Bigl)\nonumber \\
\label{eq:Qmarg-x-amp}
\end{eqnarray}
where
\begin{eqnarray}
I(x,t) & = & [\cos\varphi]\{e^{-[(x-G(t)x_{1})^{2}+(x-G(t)x_{2})^{2}]/4\sigma_{x}^{2}(t)}\nonumber \\
 &  & e^{-G(t)^{2}(x_{1}-x_{2})^{2}\sigma_{p}^{2}(t)/8\sigma_{x}^{4}(t)}\}\label{eq:int-marg}
\end{eqnarray}
We simplify for the case $x_{1}=-x_{2}$
\begin{eqnarray}
I(x,t) & = & [\cos\varphi]\{e^{-x^{2}/2\sigma_{x}^{2}(t)-G^{2}(t)x_{1}^{2}/2\sigma_{x}^{2}(t)}\}\nonumber \\
 &  & e^{-G(t)^{2}x_{1}^{2}\sigma_{p}^{2}(t)/4\sigma_{x}^{4}(t)}\}\label{eq:int-marg-1}
\end{eqnarray}
Note that where $\cos\varphi\neq0$, there is an interference peak
centered at $x=0$ which vanishes in the limit of large $G(t)$.

The boundary condition for the forward stochastic equation (\ref{eq:forwardSDE-2-1-1-1})
is determined by the marginal $Q(p,t_{0})$ {[}(\ref{eq:marg-p}){]}.
We find
\begin{eqnarray}
Q(p,t_{0}) & = & N_{t}\frac{e^{-p^{2}/2\sigma_{p}^{2}(t)}}{\sqrt{2\pi}\sigma_{p}(t)}[1+2|c_{1}c_{2}|e^{-\frac{(x_{1}-x_{2})^{2}}{8\sigma_{x}^{2}(t)}}F]\nonumber \\
\label{eq:pmarg}
\end{eqnarray}

\begin{figure}
\begin{centering}
\includegraphics[width=0.8\columnwidth]{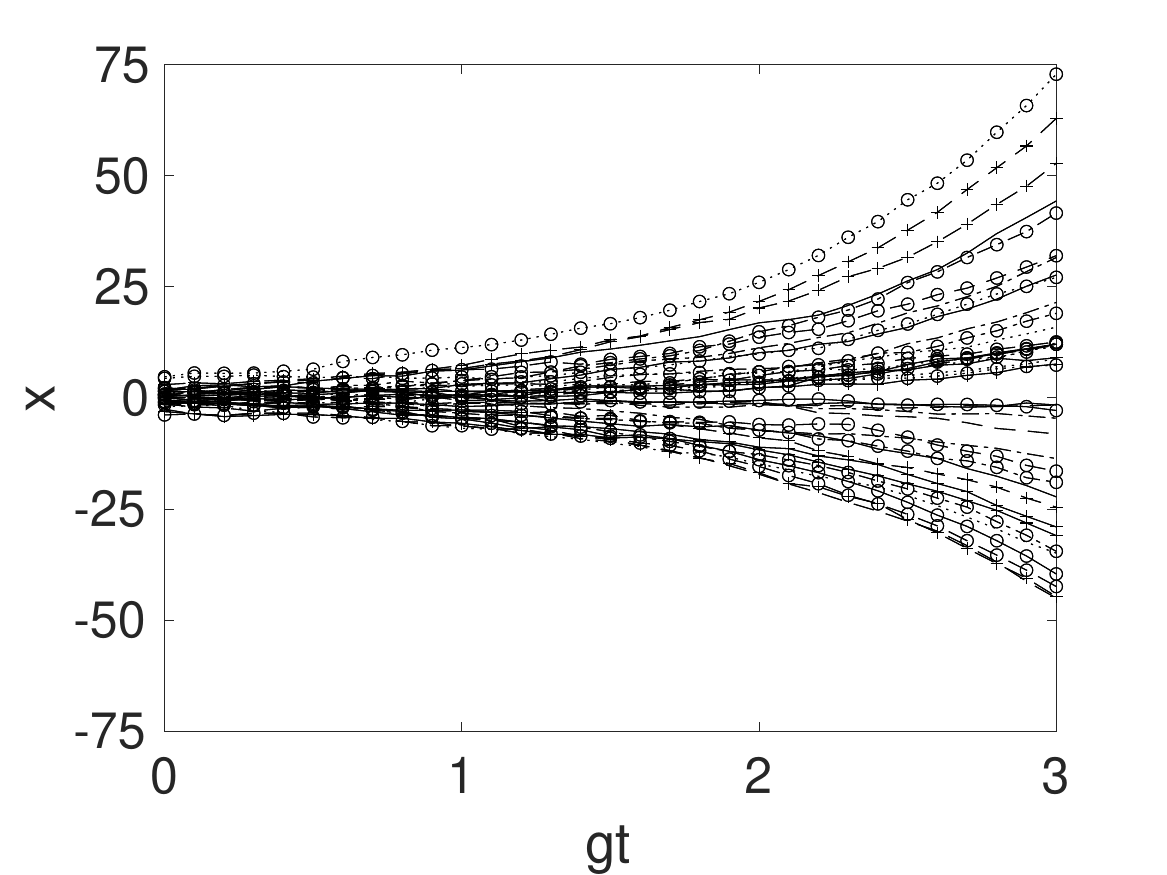}
\par\end{centering}
\begin{centering}
\includegraphics[width=0.8\columnwidth]{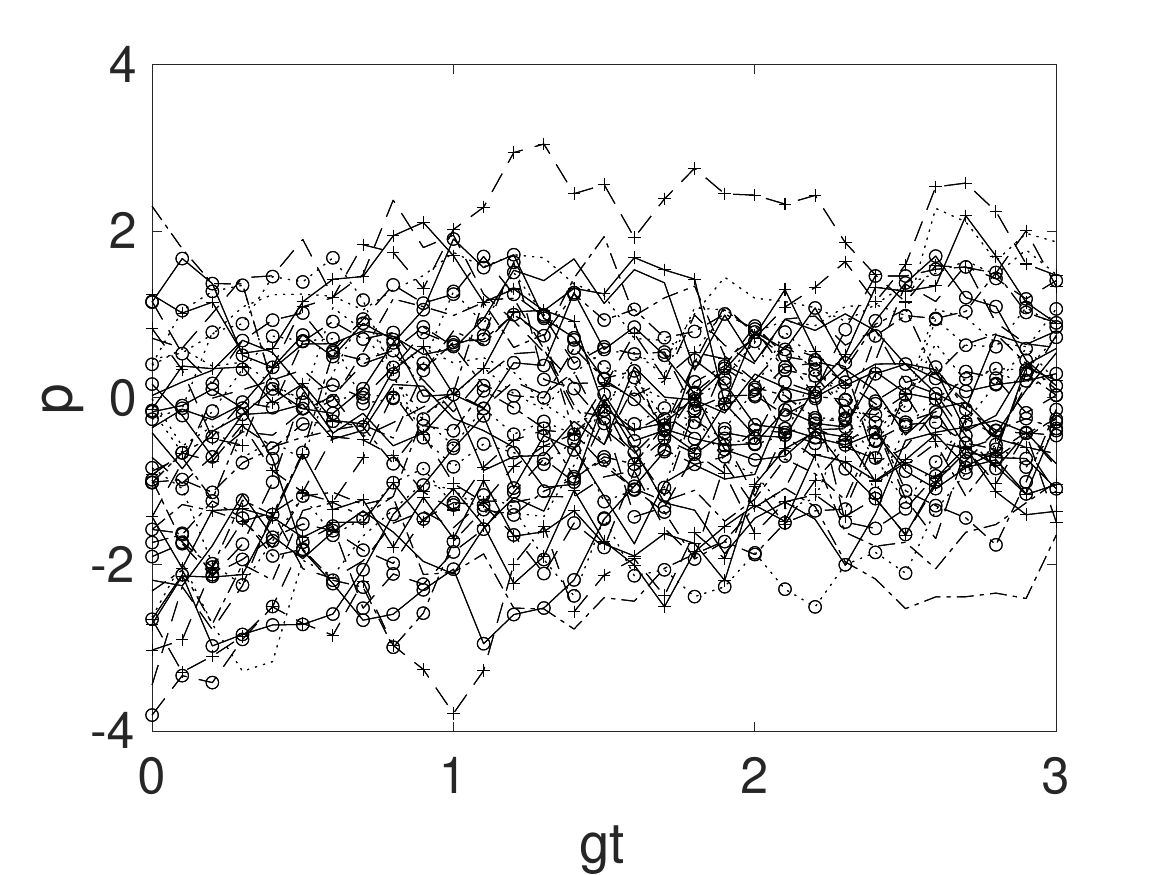}
\par\end{centering}
\caption{Forward-backward stochastic solutions modeling the measurement of
$\hat{x}$ on a system prepared in a state $|\psi_{cat}\rangle$ {[}Eq.
(\ref{eq:sup-cat}){]}, a superposition of coherent states as given
by $|\psi_{S}\rangle$ (Eq. (\ref{eq:sup-sq})) with $c_{1}=-ic_{2}=1/\sqrt{2}$,
$r=0$ and $\varphi=\pi/2$. The plots are for $\alpha_{0}=0.5$ and
$r=0$, which models measurement on a microscopic superposition of
coherent states. Plots show $10^{6}$ trajectories. $t_{f}=3/g$.
\label{fig:sup-2-1}\textcolor{green}{{} }\textcolor{red}{}}
\end{figure}

\begin{figure}
\begin{centering}
\includegraphics[width=0.8\columnwidth]{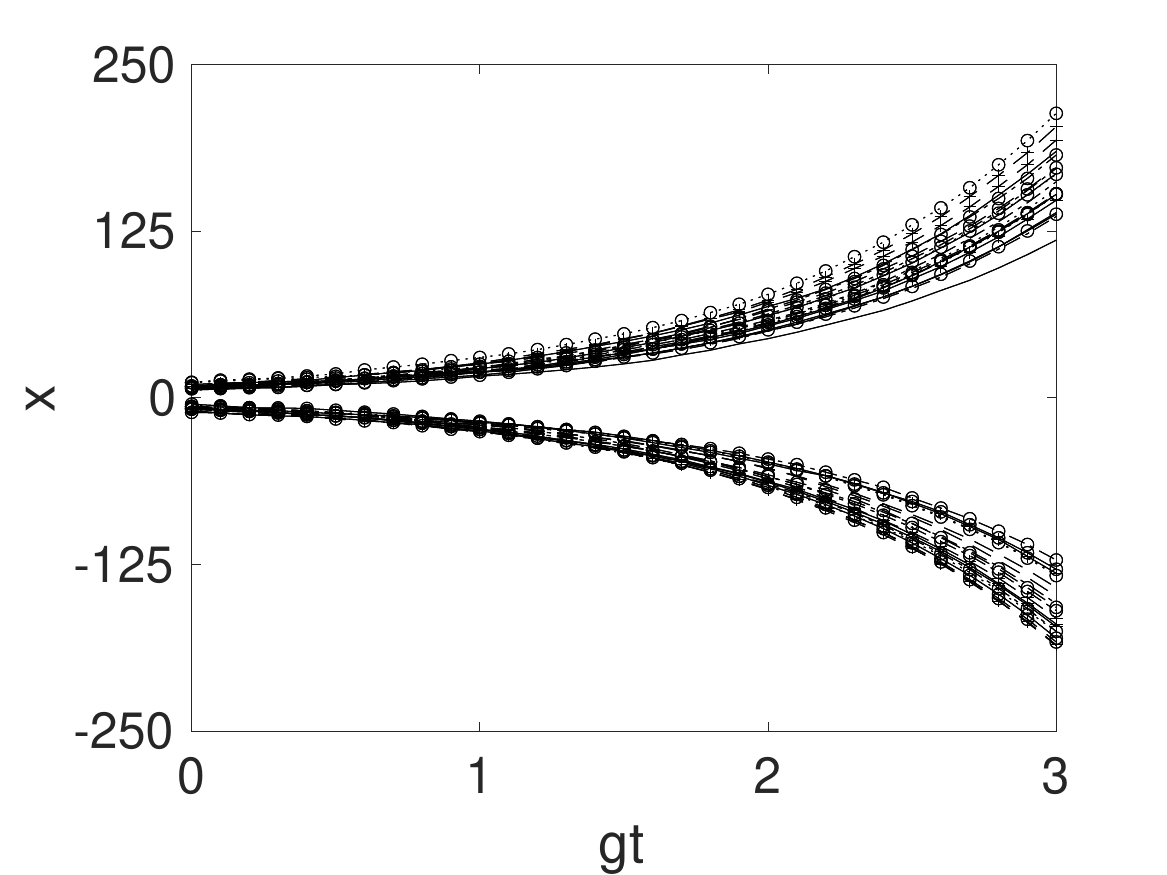}
\par\end{centering}
\begin{centering}
\includegraphics[width=0.8\columnwidth]{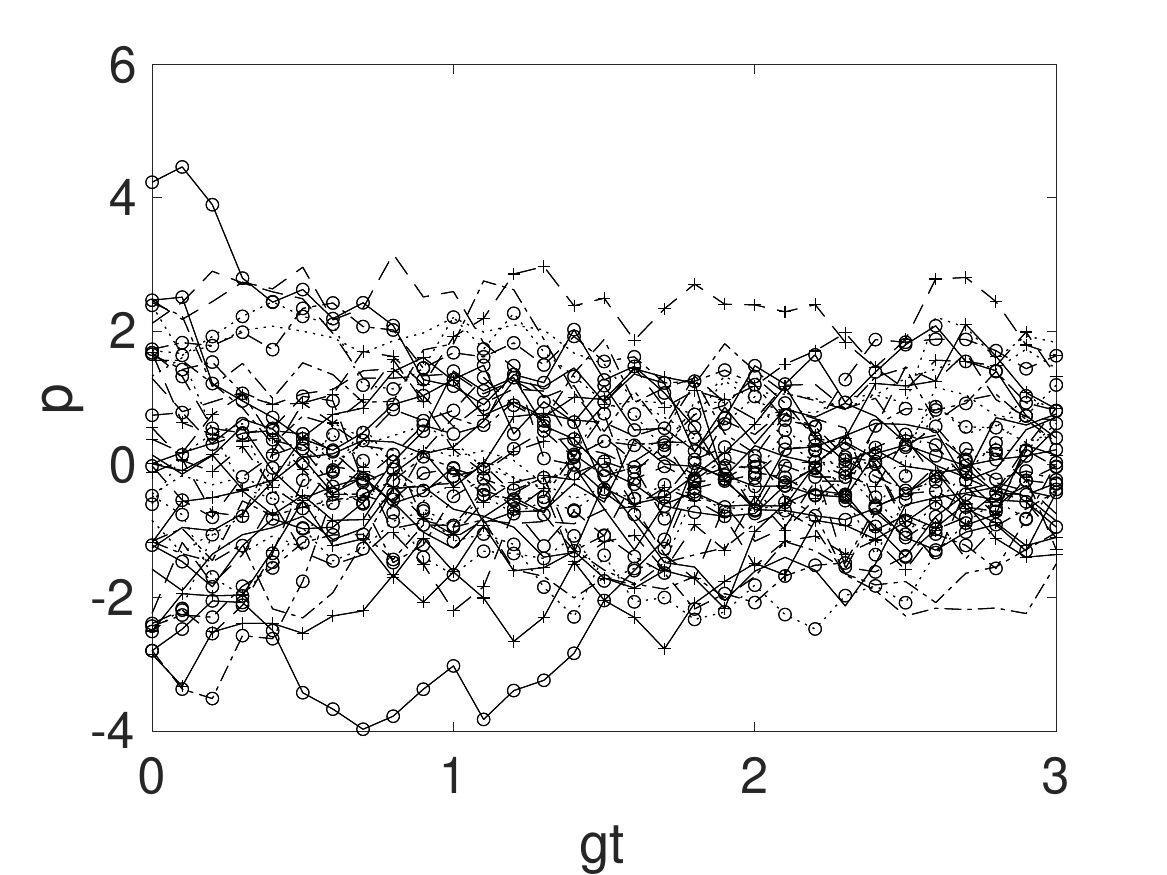}
\par\end{centering}
\caption{As for Figure \ref{fig:sup-2-1}. The plots are for $\alpha_{0}=4$
and $r=0$ and $\varphi=\pi/2$, which models measurement on a cat-state
{[}Eq. (\ref{eq:sup-cat}){]} which is a macroscopic superposition
of coherent states. Plots show $1.2\times10^{6}$ trajectories. $t_{f}=3/g$.
\label{fig:sup-2}\textcolor{green}{{} }\textcolor{red}{}\textcolor{green}{}}
\end{figure}

\begin{figure}
\begin{centering}
\includegraphics[width=0.8\columnwidth]{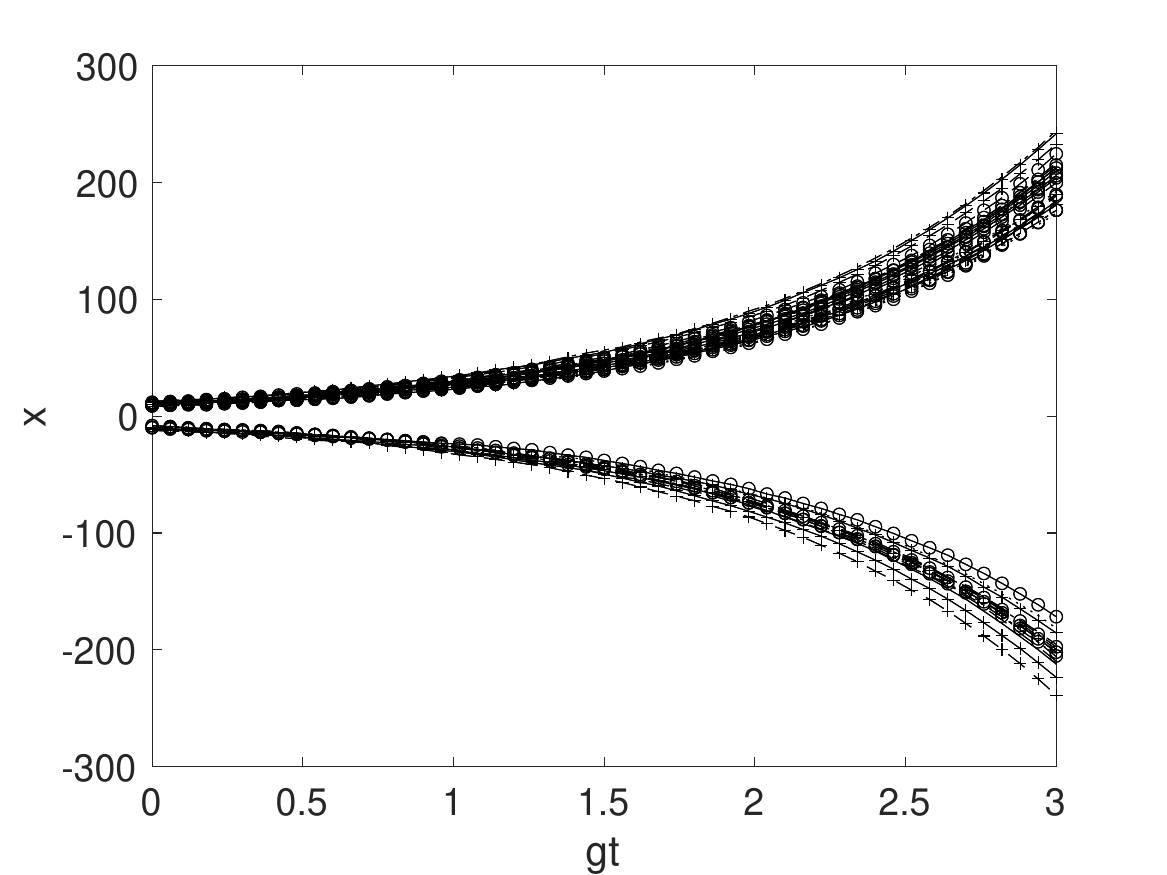}
\par\end{centering}
\begin{centering}
\includegraphics[width=0.8\columnwidth]{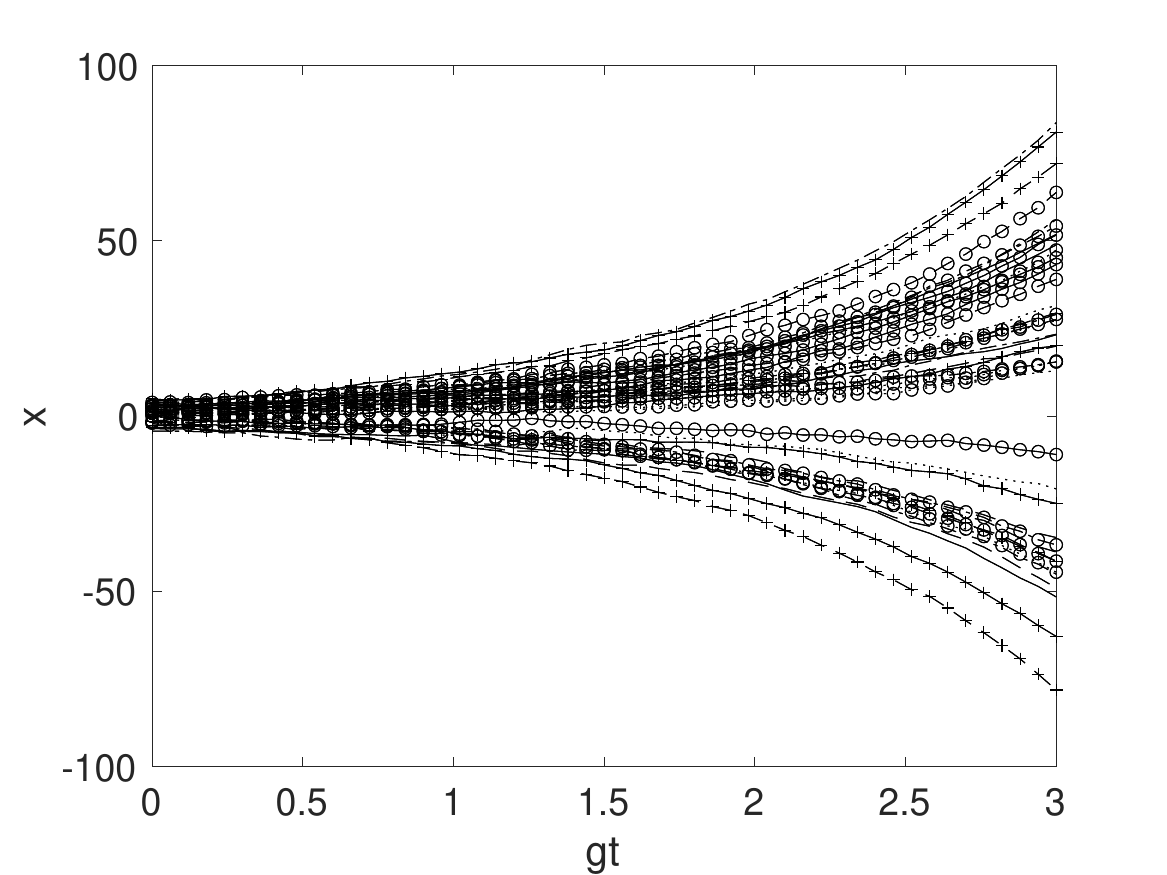}
\par\end{centering}
\caption{As for Figure \ref{fig:sup-2-1}. Measurement of $\hat{x}$ on a cat-state
{[}Eq. (\ref{eq:sup-cat}){]} (corresponding to $|\psi_{S}\rangle$
{[}Eq. (\ref{eq:sup-sq}){]} with $c_{1}=-ic_{2}=1/\sqrt{2}$, $r=0$
and $\varphi=\pi/2$) showing the transition where the separation
of the two coherent states becomes detectable. The plots show that
the variance {[}$(\Delta\hat{x})^{2}=1${]} associated with a coherent
state is amplified. The top plot shows $x_{1}=10$ ($\alpha_{0}=5$)
where the coherent states are well separated. The second plot shows
$x_{1}=2$ ($\alpha_{0}=1$), where the ``hidden'' noise $\sigma_{x}^{2}=1$
in the Q function ensures that the two peaks associated with the coherent
states $|\pm\alpha_{0}\rangle$ overlap at $t=0$. There is no overlap
however in the amplified distribution at time $t_{f}$, for this value
of $\alpha_{0}$. Here $t_{f}=3/g$.\label{fig:x-trajectories-cat-superposition-2}}
\end{figure}
Solutions of the forward-backward equations are plotted in Figures
\ref{fig:sup-2-1} and \ref{fig:sup-2}. We examine the superposition
\begin{equation}
|\psi_{S}\rangle=\frac{1}{\sqrt{2}}(|x_{1}\rangle+i|-x_{1}\rangle)\label{eq:sup-eigen}
\end{equation}
of two eigenstates $|x_{1}\rangle$ and $|-x_{1}\rangle$ by taking
$r$ to be large. The case of the cat-state (\ref{eq:cat}) defined
by \citep{yurke-stoler-1} 
\begin{equation}
|\psi_{cat}\rangle=\frac{1}{\sqrt{2}}(|\alpha_{0}\rangle+i|-\alpha_{0}\rangle)\label{eq:sup-cat}
\end{equation}
is treated by taking $r=0$. Here, $\varphi=\pi/2$ so that $\cos\varphi=0$,
and the sampling for the simulation is handled as for the mixed state
(\ref{eq:Qmix}), which has the same boundary condition.

Considering the cat-state
\begin{equation}
|\psi_{cat}\rangle=N_{0}(|\alpha_{0}\rangle+|-\alpha_{0}\rangle)\label{eq:sup-real-cat}
\end{equation}
\textcolor{red}{}where $N_{0}=\frac{1}{\sqrt{2\left(1+e^{-2|\alpha_{0}|^{2}}\right)}}$
, we see that $\cos\varphi=0$. The interference term appearing in
the boundary condition at finite $t$ needs to be taken into account.

\subsubsection*{Future boundary condition for the general superposition state: Wigner
function}

There are two equivalent approaches to solve the backward equation
(\ref{eq:backwardSDE-2-1-1-1}) that relate to the future boundary
condition (FBC) at the time\textcolor{black}{{} $t_{f}$. T}he first
is to sample from the distribution (e.g. Gaussian (\ref{eq:q-sq}))
directly \citep{q-retrocausal-model-measurement,drummond-time}. This
applies for all values of $r$, and is the method used to generate
the plots in Figures \ref{fig:fb-1}-\ref{fig:sup-2}, including for
the superposition of coherent states. The second approach, explained
in \citep{q-measurement}, is based on the Wigner function of the
state $|\psi_{S}\rangle$. This method is useful in treating the more
general superposition state, where $\varphi\neq\pi/2$, and is also
useful in establishing a causal model for the measur\textcolor{black}{ement
}\citep{q-measurement}\textcolor{black}{. Here}, we extend and apply
this method to carry out simulations of the general superposition
with $\varphi=0$ and to demonstrate the simulation of the superposition
of coherent states, the cat-state (\ref{eq:cat}).

The procedure of the second method is to expand the Q function of
the state being measured in terms of the set of Q functions that represent
the measurement basis \citep{q-measurement}. In this case, the measurement
is $\hat{x}$ and the measurement basis is the set of eigenstates
$|x_{j}\rangle$ of $\hat{x}$. The Q function $Q(x,p)$ of the eigenstate
$|x_{j}\rangle$ is
\begin{equation}
Q_{j}(x,p)=\frac{1}{2\pi\sigma_{x}\sigma_{p}}e^{-(x-x_{j})^{2}/2}e^{-p^{2}/2\sigma_{p}^{2}}\label{eq:q-eigen-3}
\end{equation}
where the variances are $\sigma_{x}^{2}=1+e^{-r}$ and $\sigma_{p}^{2}=1+e^{2r}$,
and we take $r\rightarrow\infty$. With this in mind, we write a general
state at time $t_{0}$ as 
\begin{equation}
|\psi\rangle=\sum_{i}c_{i}|x_{i}\rangle\label{eq:expsup}
\end{equation}
($c_{i}$ are probability amplitudes) for which the Q function can
be written
\begin{equation}
Q(x,p,t_{0})=\sum_{j}|c_{j}|^{2}Q_{j}(x,p)+\sum_{jk}I_{jk}\label{eq:Q-sum-sup}
\end{equation}
where $I_{jk}$ are interference terms. Now, from (\ref{eq:expsup}),
the probability for an outcome $x_{j}$ is $|c_{j}|^{2}$. In the
limit of a continuous spectrum, it is known that the distribution
for $|c_{j}|^{2}$, which becomes the probability density $P(x)$
for outcomes of $\hat{x}$, is given by the marginal $W(x)$ of the
Wigner function $W(x,p)$, so that $P(x)=W(x)\equiv\int dpW(x,p)$.

The Q function of the amplified state is denoted $Q(x,p,t_{0})$.
If we consider the marginal $Q(x,t)=\int Q(x,p,t)dp$, then it has
been pointed out that since (for the expansion with respect to the
eigenstates $|x_{j}\rangle$) $\text{\ensuremath{\sigma_{p}^{2}\rightarrow\infty} }$,
and $G(t)\rightarrow\infty$ for large $t_{f}$, the terms of type
$I$ in Eq. (\ref{eq:int-marg-1}) due to the interference $I_{jk}$
in (\ref{eq:Q-sum-sup}) will van\textcolor{black}{ish }\citep{q-measurement}\textcolor{black}{.
T}he future boundary condition is given as
\[
Q(x,t_{f})=\sum_{j}|c_{j}|^{2}Q_{j}(x,t_{f})
\]
where
\begin{equation}
Q_{j}(x,t)=\frac{1}{\sqrt{2\pi}}e^{-(x-G(t)x_{j})^{2}/2}\label{eq:QBC}
\end{equation}
is the marginal of (\ref{eq:q-eigen-2}) taking $\sigma_{x}=1$, as
for the eigenstate $|x_{j}\rangle$. Next, we define the scaled variable
\begin{equation}
x_{0}=x/G(t)\label{eq:scaled}
\end{equation}
which represents the value (\ref{eq:inferred-outcome}) inferred for
the measurement when $G(t)=e^{gt}$ is sufficiently large. In terms
of the scaled variables, the marginal $Q(x,t_{f})$ becomes
\begin{equation}
Q_{sc}(x_{0},t_{f})\rightarrow\sum_{j}|c_{j}|^{2}e^{-(x_{0}-x_{j})^{2}/2\widetilde{\sigma}^{2}}\label{eq:QBC-sc}
\end{equation}
which is a Gaussian distribution with mean $x_{j}$ and variance $\widetilde{\sigma}^{2}\rightarrow1/G^{2}(t)$
(which approaches $0$ as $t\rightarrow\infty$). In this limit, the
scaled function gives the probability of obtaining the outcome $x_{j}$
as $|c_{j}|^{2}$, which corresponds to the marginal $W(x)$ of the
Wigner function $W(x,p)$ of the state $|\psi\rangle$ \citep{wigner,wigner-dn}.
(We note here that care needs to taken with the multiple use of the
variable $x$ in the different contexts.)

Continuing, we see that the future boundary condition can hence be
obtained directly from $W(x,p)$. We establish the future boundary
condition, where $G=G(t_{f})=e^{gt_{f}}$, by rewriting the marginal
$W(x_{0})$ of the Wigner function in terms of the amplified variable
$x'=Gx_{0}$, as $W_{sc}(x')$, and evaluating the convolution with
the Gaussian function $e^{-(x-x')^{2}/2}$:
\begin{equation}
Q(x,t_{f})=\frac{1}{\sqrt{2\pi}}\int W_{sc}(x')e^{-(x-x')^{2}/2}dx'\label{eq:Q-W-FBC}
\end{equation}
This convolution brings back the ``hidden noise'' that exists in
the simulation, and the future boundary condition to be calculated
from the Wigner function of the initial state. We give examples below.

\subsubsection*{Examples}

First, we consider the coherent state $|\alpha_{0}\rangle$\textcolor{red}{{}
}\textcolor{black}{with}\textcolor{red}{{} }$x_{1}=2\alpha_{0}$. The
Wigner function is \textcolor{red}{}$W(x,p)=\frac{1}{2\pi}e^{-\frac{\left(x-x_{1}\right)^{2}}{2}}e^{-\frac{p^{2}}{2}}$.
The marginal $W(x)$ in $x$ is a simple Gaussian, which we write
as
\begin{equation}
W(x_{0})=\frac{1}{\sqrt{2\pi}}e^{-\frac{\left(x_{0}-x_{1}\right)^{2}}{2}}\label{eq:wig-coh}
\end{equation}
\textcolor{red}{}to avoid confusion with the multiples uses of the
notation $x$. Following Eq. (\ref{eq:Q-W-FBC}), we write in terms
of the scaled variable $x'=Gx_{0}$ ($G=e^{gt_{f}}$), as 
\begin{equation}
{\color{green}{\color{black}W_{sc}(x')}{\color{black}=}{\color{black}\frac{1}{\sqrt{2\pi G^{2}}}}{\color{black}e^{-\frac{\left(x^{'}/G-x_{1}\right)^{2}}{2}}}}\label{eq:wig-sc-coh}
\end{equation}
The future boundary condition is given as
\begin{align}
Q(x,t_{f}) & =\int W_{sc}(x')\frac{1}{\sqrt{2\pi}}e^{-\frac{\left(x-x^{'}\right)^{2}}{2}}dx^{'}\nonumber \\
 & =\frac{1}{\sqrt{2\pi\left(1+G^{2}\right)}}e^{-\frac{\left(x-Gx_{1}\right)^{2}}{2\left(1+G^{2}\right)}}\label{eq:q-sc-fbc}
\end{align}
where $G=e^{gt_{f}}$. We note this agrees with the solution (\ref{eq:Qmarg-x-amp})
on putting $c_{2}=0$. Since
\begin{equation}
Q(x,t_{f})=\frac{1}{\sqrt{2\pi\left(1+G^{2}\right)}}e^{-\frac{\left(x/G-x_{1}\right)^{2}}{2\left(1+1/G^{2}\right)}}\label{eq:re}
\end{equation}
we also see that
\begin{align}
\lim_{G\rightarrow\infty}Q_{sc}(x_{0},t) & =\frac{1}{\sqrt{2\pi}}e^{-\frac{(x_{0}-x_{1})^{2}}{2}}\nonumber \\
 & =W(x_{0})\label{eq:lim}
\end{align}
As required, the limiting behavior of the Q function in terms of the
scaled variable $x_{0}=x/G$ gives the Wigner function.

\begin{figure}
\begin{centering}
\negthickspace{}
\par\end{centering}
\begin{centering}
\includegraphics[width=0.8\columnwidth]{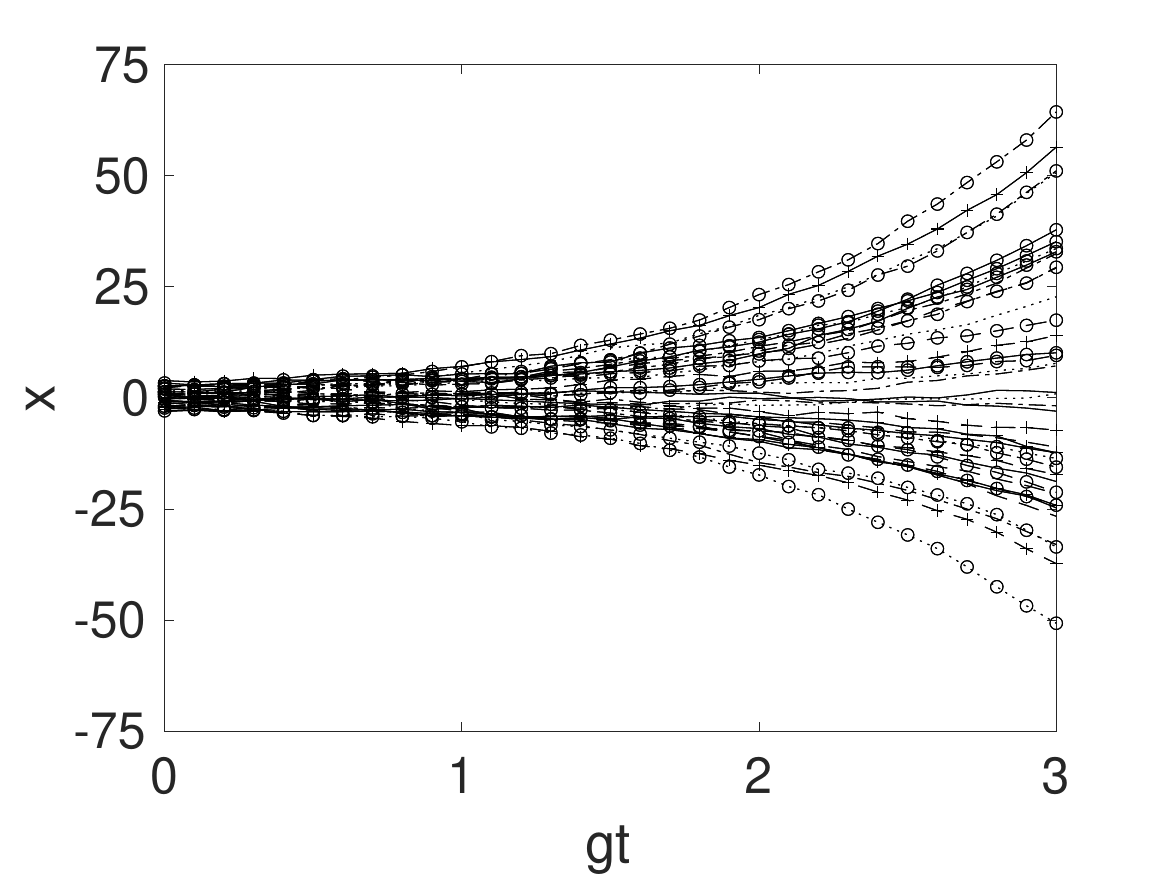}
\par\end{centering}
\begin{centering}
\includegraphics[width=0.8\columnwidth]{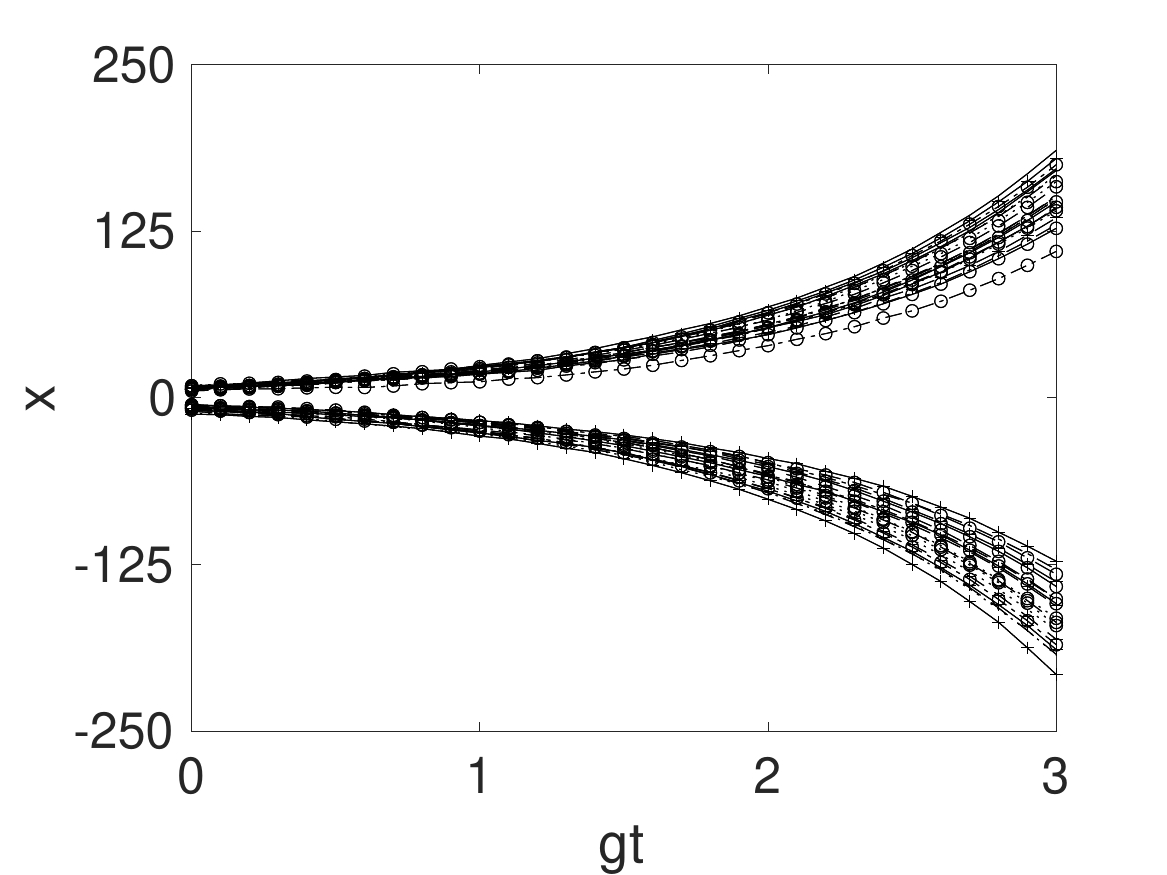}
\par\end{centering}

\caption{Forward-backward stochastic solutions modeling the measurement of
$\hat{x}$ on a system prepared in a superposition of coherent states,
as given by $|\psi_{S}\rangle$ (Eq. (\ref{eq:sup-sq})) with $c_{1}=-ic_{2}=1/\sqrt{2}$,
$r=0$ and $\varphi=\pi/2$. The top plot shows $\alpha_{0}=0.5$
and $r=0$, which models measurement on a microscopic superposition
of coherent states. The lower plot shows $\alpha_{0}=4$ and $r=0$
which models measurement on a macroscopic superposition of coherent
states. Plots show $1.2\times10^{6}$ trajectories and $t_{f}=3/g$.\textcolor{green}{\label{fig:wign-cat-im}}}
\end{figure}

Similarly, we consider the cat state $N_{0}(|\alpha_{0}\rangle+|-\alpha_{0}\rangle)$
of Eq. (\ref{eq:sup-real-cat}) where $\varphi=0$. The Wigner function
is
\begin{align}
W(x,p) & =\frac{e^{-\frac{p^{2}}{2}}}{4\pi\left(1+e^{-\frac{x_{1}^{2}}{2}}\right)}\left(e^{-\frac{\left(x-x_{1}\right)^{2}}{2}}\right.\nonumber \\
 & \left.+e^{-\frac{\left(x+x_{1}\right)^{2}}{2}}+2e^{-\frac{x^{2}}{2}}\cos\left(px_{1}\right)\right)\label{eq:wig-cat}
\end{align}
 which shows an interference term. The marginal is
\begin{align}
W(x_{0}) & =\frac{1}{2\sqrt{2\pi}\left(1+e^{-\frac{x_{1}^{2}}{2}}\right)}\left(e^{-\frac{\left(x_{0}-x_{1}\right)^{2}}{2}}\right.\nonumber \\
 & \left.+e^{-\frac{\left(x_{0}+x_{1}\right)^{2}}{2}}+2e^{-\frac{x_{0}^{2}}{2}}e^{-\frac{x_{1}^{2}}{2}}\right)\label{eq:wig-cat-marg}
\end{align}
which also shows an interference term. Following Eq. (\ref{eq:Q-W-FBC}),
we define
\begin{align}
W_{sc}(x') & =\frac{1}{2\sqrt{2\pi}\left(1+e^{-\frac{x_{1}^{2}}{2}}\right)}\left(e^{-\frac{\left(x'/G-x_{1}\right)^{2}}{2}}\right.\nonumber \\
 & \left.+e^{-\frac{\left(x'/G+x_{1}\right)^{2}}{2}}+2e^{-\frac{x'^{2}/G^{2}}{2}}e^{-\frac{x_{1}^{2}}{2}}\right)\label{eq:wig-cat-sc}
\end{align}
and evaluate\textcolor{green}{}
\begin{eqnarray}
Q(x,t_{f}) & = & \frac{1}{\sqrt{2\pi}}\int W_{sc}(x')e^{-(x-x')^{2}/2}dx'\nonumber \\
 & = & \frac{1}{2\sqrt{2\pi\left(1+G^{2}\right)}\left(1+e^{-\frac{x_{1}^{2}}{2}}\right)}\left\{ e^{-\frac{\left(x-Gx_{1}\right)^{2}}{2\left(1+G^{2}\right)}}\right.\nonumber \\
 &  & \left.+e^{-\frac{\left(x+Gx_{1}\right)^{2}}{2\left(1+G^{2}\right)}}+2e^{\frac{-x^{2}}{2\left(1+G^{2}\right)}}e^{\frac{-x_{1}^{2}}{2}}\right\} \label{eq:Q-FBC}
\end{eqnarray}
We see that $\lim_{G\rightarrow\infty}Q(x_{0},t_{f})=W(x_{0})$ as
required. \textcolor{green}{}

\begin{figure}
\begin{centering}
\par\end{centering}
\begin{centering}
\includegraphics[width=0.8\columnwidth]{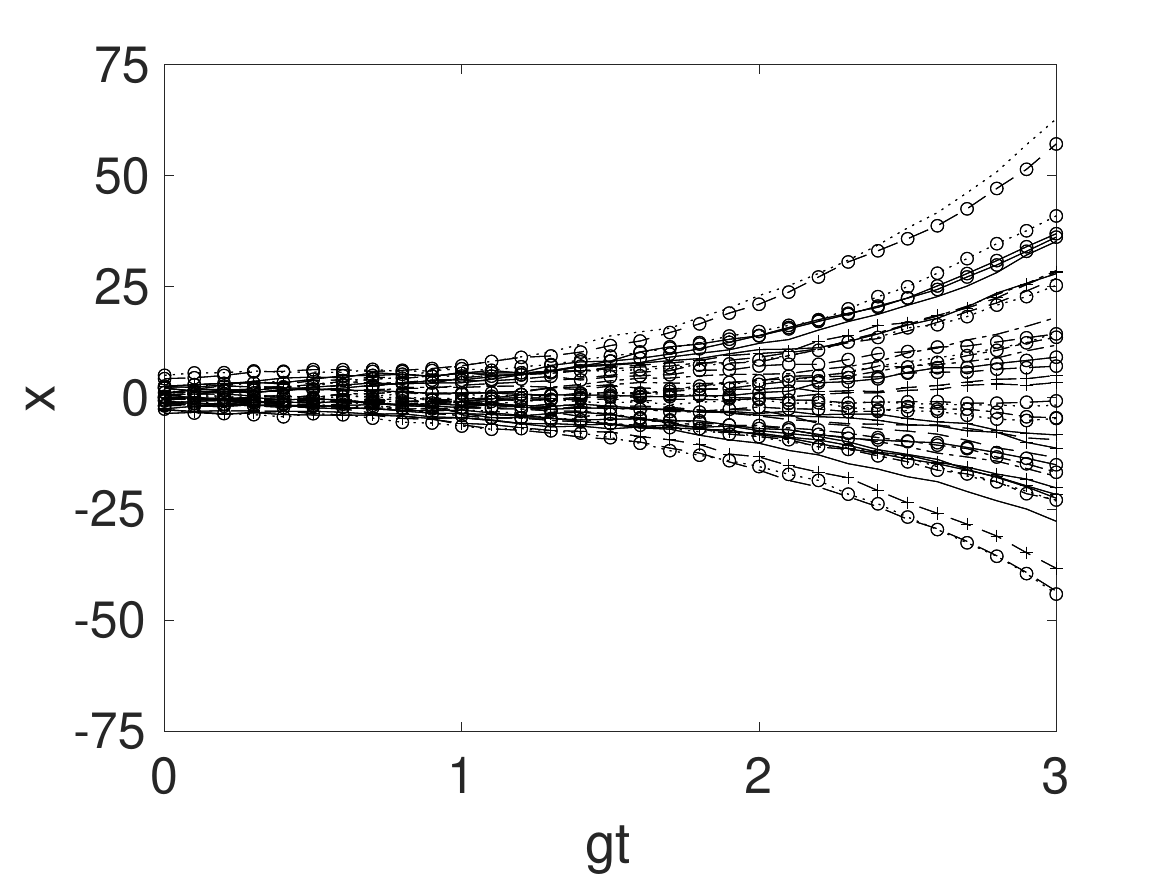}
\par\end{centering}
\begin{centering}
\includegraphics[width=0.8\columnwidth]{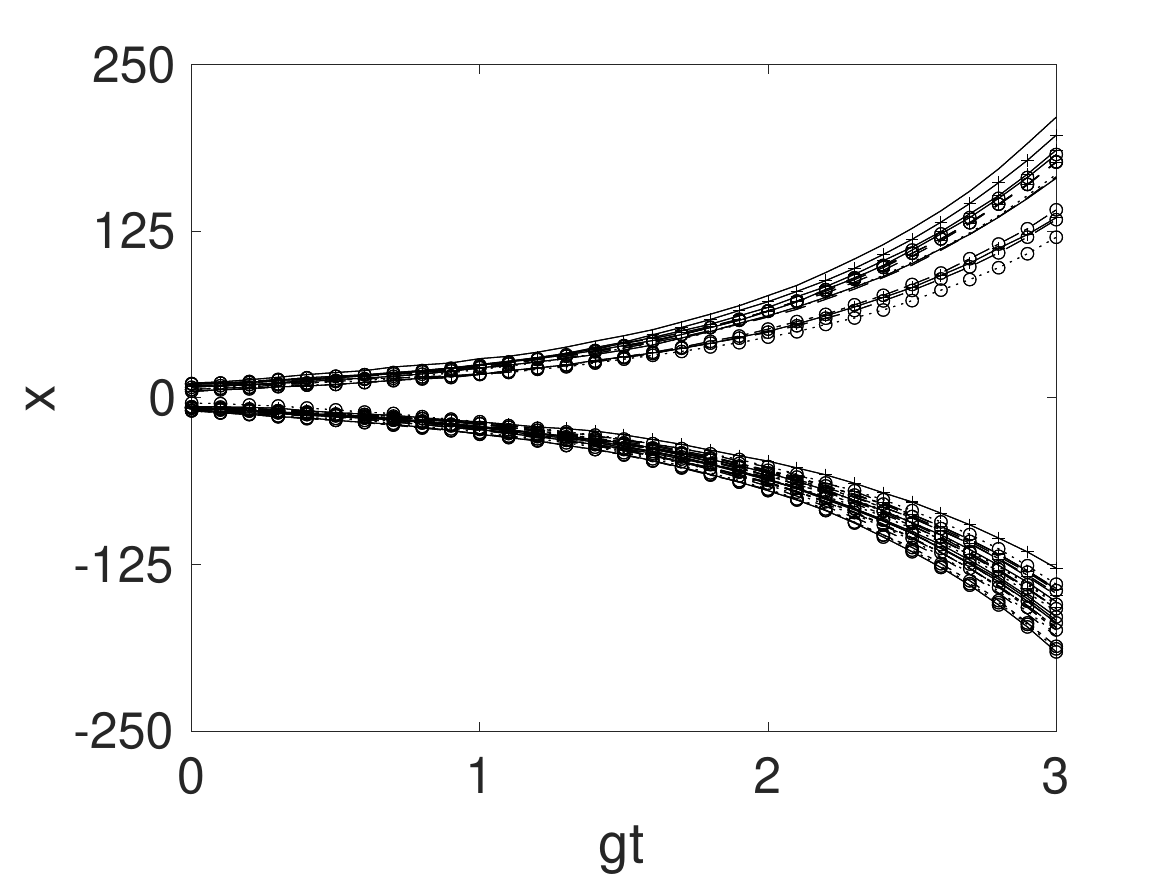}
\par\end{centering}
\caption{Forward-backward stochastic solutions modeling the measurement of
$\hat{x}$ on a system prepared in a superposition of coherent states,
as given by $|\psi_{S}\rangle$ (Eq. (\ref{eq:sup-sq})) with $c_{1}=c_{2}=1/\sqrt{2}$,
$r=0$ and $\varphi=0$. The top plot shows $\alpha_{0}=0.5$ and
$r=0$, which models measurement on a microscopic superposition of
coherent states. The lower plot shows $\alpha_{0}=4$ and $r=0$ which
models measurement on a macroscopic superposition of coherent states.
Plots show $1.2\times10^{6}$ trajectories and $t_{f}=3/g$.\textcolor{green}{{}
}These plots are generated using the Wigner function to evaluate the
boundary condition.\textcolor{green}{{} \label{fig:wignfn-cat-real}}}
\end{figure}

\begin{figure}
\begin{centering}
\includegraphics[width=0.8\columnwidth]{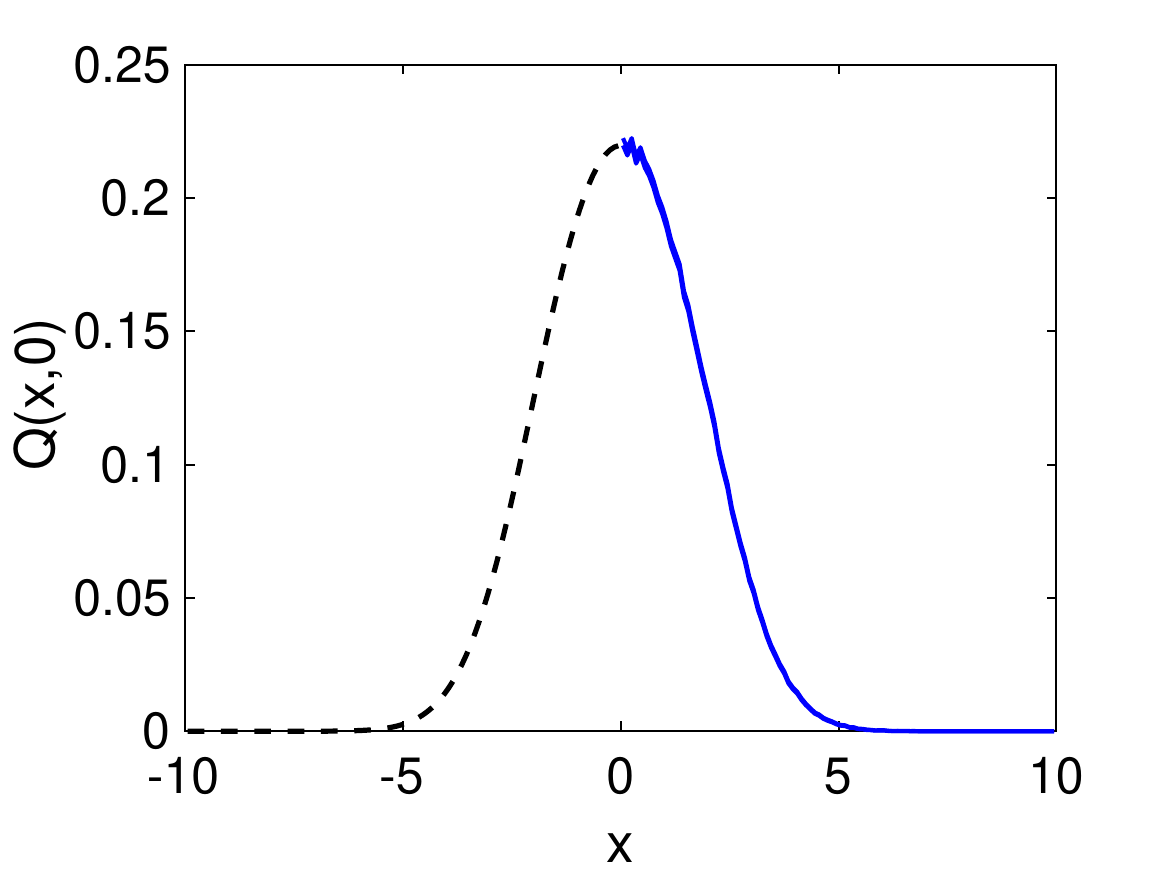}
\par\end{centering}
\begin{centering}
\includegraphics[width=0.8\columnwidth]{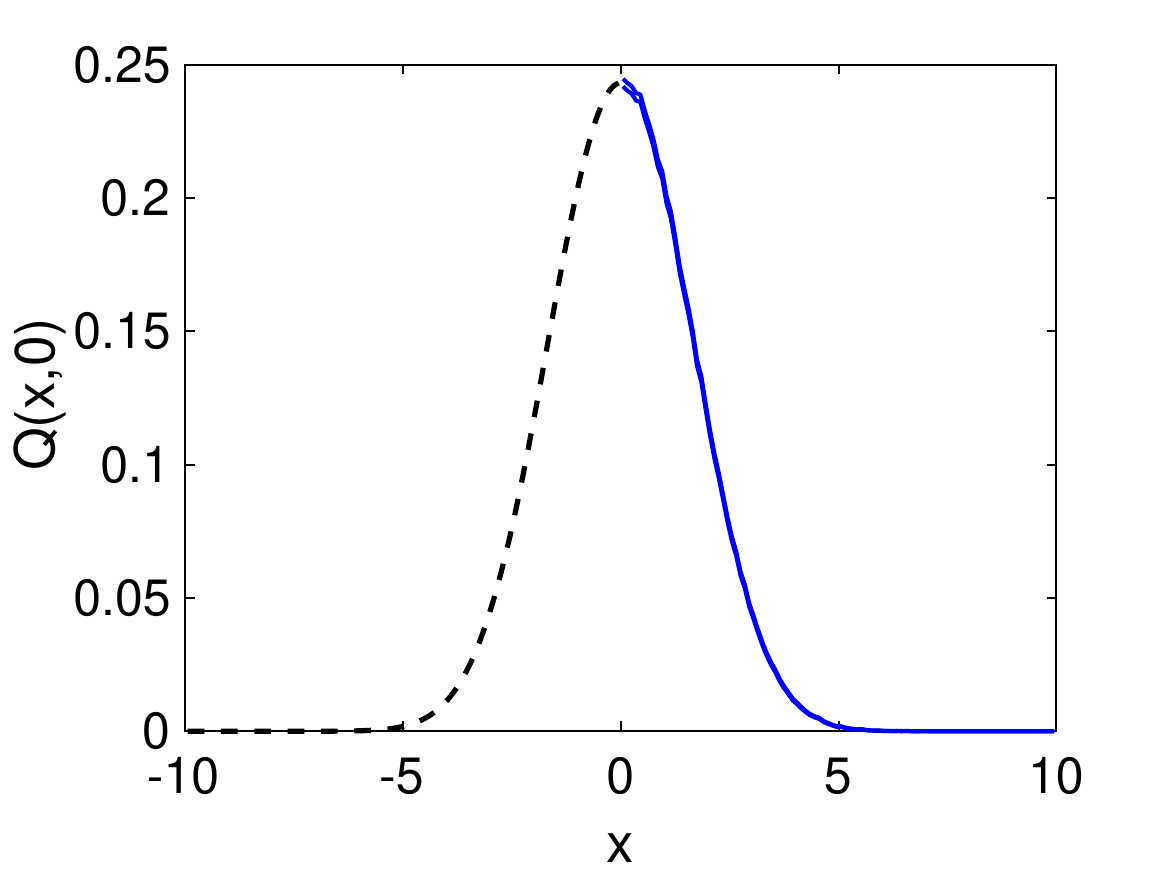}
\par\end{centering}
\caption{As for Figures \ref{fig:wign-cat-im} and \ref{fig:wignfn-cat-real},
but here we plot the marginal distribution $Q(x,t)$ for $t=0$, evaluated
from the simulation (blue solid line to the right of the curve), for
$\alpha_{0}=0.5$ and $r=0$.\textcolor{green}{{} }The black dashed
line is the actual $Q(x,0)$, given by Eq. (\ref{eq:Q-FBC}). The
curves are symmetrical about $x=0$ in each case. The top plot is
for $c_{1}=-ic_{2}=1/\sqrt{2}$.\textcolor{green}{{} }The lower plot
is for $c_{1}=c_{2}=1/\sqrt{2}$. \textcolor{green}{\label{fig:checkint}}}
\end{figure}

The plots of the simulation using the Wigner-function method to establish
the future boundary condition are shown in Figures \ref{fig:wign-cat-im}
and \ref{fig:wignfn-cat-real}. The central peak due to the interference
term $I$ is evident in the marginal $Q(x,0)$, which is $Q(x,t)$
at the initial time when $t=0$, for the cat state where $c_{1}=c_{2}=1/\sqrt{2}$.
We see from (\ref{eq:Q-FBC}) that this term is
\[
I=\frac{1}{4\sqrt{\pi}\left(1+e^{-\frac{x_{1}^{2}}{2}}\right)}2e^{-x_{1}^{2}/2}e^{-x^{2}/4}
\]
and that
\begin{eqnarray}
Q(x,0) & = & \frac{1}{4\sqrt{\pi}\left(1+e^{-\frac{x_{1}^{2}}{2}}\right)}\Biggl(e^{-\frac{\left(x-x_{1}\right)^{2}}{4}}\nonumber \\
 &  & +e^{-\frac{\left(x+x_{1}\right)^{2}}{4}}+2e^{\frac{-x^{2}}{4}}e^{\frac{-x_{1}^{2}}{2}}\Biggl)\label{eq:q-an}
\end{eqnarray}
Hence, for $x_{1}=1$, we find that $Q(x,0)=0.2432$ for $x=0$. By
comparison, for $c_{1}=-ic_{2}$,
\begin{equation}
Q(x,0)=\frac{1}{4\sqrt{\pi}}\left(e^{-\frac{\left(x-x_{1}\right)^{2}}{4}}+e^{-\frac{\left(x+x_{1}\right)^{2}}{4}}\right)\label{eq:qan2}
\end{equation}
implying for $x_{1}=1$, $Q(x,0)=0.2197$ for $x=0$. The function
$Q(x,0)$ is evaluated from the simulation in Figure \ref{fig:checkint},
for $x_{1}=1$, showing agreement with the predicted analytical values
given by (\ref{eq:q-an}) and (\ref{eq:qan2}).\textcolor{green}{}

\subsection{Born's rule}

\textcolor{red}{}The derivation of the relationship between the Q
function of the amplified state and the Wigner function, as summarized
in Eqs. (\ref{eq:expsup})-(\ref{eq:Q-W-FBC}), is a proof of Born's
rule for the measurement of $\hat{x}$. We now examine the example
of the measurement on a cat state more closely, including measurements
of both $\hat{x}$ and $\hat{p}$, confirming compatibility with Born's
rule. For simplicity, we consider the system prepared in a superposition
$\frac{e^{-i\pi/4}}{\sqrt{2}}\{|\alpha_{0}\rangle+e^{i\pi/2}|-\alpha_{0}\rangle\}$
of two coherent states \citep{yurke-stoler-1} which for large $\alpha_{0}$
is the ``cat state''. This corresponds to the cat state (\ref{eq:sup-cat}),
where $r=0$ and $\varphi=\pi/2$ in the expression (\ref{eq:sq-sup})
for the Q function, with $x_{1}=2\alpha_{0}$. We take $\alpha_{0}$
real.

\begin{figure}
\begin{centering}
\includegraphics[width=0.8\columnwidth]{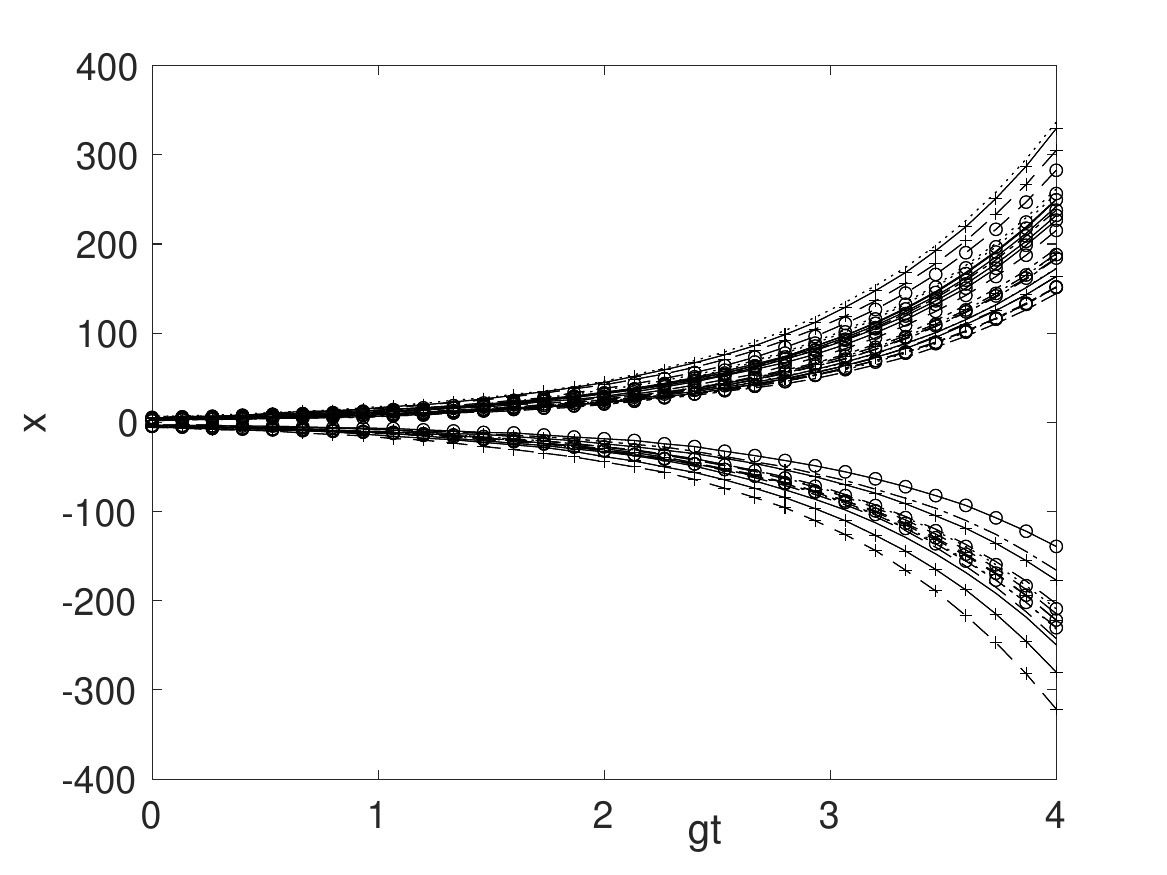}
\par\end{centering}
\begin{centering}
\includegraphics[width=0.8\columnwidth]{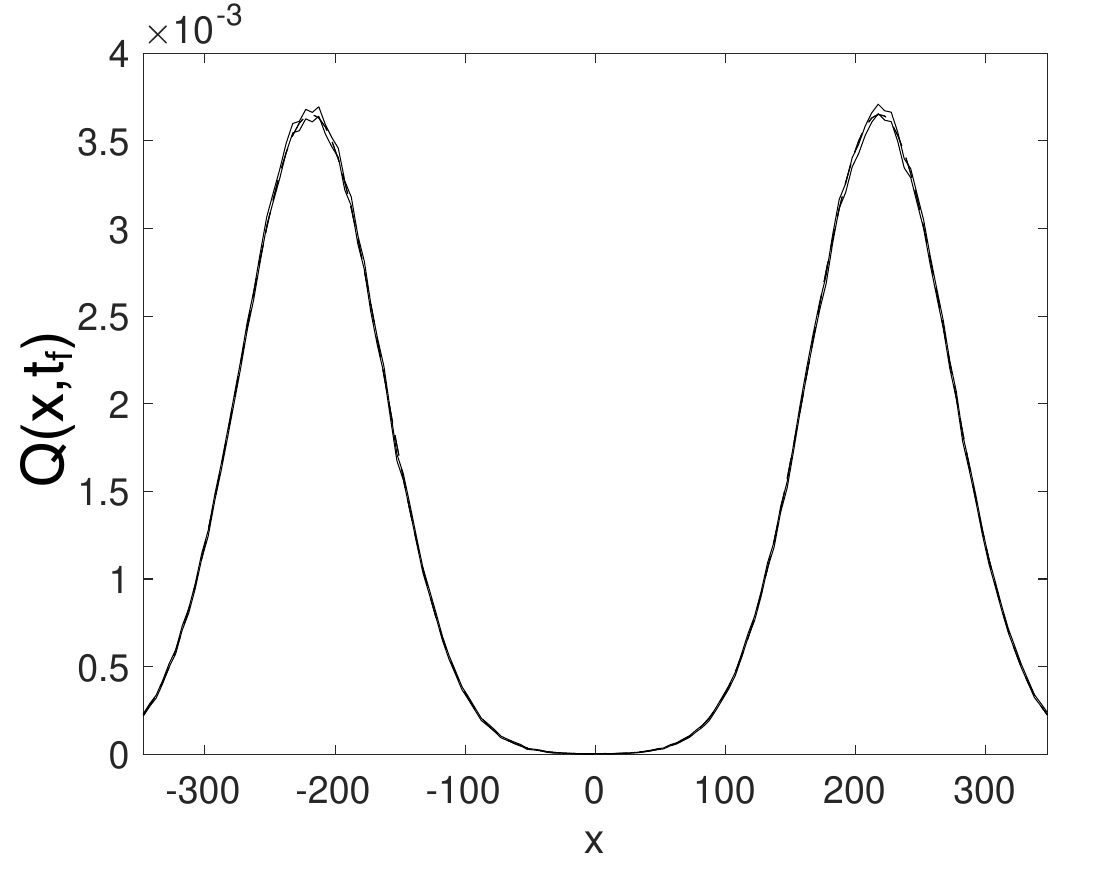}
\par\end{centering}
\caption{Measurement of $\hat{x}$ on a cat state. Plot of backward-propagating
trajectories for the amplified variable $x$ versus $gt$, for the
system prepared in a cat state $|\psi_{cat}\rangle$ (Eq. (\ref{eq:sup-cat}))
with $x_{1}=4$ corresponding to $\alpha_{0}=2$. Also plotted is
the marginal distribution $Q(x,t_{f})$. This plot agrees with the
Born-rule distribution $P_{B}(x)=|\langle x|\psi_{cat}\rangle|^{2}$
predicted by quantum mechanics for the cat state. Here, $t_{f}=4/g$.
\label{fig:x-trajectories-cat-superposition-1}}
\end{figure}

\subsubsection{Measurement of $\hat{x}$}

The quantum noise associated with the Q function for the coherent
state $|\alpha_{0}\rangle$ has two contributions: The first noise
contribution is the ``hidden'' vacuum noise which exists for the
eigenstate $|x_{1}\rangle$ itself, and is not amplified. The second
noise contribution corresponds to the measured vacuum noise level
\textcolor{black}{$(\Delta\hat{x})^{2}=\langle\hat{x}^{2}\rangle-\langle\hat{x}\rangle^{2}=1$}
of a coherent state. This noise level, being measurable, is amplified
by the measurement interaction $H_{amp}$\textcolor{black}{. The additional
vacuum noise appears in the simulations as extra noise in the final
amplified outputs at time $t_{f}$, evident from Figures \ref{fig:x-trajectories-cat-superposition-2}
and \ref{fig:x-trajectories-cat-superposition-1}.}

We demonstrate the effectiveness of the model $H_{amp}$ for the measurement
of $\hat{x}$ by evaluating\textcolor{black}{{} the final marginal distribution
$Q(x,t_{f})$ (in the large amplification limit). This corresponds
to $P_{B}(x)=|\langle x|\psi_{cat}\rangle|^{2}$ as predicted by quantum
mechanics (Born's rule). Here, $|x\rangle$ is the eigenstate for
$\hat{x}$. }This is demonstrated analytically, since the marginal
for $x$ where $gt\rightarrow\infty$ written in terms of the scaled
variable \textcolor{black}{$x_{0}=x/e^{gt}$ is}\textbf{\textcolor{black}{{}
\begin{eqnarray}
Q(x_{0},t) & = & \frac{1}{2\sqrt{2\pi}}\left\{ e^{-\left(x_{0}-x_{1}\right){}^{2}/2}\left.+e^{-\left(x_{0}+x_{1}\right)^{2}/2}\right\} \right.\nonumber \\
\label{eq:catpx}
\end{eqnarray}
}}\textcolor{black}{where we use the result that $\sigma_{x}^{2}(t)\rightarrow e^{2gt}$.
This agrees with $P_{B}(x)=|\langle x|\psi_{cat}\rangle|^{2}$ as
predicted by quantum mechanics, and evaluated in \citep{yurke-stoler-1}
using $x_{1}=2\alpha_{0}$. The equivalence with $P_{B}(x)$ is shown
in Figure \ref{fig:x-trajectories-cat-superposition-1}, for $\alpha_{0}=2$.}

\begin{figure}
\begin{centering}
\includegraphics[width=0.8\columnwidth]{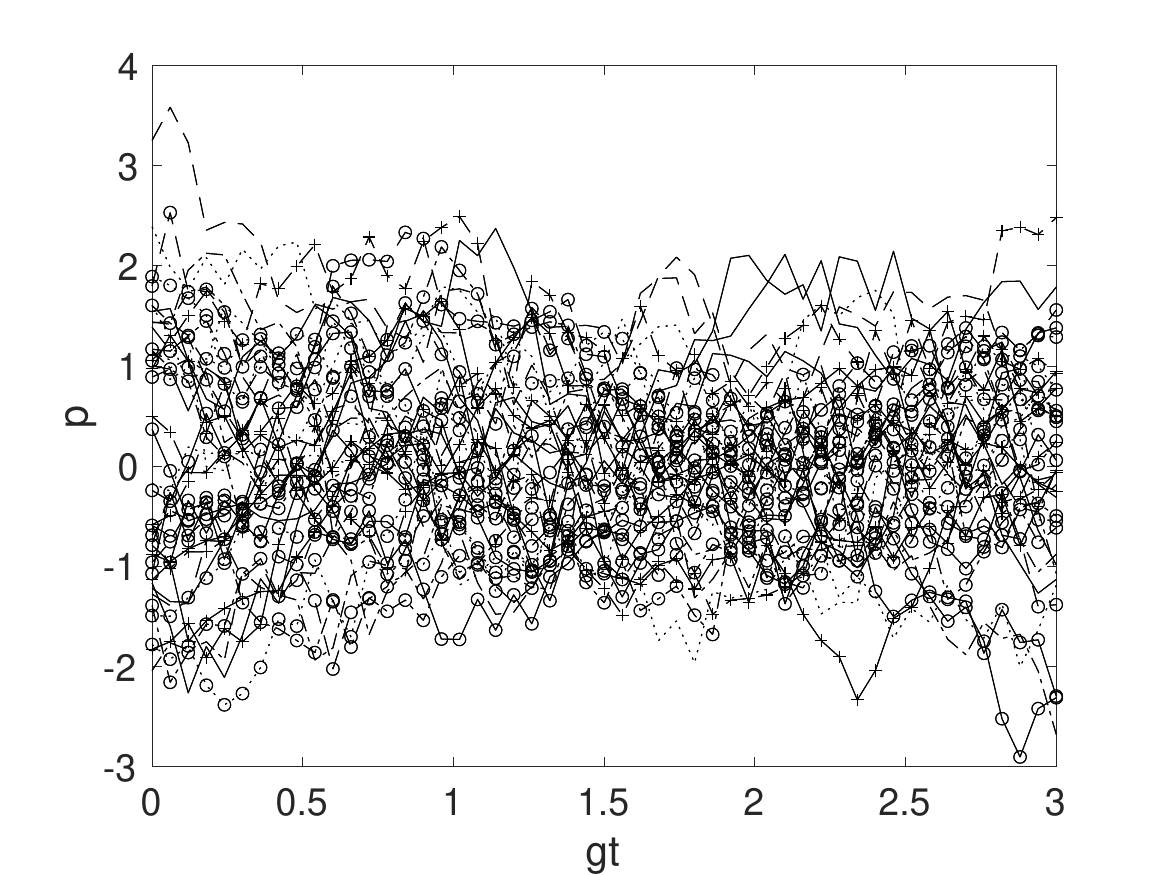}
\par\end{centering}
\begin{centering}
\includegraphics[width=0.8\columnwidth]{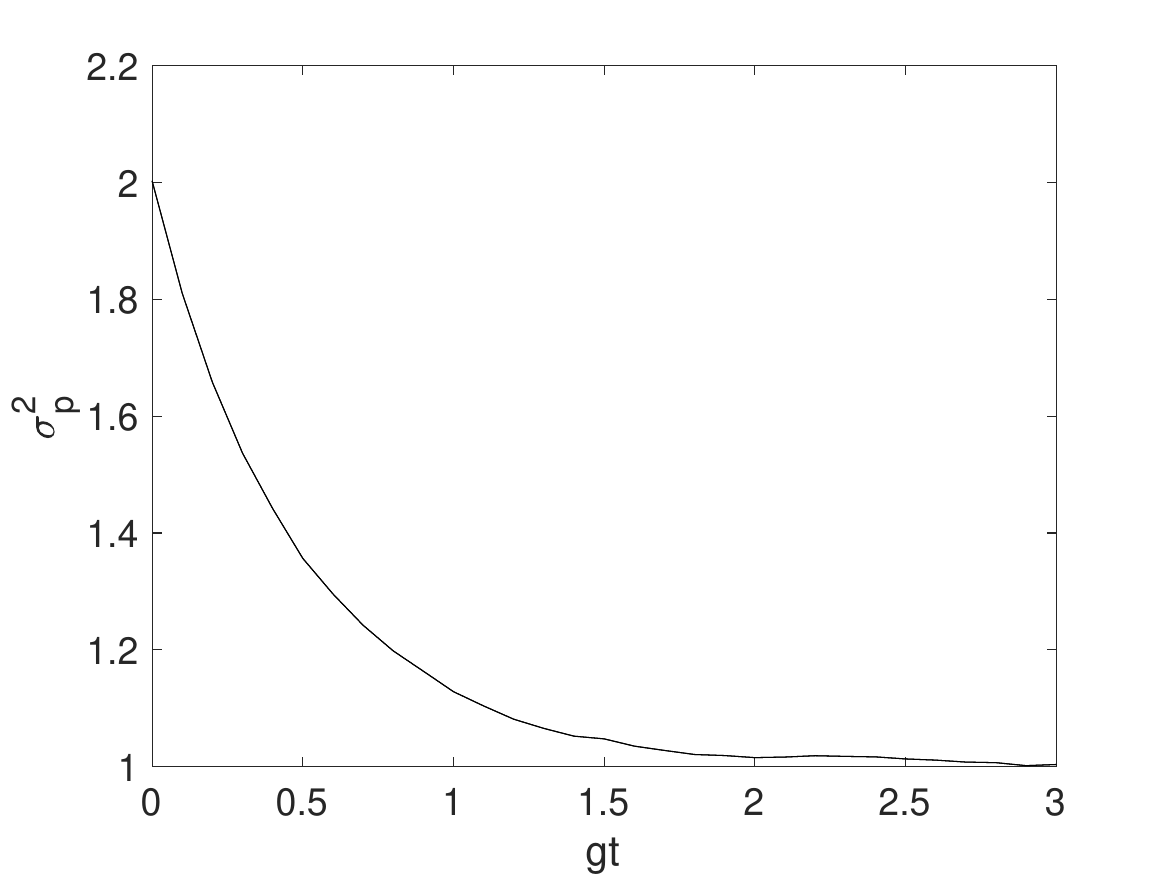}
\par\end{centering}
\caption{Measurement of $\hat{x}$ on a cat state. Plot of forward propagating
trajectories for the attenuated variable $p$ versus time $gt$, for
the system prepared in a cat state $|\psi_{cat}\rangle$ (Eq. (\ref{eq:sup-cat})).
The top plot shows $x_{1}=10$ where the coherent states are well
separated. Here, $t_{f}=3/g$. The lower plot shows the variance $\sigma_{p}^{2}$
evaluated at each time $gt$, evaluated over a large sample of trajectories.
The final variance $\sigma_{p}^{2}(t)$ decays to $1$. \label{fig:x-trajectories-superposition-1-2-1}}
\end{figure}

\textcolor{black}{From Figure \ref{fig:x-trajectories-superposition-1-2-1},
the trajectories for $p$ when $\hat{x}$ is measured are attenuated.
The effect is less pronounced compared to that of the superposition
of position eigenstates (Figure \ref{fig:sup}), because there is
a reduced noise in $p$ at the initial time in this case. The measurement
$H_{amp}$ of $\hat{x}$ amplifies $\hat{x}$ and squeezes $\hat{p}$.
The noise levels for the initial cat state are approximately at the
vacuum level $(\Delta\hat{p})^{2}\sim1$, and $H_{amp}$ has the effect
of squeezing the fluctuations in $\hat{p}$, as shown by the variance
in $p$ in Figure \ref{fig:x-trajectories-superposition-1-2-1}.}

\subsubsection{Measurement of $\hat{p}$}

We now consider the complementary measurement, $\hat{p}$. This is
incompatible with an $\hat{x}$ measurement because it requires a
different meter setting, which implies a different measurement Hamiltonian.
The resulting outputs have the complementary feature of interference
fringes.

The $\hat{p}$ measurement requires amplification of $p$. We use
\begin{equation}
H_{amp}=\frac{i\hbar g}{2}\left[\hat{a}^{\dagger2}-\hat{a}^{2}\right]\label{eq:ham-1}
\end{equation}
where $g$ is real and $g<0$. The dynamics from the standard operator
Heisenberg equations gives the solutions
\begin{eqnarray}
\hat{x}\left(t\right) & = & \hat{x}\left(0\right)e^{-|g|t}\nonumber \\
\hat{p}\left(t\right) & = & \hat{p}\left(0\right)e^{|g|t}\label{eq:amp-1}
\end{eqnarray}
and we see that $\hat{p}$ is amplified. The solutions for the dynamics
of the $x$ and $p$ variables of the Q function are as above, except
that $x$ and $p$ exchange roles. The trajectories for $p$ are amplified
and propagate back in time. Those for $x$ are attenuated and propagate
forward in time.

\begin{figure}
\begin{centering}
\includegraphics[width=0.8\columnwidth]{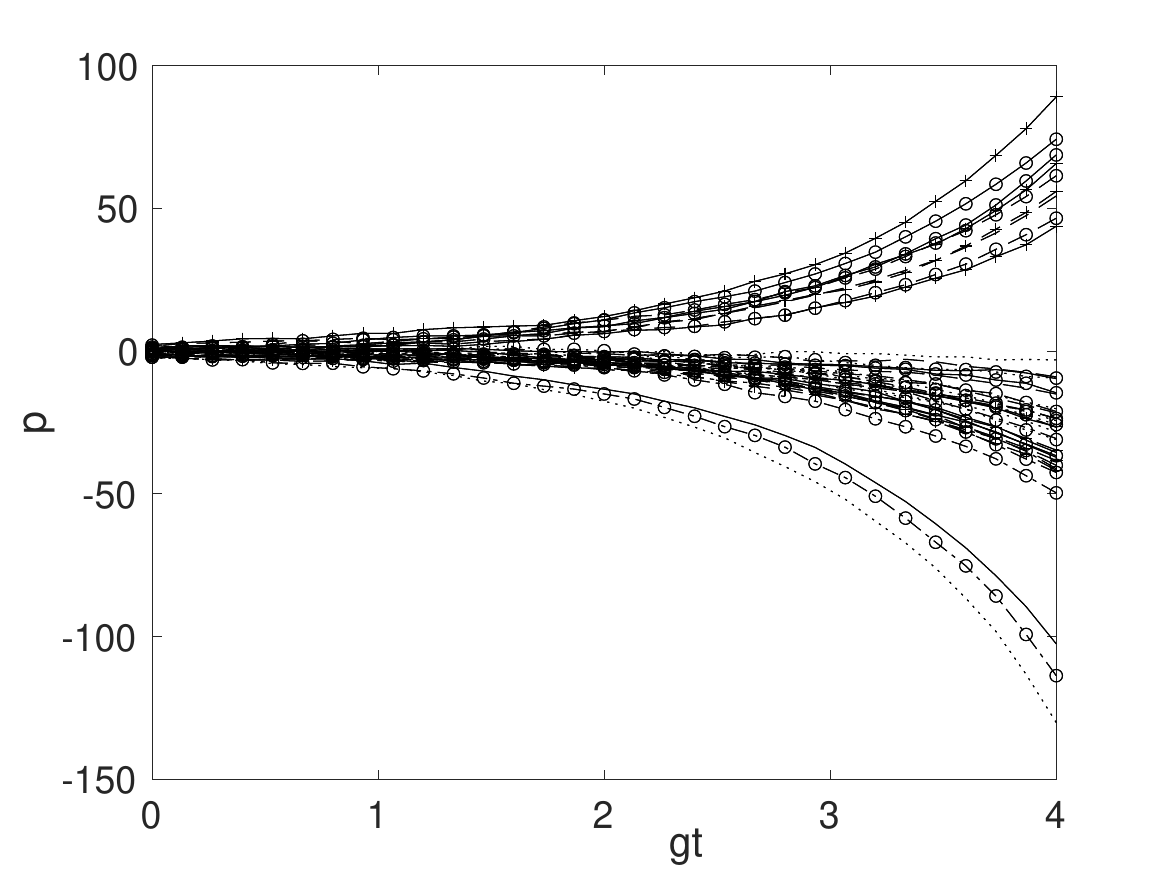}
\par\end{centering}
\begin{centering}
\includegraphics[width=0.8\columnwidth]{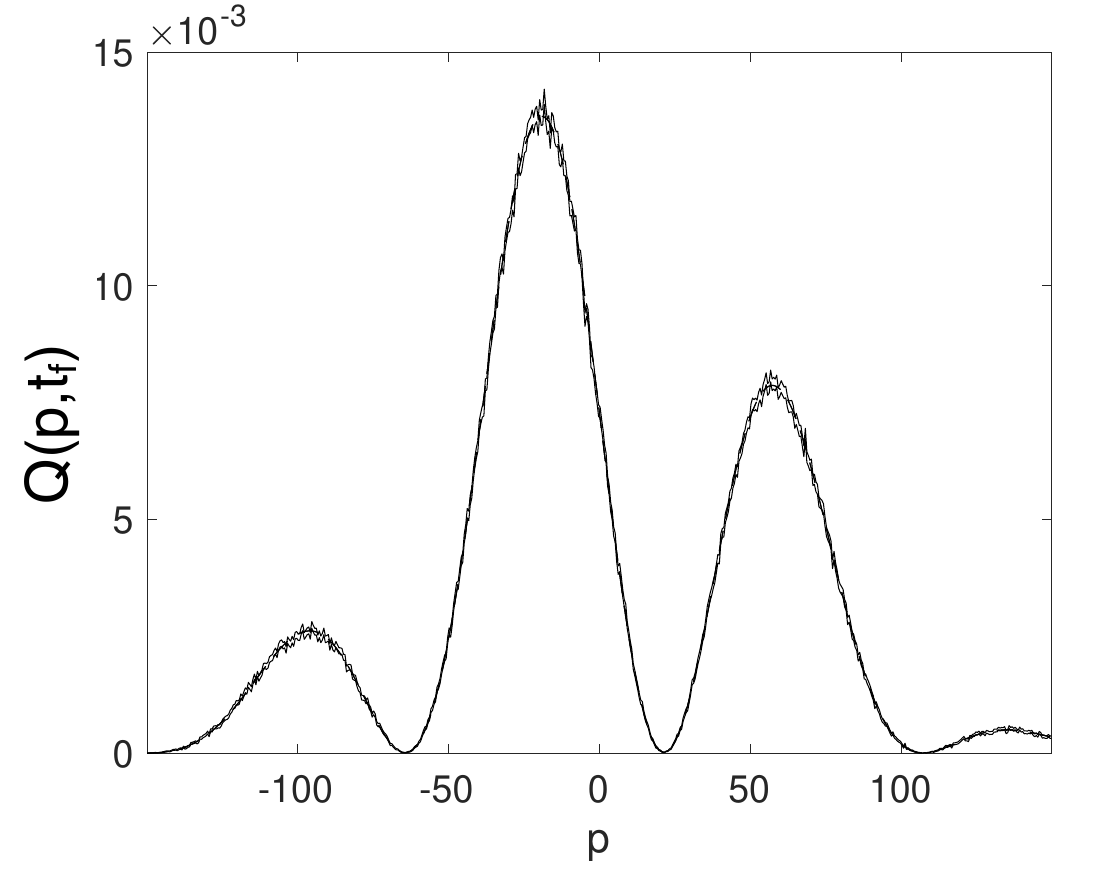}
\par\end{centering}
\caption{Plotted are the backward trajectories for $p$ for the measurement
of $\hat{p}$, using the Hamiltonian $H_{amp}$ where $g<0$. We consider
the cat state $|\psi_{cat}\rangle$ given by Eq. (\ref{eq:sup-cat}),
(corresponding to $|\psi_{S}\rangle$ {[}Eq. (\ref{eq:sup-sq}){]}
with $c_{1}=-ic_{2}=1/\sqrt{2}$, $r=0$ and $\varphi=\pi/2$) with
well-separated coherent states given by $\alpha_{0}=2$ ($x_{1}=4$).
Here, $t_{f}=4/g$. The final marginal distribution $Q(p,t_{f})$
is given in the lower plot. The fringes are sharply defined in agreement
with $P_{B}(p)$ as predicted by quantum mechanics. Graphs show the
upper and lower sampling error values, and the exact result. \label{fig:measure-p-backwardtraj}}
\end{figure}

If we measure $\hat{p}$ by amplifying the $\hat{p}$ quadrature so
that $g<0$, then the full state at the later time is evaluated directly
as before. The solution for the initial state (\ref{eq:sup-sq}) is
given by (\ref{eq:Q-sup-1-1}), which becomes (for $\varphi=\pi/2$
and $|c_{1}|=|c_{2}|=1/\sqrt{2}$) 
\begin{eqnarray}
Q(x,p,t) & = & \frac{e^{-p^{2}/2\sigma_{p}^{2}(t)}}{4\pi\sigma_{x}(t)\sigma_{p}(t)}\Bigl\{ e^{-\left(x-G(t)x_{1}\right){}^{2}/2\sigma_{x}^{2}(t)}\nonumber \\
 &  & +e^{-\left(x+G(t)x_{1}\right){}^{2}/2\sigma_{x}^{2}(t)}\nonumber \\
 &  & -2e^{-(x^{2}+G^{2}(t)x_{1}^{2})/2\sigma_{x}^{2}(t)}\sin\bigr(\frac{pG(t)x_{1}}{\sigma_{x}^{2}(t)}\bigr)\Bigl\}\nonumber \\
\label{eq:qfuture-1-1}
\end{eqnarray}
where $G(t)=e^{gt}$ and $\sigma_{x/p}^{2}(t)=2(1\pm\tanh(r-gt))^{-1}=1+e^{\pm2(gt-r)}$,
except that now $g<0$. Therefore $G(t)=e^{-|g|t}\rightarrow0$ and
$\sigma_{x}^{2}(t)=1+e^{-2(|g|t+r)}\rightarrow1$, in the limit of
$|g|t\rightarrow\infty$.  Hence, the future marginal in $p$ at
time $t_{f}$ is
\begin{eqnarray}
Q(p,t) & = & \frac{e^{-p^{2}/2\sigma_{p}^{2}(t)}}{\sqrt{2\pi}\sigma_{p}}\Bigl\{1-e^{-\frac{G^{2}(t)x_{1}^{2}}{2\sigma_{x}^{2}(t)}}\sin\bigr(\frac{pG(t)x_{1}}{\sigma_{x}^{2}(t)}\bigr)\Bigl\}\nonumber \\
\nonumber \\
 & \rightarrow & \frac{e^{-p^{2}/2\sigma_{p}^{2}(t)}}{\sqrt{2\pi}\sigma_{p}}\Bigl\{1-\sin(pG(t)x_{1})\Bigl\}\label{eq:measure-p}
\end{eqnarray}
We may write the solution as
\begin{equation}
Q_{sc}(p_{0},t)\rightarrow\frac{e^{-p_{0}^{2}/(2e^{2r})}}{\sqrt{2\pi}}\Bigl\{1-\sin(p_{0}x_{1})\Bigl\}\label{eq:P-function}
\end{equation}
using the scaled variable $p_{0}=p/e^{|g|t}$ and noting that $\sigma_{p}^{2}(t)\rightarrow e^{2|g|t}e^{2r}$
for large $|gt|$, with $g<0$.

\begin{figure}
\begin{centering}
\includegraphics[width=0.8\columnwidth]{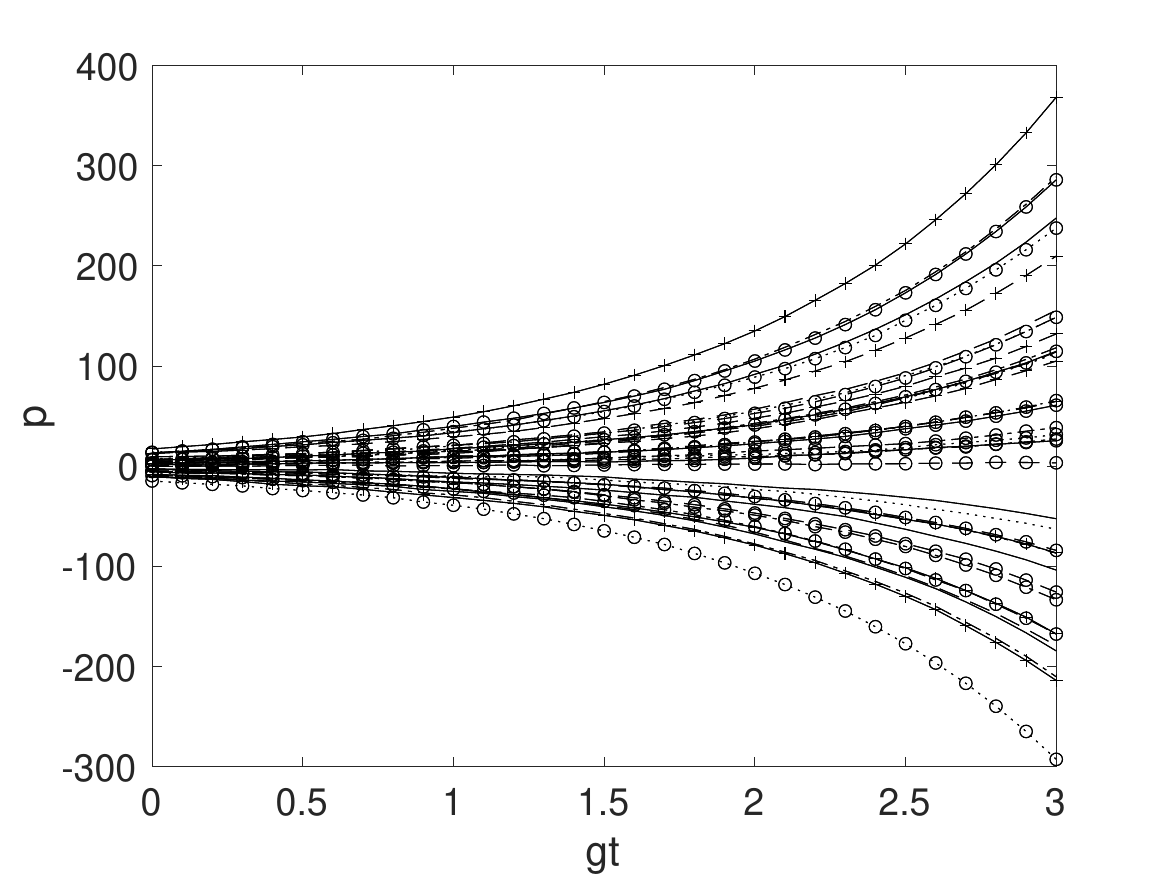}
\par\end{centering}
\begin{centering}
\includegraphics[width=0.8\columnwidth]{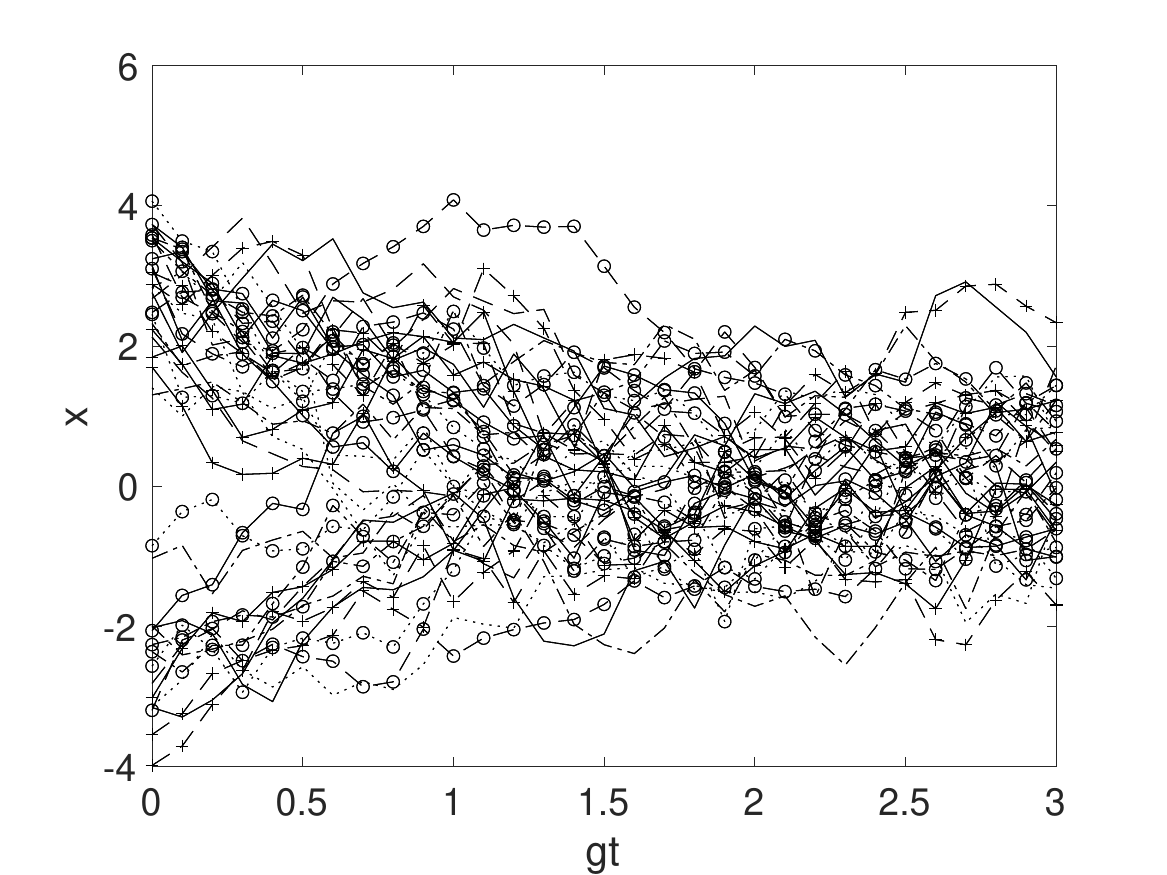}
\par\end{centering}
\caption{Measurement of $\hat{p}$ for a system prepared in a superposition
$|\psi_{S}\rangle$ of two eigenstates $|x_{1}\rangle$ and $|x_{2}\rangle$
of $\hat{x}$ with $x_{1}=-x_{2}=3$, modeled as (\ref{eq:sq-sup})
with $r=2$ and $\varphi=\pi/2$. The top plot shows trajectories
for $p$ propagating according to $H_{amp}$ with $g<0$. The final
values of $p$ at $t_{f}$ are amplified. The trajectories for $p$
propagate backwards, from the future boundary at $t_{f}$. The lower
plot shows forward-propagating trajectories for the complementary
variable $x$, which are attenuated to the vacuum noise level $\sigma_{x}^{2}=1$.
\label{fig:summary-1-1-1}}
\end{figure}

We find agreement with the quantum prediction for the distribution
\[
P_{B}(p)=\frac{e^{-p^{2}/(2e^{2r})}}{\sqrt{2\pi}}\Bigl\{1-\sin(px_{1})\Bigl\}
\]
for the outcome $p$ upon measurement of $\hat{p}$, given as $P_{B}(p)=|\langle p|\psi_{S}\rangle|^{2}$,
where $|p\rangle$ is the eigenstate of $\hat{p}$. We consider the
cat state $|\psi_{cat}\rangle$ which implies $r=0$. The Born-rule
solution
\begin{equation}
P_{B}(p)=\frac{e^{-p^{2}/2}}{\sqrt{2\pi}}\{1-\sin(2\alpha_{0}p)\}\label{eq:p(p)fringe-cat}
\end{equation}
given in \citep{yurke-stoler-1} is in agreement with (\ref{eq:P-function}),
upon noting that $x_{1}=2\alpha$. Figure \ref{fig:measure-p-backwardtraj}
shows the future marginal and the trajectories for $p$, for large
$|gt|$. As expected, for the cat state, the fringes are prominent.
The comparison is also done for the superposition $|\psi_{S}\rangle$
of two eigenstates, in Figure \ref{fig:summary-1-1-1}, giving agreement
with the quantum prediction.

\section{Q-based, objective-field model of reality}

A model of reality has been proposed based on the simulations. We
refer to the model as the Q-based, objective-field interpretation
of quantum mechanics. The model of reality is simply that the system
exists at a time $t$ (prior to and during the measurement) in a state
given by the amplitudes $x$ and $p$, with a probability $Q(x,p,t)$.
The measurement problem is addressed in the following way.

First, there is consistency with macroscopic realism (MR) (Conclusion
(1) of this paper). The simulations of the superposition $\sum_{j}c_{j}|x_{j}\rangle$
{[}Eq. (\ref{eq:expsup}){]} of eigenstates of $\hat{x}$ reveal branches
forming for the amplitudes $x(t)$, at a time $t_{m}$, as $t\rightarrow\infty$
(Figure \ref{fig:sup}). Each branch $B_{j}$ is a distinct group
of amplitudes $x(t)$, corresponding to a distinct eigenvalue $x_{j}$.
In the model, the measurement is completed by a detection of the macroscopic
amplitude $x(t_{f})$. The outcome of the measurement of $\hat{x}$
accounts for amplification and is $x(t_{f})/G$ where $G=e^{gt_{f}}$,
as given by Eq. (\ref{eq:inferred-outcome}). For the system at time
$t_{m}$, when in a superposition $\sum_{j}c_{j}e^{-iH_{amp}t_{m}/\hbar}|x_{j}\rangle$
of macroscopically distinct states, macroscopic realism (MR) holds,
since the outcome of the measurement $\hat{x}$ is determined by the
value of $x(t_{m})$ (\citep{q-measurement}. This is also true of
the cat state $|\psi_{cat}\rangle$, where branches form for large
$\alpha_{0}$, so that MR holds (Figures \ref{fig:sup-2}-\ref{fig:x-trajectories-cat-superposition-2}).

We see from the Q function (\ref{eq:Q-sup-1-1}) that as the system
is amplified, the means $x_{j}$ in $x$ of the Gaussians corresponding
to each eigenstate $|x_{j}\rangle$ are amplified, so that
\begin{equation}
x_{j}\rightarrow Gx_{j}\label{eq:det}
\end{equation}
However, the variance $\sigma_{x}$ of $x$ in the Gaussian distributions
corresponding to the eigenstate $|x_{j}\rangle$ is not amplified.
The fluctuations $\delta x$ about the mean amplitude $x_{j}$ of
each branch $B_{j}$ remain of constant average magnitude, as evident
from Figure \ref{fig:sup}. Hence, it is the eigenvalue that is measured.
In the model, the probability density $P(x)$ for an outcome $x$
is given by the density of amplitudes of each branch $B_{j}$, which
explains Born's rule.

The amplification (\ref{eq:det}) is a deterministic relation. A causal
model (as in a set of cause-and-effect relations) for measurement
has been put forward, motivated by the $Q$-based model and simulations,
in Ref. \citep{q-measurement}. In the simplest causal model, the
system at time $t_{0}$ exists in a state (as defined by a distribution
of $x$ and $p$) with a definite probability ($|c_{j}|^{2}$) that
the outcome will be $x_{j}$. Alternatively, the system is regarded
to be, with probability $|c_{j}|^{2}$, in a state with a predetermined
outcome, $x_{j}$. This motivates the analysis of the postselected
state, as in the next section.

\section{Postselected state and Hidden variables: single-mode system}

\textcolor{red}{}Here, we examine the distribution $Q(x,p,t_{0}|x_{j})$
for the amplitudes at the initial time $t_{0}$ \emph{conditioned}
on a given outcome $x_{j}$ for $\hat{x}$. We extend the treatment
given in \citep{q-measurement} to examine more general superposition
states, including the cat state (\ref{eq:sup-cat}). We show that
this postselected distribution cannot correspond to the Q function
of a quantum state. The variables $x_{A}$ and $p_{A}$ are hence
seen to be ``hidden variables'' in the reality model associated
with the measurement (Conclusion (3) of this paper).

\subsection{Conditional distribution at time $t_{0}$}

A given $x(t_{f})$ from the time $t_{f}$ propagates back to a single
value of $x(0)$ at the initial time, $t_{0}=0$. For each such $x(0)$,
there is a set of $p(0)$ at time $t_{0}=0$. This set is determined
by the conditional distribution \citep{q-measurement}
\begin{equation}
Q(p|x)=Q(x,p,t_{0})/Q(x,t_{0})\label{eq:cond-1}
\end{equation}
evaluated from the Q function $Q(x,p,t_{0})$ at $t_{0}=0$.

Here, we consider the system in the superposition where $Q(x,p,t_{0})$
is given by Eq. (\ref{eq:Q-sup-1}). We select $x_{2}=-x_{1}$, and
$x_{1}>0$. Hence, for general $\varphi$, the marginal in $x$
reduces to
\begin{eqnarray}
Q(x,0) & = & \frac{1}{\sqrt{2\pi}\sigma_{x}f(\varphi)}\Bigl(|c_{1}|^{2}e^{-(x-x_{1})^{2}/2\sigma_{x}^{2}}\nonumber \\
 &  & +|c_{2}|^{2}e^{-(x+x_{1})^{2}/2\sigma_{x}^{2}}\nonumber \\
 &  & +2|c_{1}c_{2}|e^{-x^{2}/2\sigma_{x}^{2}}e^{-x_{1}^{2}(1+\sigma_{p}^{2}/2\sigma_{x}^{2})/2\sigma_{x}^{2}}\cos\varphi\Bigl)\nonumber \\
\end{eqnarray}
where $f(\varphi)=1+[\cos\varphi]e^{-x_{1}^{2}\left(1+\sigma_{p}^{2}/\sigma_{x}^{2}\right)/2\sigma_{x}^{2}}$
and $\sigma_{x}^{2}=1+e^{-2r}$ and $\sigma_{p}^{2}=1+e^{2r}$. As
shown in \citep{q-measurement}, for the superposition of eigenstates
$|x_{j}\rangle$, where $\text{\ensuremath{\sigma_{p}^{2}\rightarrow\infty}, }$
the term
\begin{equation}
T_{e}=\frac{2|c_{1}c_{2}|e^{-x^{2}/2\sigma_{x}^{2}}}{\sqrt{2\pi}\sigma_{x}f(\varphi)}e^{-x_{1}^{2}(1+\sigma_{p}^{2}/2\sigma_{x}^{2})/2\sigma_{x}^{2}}\cos\varphi\label{eq:Te}
\end{equation}
becomes negligible. In fact, $T_{e}$ vanishes exactly, if $\varphi=\pm\pi/2$.
In these cases,
\begin{equation}
Q(x,0)=\frac{1}{2}\{Q_{+}(x)+Q_{-}(x)\}\label{eq:marg}
\end{equation}
where $Q_{\pm}(x)=\frac{1}{\sqrt{2\pi}\sigma_{x}}e^{-\left(x\mp x_{1}\right){}^{2}/2\sigma_{x}^{2}}$
\textcolor{black}{. For general $r$ but where $\varphi=\pi/2$,
we find}
\begin{eqnarray}
Q(p|x) & = & \frac{e^{-p^{2}/2\sigma_{p}^{2}}}{\sigma_{p}\sqrt{2\pi}}\Bigr(1-\frac{\sin(p|x_{1}|/\sigma_{x}^{2})}{\cosh(x|x_{1}|/\sigma_{x}^{2})}\Bigl)\label{eq:pcond-1}
\end{eqnarray}
Fringes are present, becoming finer for large $x_{1}$ (which is the
separation between the states of the superposition) and also increasingly
damped, provided $x\neq0$.

\subsection{Inferred state at time $t_{0}$ given an outcome for $\hat{x}$}

The conditional distribution (\ref{eq:pcond-1}) implies that the
trajectories for $x$ and $p$ are correlated. For a set of values
of $x(t_{f})$ at the time $t_{f}$, we can match the set with a set
of $p$ trajectories, by propagating each given $p(0)$ at time $t_{0}=0$
from the sample generated by $Q(p|x)$, back to the time $t_{f}$.
This creates a ``loop'' associated with the set $\{x(t_{f})\}$,
which determines the detected outcome, or set of outcomes. We define
the postselected distribution $Q_{loop}(x,p,t_{0}|\{x(t_{f})\})$
for the values $x$ and $p$ at the time $t_{0}=0$ that connect to
the set $\{x(t_{f})\}$.

Here, as in \citep{q-measurement}, we consider a superposition
\begin{equation}
|\psi_{S}\rangle=c_{1}|\frac{x_{1}}{2},r\rangle+c_{2}|-\frac{x_{1}}{2},r\rangle\label{eq:sup-eigenstates}
\end{equation}
where $x_{1}>0$, and $x_{2}=-x_{1}$. We consider the set of outcomes
for $\hat{x}$ that are positive. For large $gt_{f}$, the positive
values of $x(t_{f})$ imply that the outcome for $\hat{x}$ is positive.
We denote the set of outcomes $\{x(t_{f}):x(t_{f})>0\}$ by the symbol
$+$. In the limit of large $gt_{f}$, the postselected distribution
is denoted
\begin{equation}
Q_{+}(x,p,t_{0})\equiv Q_{loop}(x,p,t_{0}|+)\label{eq:qloop}
\end{equation}
The distribution $Q_{+}(x,p,t_{0})$ can be interpreted in the model
for reality that we give for measurement as describing the ``state''
inferred at time $t_{0}$, given a positive outcome for $\hat{x}$.

\subsubsection*{Superpositions of eigenstates of $\hat{x}$, or where $\varphi=\pi/2$}

To gain some analytical insight, we first review Ref. \citep{q-measurement}
and restrict to the case of the superposition of two eigenstates of
$\hat{x}$, as in (\ref{eq:sup-eigenstates}) for $r\rightarrow\infty$.
For sufficiently large $gt$, each $x(t_{f})$ is either positive
or negative, associated with the outcome $x_{1}$ or $-x_{1}$ which
we denote by $+$ or $-$. Then $Q_{+}(x,p,t_{0})$ is written $Q(x,p,t_{0}|x_{1})$
to denote the postselected distribution given the outcome $x_{1}$.
The equality follows for large $gt_{f}$, because the trajectories
of two eigenstates with different eigenvalues will always separate
with sufficient amplification (Figure \ref{fig:sup}).

The future boundary condition for the superposition is determined
by the probabilistic mixture
\begin{equation}
|c_{1}|^{2}Q_{+}(x,t_{f})+|c_{2}|^{2}Q_{-}(x,t_{f})\label{eq:fbc-prob-mix}
\end{equation}
 of the two Gaussian distributions $Q_{\pm}(x,t_{f})$, defined as
\begin{equation}
Q_{\pm}(x,t_{f})=\frac{1}{\sqrt{2\pi}\sigma_{x}}e^{-(x\mp G(t_{f})x_{1})^{2}/2\sigma_{x}}\label{eq:fbc}
\end{equation}
where for the superposition of eigenstates of $\hat{x}$, $\sigma_{x}^{2}=1+e^{-2r}\rightarrow1$.
We define the set of amplitudes propagating from the amplitude $G(t_{f})x_{1}$
by $B_{+}$, recognizing that for $r\rightarrow\infty$, this set
corresponds to the branch $\mathcal{B}_{+}$ defined for the outcome
$x_{1}$. Hence, in the limit of large $gt_{f}$, the values of $B_{+}$
emanating from $Q_{+}(x,t_{f})$ correspond to the set defined by
the Gaussian 
\begin{equation}
Q_{+}(x)=\frac{1}{\sqrt{2\pi}\sigma_{x}}e^{-(x-|x_{1}|)^{2}/2\sigma_{x}}
\end{equation}
at time $t_{0}=0$. For each trajectory beginning from $x(t_{f})$,
there is a single value at $x(t_{0})$. Summed over all trajectories,
this defines the distribution $Q(x,p,t_{0}|B_{+})$ for the branch
$B_{+}$ at the initial time. With the notation introduced, the postselected
distribution $Q_{+}(x,p,t_{0})$ is denoted
\begin{equation}
Q(x,p,t_{0}|B_{+})\equiv Q(x,p,t_{0}|x_{1})\equiv Q_{+}(x,p,t_{0})\label{eq:qdef}
\end{equation}

In this paper, we extend the treatment of Ref. \citep{q-measurement},
to consider the superposition $|\psi_{S}\rangle$ {[}Eq. (\ref{eq:sup-eigenstates}){]}
with general $r$. We treat the simple case where $\varphi=\pi/2$,
where the future boundary condition for the superposition is also
exactly the probabilistic mixture of the two Gaussian functions, $Q_{\pm}(x,t_{f})$,
given by Eq. (\ref{eq:fbc-prob-mix}). In this case, the variance
$\sigma_{x}^{2}=1+e^{-2r}$ is for general $r$. One can define a
set of amplitudes $\mathcal{S}_{+}=\{x(t)\}$ that emanate from the
positive Gaussian $Q_{+}(x,t_{f})$ and return to the Gaussian $Q_{+}(x)$,
as in the simulation of the state $|\frac{x_{1}}{2},r\rangle$. This
is shown in Figure \ref{fig:fb-1-1} for the coherent state, where
$r=0$. Hence, one can define the postselected distribution conditioned
on the system detected in the state $|\frac{x_{1}}{2},r\rangle$ with
the positive mean $x_{1}/2$. However, the sign of the detected amplitude
$x(t_{f})$ can only distinguish the two states, $|\frac{x_{1}}{2},r\rangle$
and $|-\frac{x_{1}}{2},r\rangle$ where $x_{1}$ (or $r$) is very
large. Nonetheless, one can define $B_{+}$ as the group of amplitudes
that are positive at time $t_{f}$, and recognize that this corresponds
exactly to the Gaussian $Q_{+}(x,t_{f})$ in the limit of large $x_{1}$
and $r$.

\paragraph*{Evaluation of the postselected state}

In order to evaluate the distribution $Q(x,p,t_{0}|B_{+})$, we apply
the conditional distribution $Q(p|x)$.\textcolor{red}{} Hence, we
write
\begin{eqnarray}
Q(x,p,t_{0}|B_{+}) & = & Q_{+}(x)Q(p|x)\nonumber \\
 & = & \frac{Q_{+}(x)Q(x,p,0)}{(Q_{+}(x)+Q_{-}(x))/2}\label{eq:qrel}
\end{eqnarray}
where $Q(p|x)$, $Q(x,p,0)$ and $Q(x,0)$ are defined by Eqs. (\ref{eq:cond-1}),
(\ref{eq:Q-sup-1}) and (\ref{eq:marg}), and we have taken the limit
of large $r$, or where $\varphi=\pi/2$, so that $T_{e}\rightarrow0$.
\textcolor{green}{} Inserting the solutions, the state inferred
at the time $t_{0}$, given a positive outcome for $\hat{x}$, is
\begin{eqnarray}
Q(x,p,t_{0}|B_{+}) & = & \frac{e^{-(x-|x_{1}|)^{2}/2\sigma_{x}^{2}}}{\sqrt{2\pi}\sigma_{x}}\frac{e^{-p^{2}/2\sigma_{p}^{2}}}{\sqrt{2\pi}\sigma_{p}f(\varphi)}\nonumber \\
 &  & \times\Bigl[1+\frac{\cos(\varphi+p|x_{1}|/\sigma_{x}^{2})}{\cosh\left(x|x_{1}|/\sigma_{x}^{2}\right)}\Bigl]\label{eq:ps-1}
\end{eqnarray}
\textcolor{black}{This solution is valid for superpositions of type
(\ref{eq:sup-eigenstates}) in the limit of $r\rightarrow\infty$,
and exact for superpositions (\ref{eq:sup-eigenstates}) where $\varphi=\pi/2$.
As deduced in Ref. \citep{q-measurement} for large $r$, the postselected
state has both the bivariate Gaussian with positive mean value $x_{1}$,
and the fringe term.}

\subsection{Hidden variables}

We examine the postselected distribution to show that it cannot correspond
to the Q function of a quantum state. The variables $x_{A}$ and $p_{A}$
are hence be referred to as ``hidden variables'' (Conclusion (3)
of this paper).

\subsubsection{Variances of the postselected distribution: Analytical analysis}

To allow exact solutions, we examine the case $\varphi=\pi/2$. We
find the distribution $Q(p,t_{0}|B_{+})$ for $p$ at the initial
time $t_{0}$, conditioned on the branch $B_{+}$ for a positive outcome,
by integrating (\ref{eq:ps-1}) over $x$. We find\textcolor{black}{
\begin{align}
Q(p,t_{0}|B_{+}) & =\frac{e^{-p^{2}/2\sigma_{p}^{2}}}{\sqrt{2\pi}\sigma_{p}}\Bigl[1-e^{-x_{1}^{2}/2\sigma_{x}^{2}}\sin(\frac{p|x_{1}|}{\sigma_{x}^{2}})\Bigl]\nonumber \\
\label{eq:qp}
\end{align}
}\textcolor{green}{}\textcolor{black}{where we have used that
\begin{equation}
\intop dx\frac{e^{-x^{2}/2\sigma_{x}^{2}}}{1+e^{-2x|x_{1}|/\sigma_{x}^{2}}}=\sqrt{\frac{\pi}{2}}\sigma_{x}\label{eq:int}
\end{equation}
}\textcolor{red}{}\textcolor{red}{}The distribution $Q(p,t_{0}|B_{-})$
conditioned on the negative outcome $-x_{1}$ (the negative branch)
is the same as $Q(p,t_{0}|B_{+})$. For a fixed superposition $|\psi_{S}\rangle$
where $|x_{1}|$ is fixed, the distribution for $p$ is constant,
independent of the branch, or set of outcomes $\{x(t_{f})\}$, being
considered. The sign associated with the fringe pattern is changed
by the phase $\varphi$ (whether $\pi/2$ of $-\pi/2$) associated
with the superposition, not by the branch.

\begin{figure}
\centering{}\includegraphics[width=0.8\columnwidth]{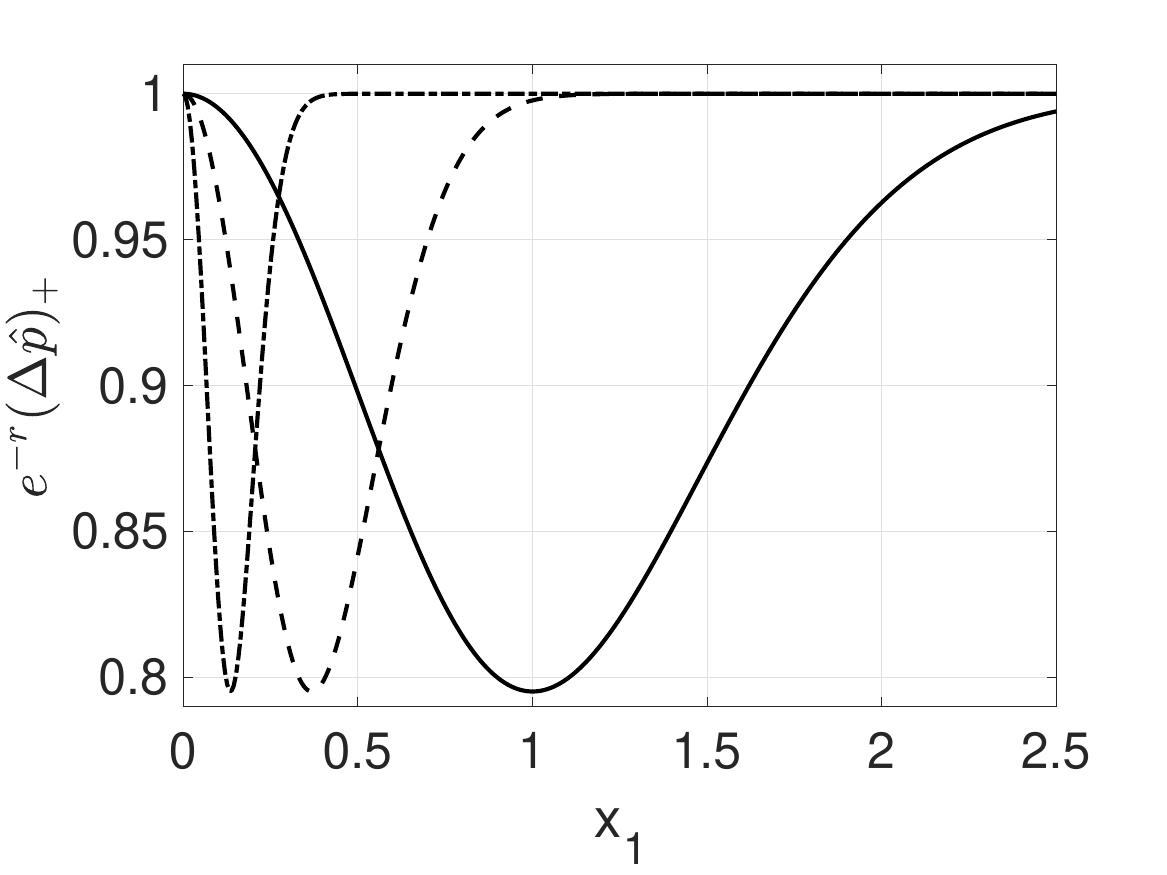}\caption{\textcolor{black}{Here we plot the $e^{-r}(\Delta\hat{p})_{+}$ where
$(\Delta\hat{p})_{+}^{2}$ {[}Eq. (\ref{eq:anp}){]} is the analytical
variance of the distribution for $x$ and $p$ conditioned on a positive
outcome $x_{1}$ for $\hat{x}$, for the system prepared in a superposition
$|\psi_{S}\rangle$ {[}Eq. (\ref{eq:sup-sq}){]} with $\varphi=\pi/2$.
The dashed-dotted line is for a superposition $|\psi_{S}\rangle$
of two eigenstates of $\hat{x}$ with $r=2$. The dashed line is for
$|\psi_{S}\rangle$ with $r=1$. The solid line is for $r=0$.\label{fig:analytical-var-p}}}
\end{figure}

The variance $(\Delta p)_{+}^{2}$ in $p$ of the postselected distribution
$Q(x,p,t_{0}|B_{+})$ is deduced.
\begin{eqnarray}
\langle p^{2}\rangle_{+} & = & \sigma_{p}^{2}\nonumber \\
\langle p\rangle_{+} & = & \intop dp\frac{e^{-p^{2}/2\sigma_{p}^{2}}}{\sqrt{2\pi}\sigma_{p}}\Bigl[1-e^{-x_{1}^{2}/2\sigma_{x}^{2}}\sin(\frac{p|x_{1}|}{\sigma_{x}^{2}})p\Bigl]\nonumber \\
 & = & -\frac{\sigma_{p}^{2}x_{1}}{\sigma_{x}^{2}}e^{-x_{1}^{2}\left(1+\sigma_{p}^{2}/\sigma_{x}^{2}\right)/2\sigma_{x}^{2}}\label{eq:relp}
\end{eqnarray}
Hence
\begin{align}
(\Delta p)_{+}^{2} & =1+e^{2r}-x_{1}^{2}\frac{\left(1+e^{2r}\right)^{2}}{\left(1+e^{-2r}\right)^{2}}e^{-\frac{x_{1}^{2}\left(1+\frac{1+e^{2r}}{1+e^{-2r}}\right)}{1+e^{-2r}}}\label{eq:anp}\\
 & \rightarrow1+e^{2r}-x_{1}^{2}e^{4r}e^{-x_{1}^{2}e^{2r}}
\end{align}
where the limit in the last line is for large $r$. If the distribution
corresponds to a Q function, the measured variances in $\hat{x}_{B}$
and $\hat{p}_{B}$ for a given state will be
\begin{eqnarray}
(\Delta\hat{x})_{+}^{2} & = & (\Delta x)_{+}^{2}-1\nonumber \\
(\Delta\hat{p})_{+}^{2} & = & (\Delta p)_{+}^{2}-1\label{eq:varp}
\end{eqnarray}
The variance $(\Delta\hat{p})_{+}^{2}$ reduces below $e^{2r}$, as
plotted in Figure \ref{fig:analytical-var-p}.

We also calculate the distribution $Q(x,t_{0}|B_{+})$ inferred for
$x$, given the outcome $x_{1}$. In this case for $\varphi=\pi/2$,
we find\textcolor{green}{}\textcolor{black}{
\begin{equation}
Q(x,t_{0}|B_{+})=\frac{1}{\sqrt{2\pi}\sigma_{x}}e^{-(x-x_{1})^{2}/2\sigma_{x}^{2}}\label{eq:disx}
\end{equation}
}Hence
\begin{align}
(\Delta x)_{+}^{2} & =\sigma_{x}^{2}=1+e^{-2r}\label{eq:varx}
\end{align}
The measured variance in $\hat{x}_{A}$ is $(\Delta\hat{x})_{+}^{2}\rightarrow e^{-2r}$
\citep{q-measurement}.

We conclude that the uncertainty product associated with the postselected
distribution $Q(x,p,t_{0}|B_{+})$ is reduced below that given by
the Heisenberg uncertainty relation $(\Delta\hat{x})(\Delta\hat{p})\geq1$:
\begin{equation}
(\Delta\hat{x})_{+}(\Delta\hat{p})_{+}<1\label{eq:hp}
\end{equation}
The postselected distribution $Q(x,p,t_{0}|B_{+})$ {[}Eq. (\ref{eq:ps-1}){]}
cannot correspond to the Q function of a quantum state.

\begin{figure}
\begin{centering}
\includegraphics[width=0.8\columnwidth]{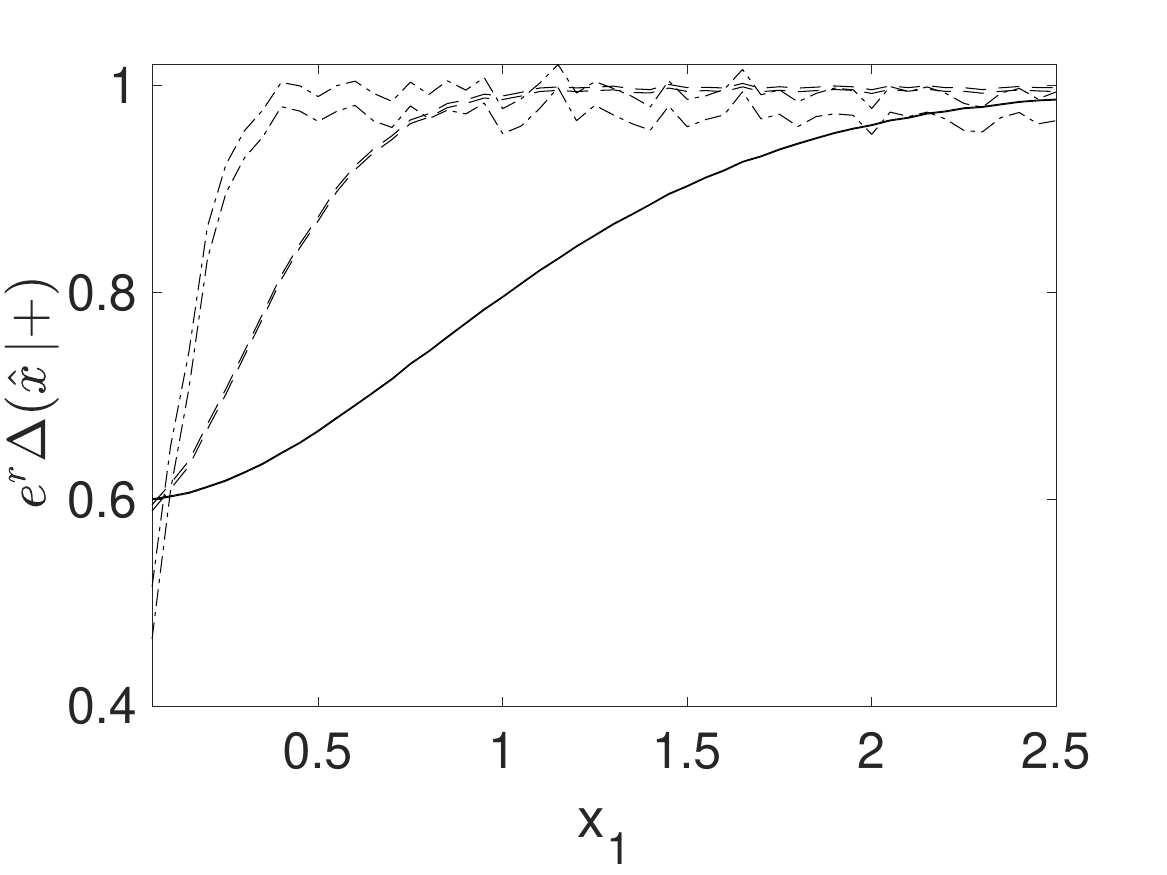}
\par\end{centering}
\begin{centering}
\includegraphics[width=0.8\columnwidth]{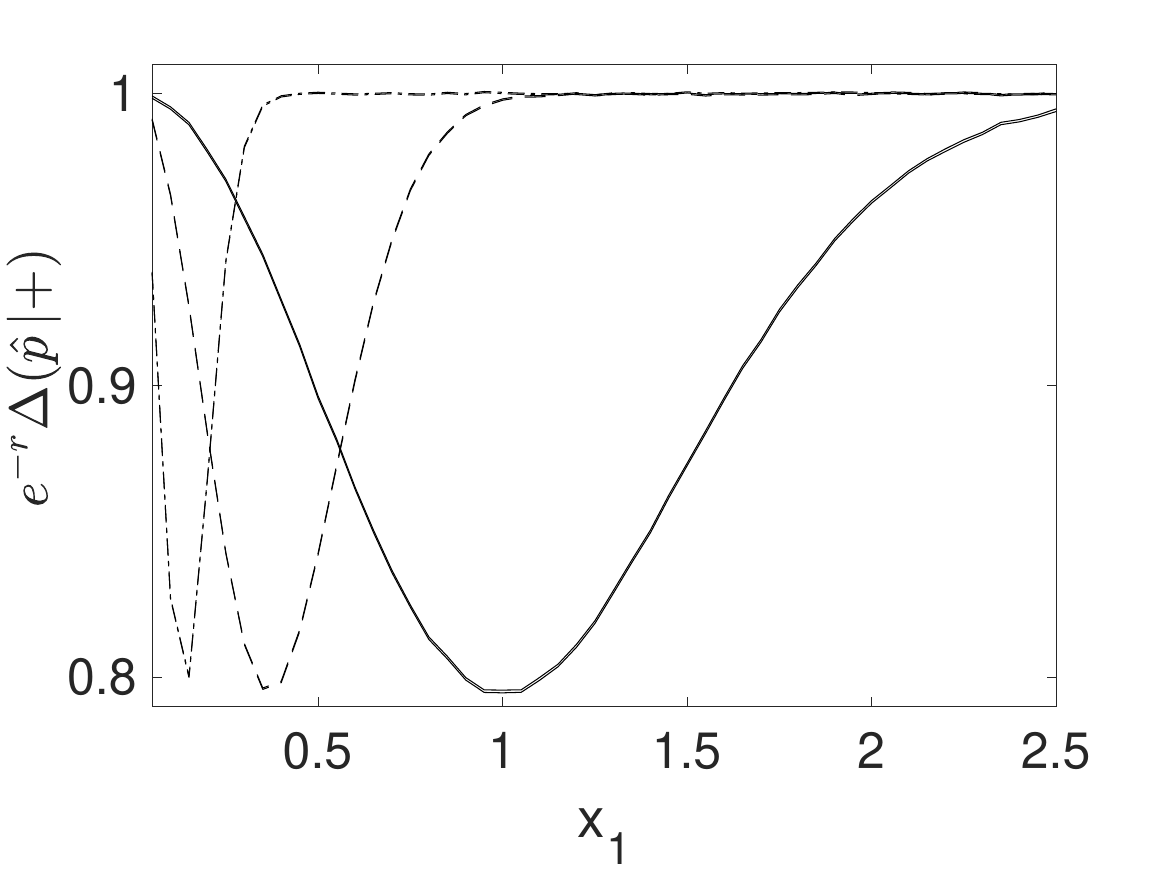}
\par\end{centering}
\begin{centering}
\includegraphics[width=0.8\columnwidth]{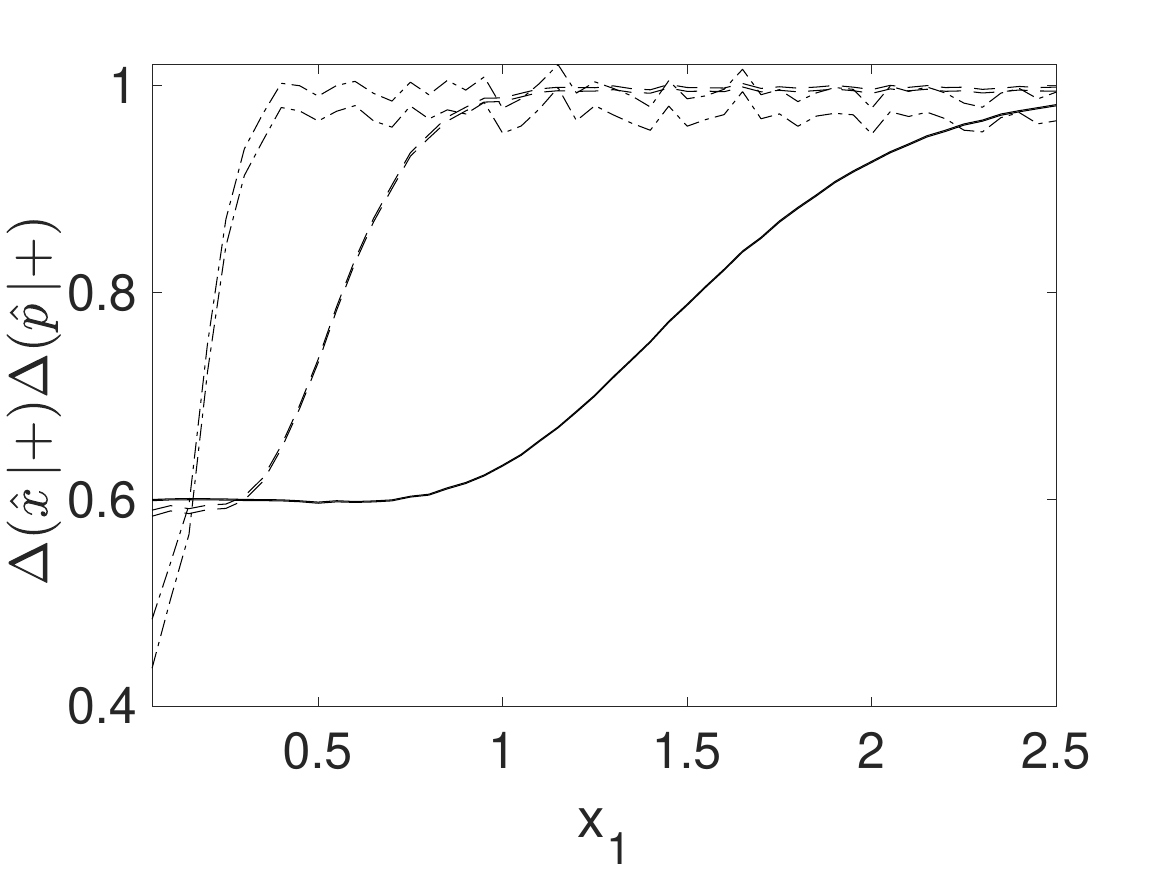}
\par\end{centering}
\caption{Here we plot the variances $\Delta(x|+)$, $\Delta(p|+)$ and the
uncertainty product $\epsilon=\Delta(x|+)\Delta(p|+)$ of the postselected
distribution as conditioned on a positive outcome $x_{1}$ for $\hat{x}$,
\textcolor{black}{for the system prepared in a superposition $|\psi_{S}\rangle$
{[}Eq. (\ref{eq:sup-sq}){]} with $\varphi=\pi/2$.} The variances
are evaluated numerically from the forward-backward stochastic solutions.
The upper dashed-dotted line is for a superposition $|\psi_{S}\rangle$
of two eigenstates of $\hat{x}$ with $r=2$. The dashed line is
for $|\psi_{S}\rangle$ with $r=1$. The solid line is for $r=0$.
We choose $gt_{f}=3$. The two parallel lines indicate the upper
and lower error bounds from sampling errors, with $1.2\times10^{7}$
trajectories. \label{fig:cond-var-inferred}\textcolor{green}{ }\textcolor{red}{}}
\end{figure}

\subsubsection{Variances: Numerical results}

We may deduce the postselected state (\ref{eq:ps-1}) numerically.
We trace a given set of trajectories in $x$ back to the time $t_{0}=0$
given the post-selection of $x(t_{f})>0$, and construct the distribution
of $x$ at time $t_{0}=0$ for all such trajectories. At $t_{0}=0$,
each value of $x$ is connected to a set of trajectories in $p$,
according to the conditional distribution $Q(p|x)$.

We construct the joint distribution $Q_{+}(x,p,t_{0})\equiv Q(x,p,t_{0}|+)$
given by Eq. (\ref{eq:qloop}), describing the values $x$ and $p$
at the time $t_{1}=0$, conditioned on the positive outcome for $\hat{x}$,
meaning that we take the subset of trajectories with $x(t_{f})>0$,
assuming $gt_{f}$ is sufficiently large. We carry out the procedure
for arbitrary $r$, which both tests and extends the analytical results
above. We then determine the variances $[\Delta(x|+)]^{2}$ and $[\Delta(p|+)]^{2}$
for $x$ and $p$ for this distribution, and define the observed variances
once anti-normal ordering is accounted for \citep{q-measurement}:
\begin{eqnarray}
[\Delta(\hat{x}|+)]^{2} & = & [\Delta(x|+)]^{2}-1\nonumber \\{}
[\Delta(\hat{p}|+)]^{2} & = & [\Delta(p|+)]^{2}-1\label{eq:infvar}
\end{eqnarray}
Similar variances $[\Delta(\hat{x}|-)]^{2}$ and $[\Delta(\hat{p}|-)]^{2}$
could be determined for the trajectories postselected on the $x(t_{f})<0$
corresponding to the outcome $-x_{1}$. This tells us what we infer
about the original state (in the reality model) at time $t=0$ based
on the measurement outcome, whether $+$ or $-$.

The variances are given in Figure \ref{fig:cond-var-inferred} versus
$x_{1}$ for a large value of $gt$, for the superposition $|\psi_{S}\rangle$
where $\varphi=\pi/2$. We also define the uncertainty product for
the inferred initial state:
\begin{equation}
\epsilon=\Delta(\hat{x}|+)\Delta(\hat{p}|+).\label{eq:epr-prod}
\end{equation}
From the Figures \ref{fig:cond-var-inferred}, see that $\epsilon<1$
for all $x_{1}$ (and $\alpha_{0}$) although $\epsilon\rightarrow1$
as $x_{1}\rightarrow\infty$.

The figures show what happens if we postselect on the positive outcome
for a measurement of $\hat{x}$. In the limit of $x_{1}$ large, or
$r$ large, we expect consistency with the predictions above, that
\begin{align}
[\Delta(\hat{x}|+)]^{2}\rightarrow(\Delta\hat{x})_{+}^{2} & =e^{-2r}\label{eq:varx-1}
\end{align}
which implies $[\Delta(\hat{x}|+)]^{2}e^{2r}=(\Delta\hat{x})_{+}^{2}e^{2r}=1$.
This is because where $x_{1}$ is large, even for smaller $r$, the
two states $|x_{1}/2,r\rangle$ and $|-x_{1}/2,r\rangle$ of the superposition
are well separated. The superposition of two eigenstates of $\hat{x}$
is modeled by choosing $r\geq2$. Provided $t_{f}$ is sufficiently
large, there are always two branches, one for $x_{1}$ and one for
$-x_{1}$, that are distinguished by the sign of $x(t_{f})$. The
variance $[\Delta(x|+)]^{2}$ is hence always $1$, for $r$ sufficiently
large, as shown in Figures \ref{fig:cond-var-inferred}, where we
choose $r=2$, and see that $[\Delta(\hat{x}|+)]^{2}\rightarrow e^{-2r}$.
This is not the case for a given $gt_{f}$, when $x_{1}$ is sufficiently
small, however. By binning the $x(t_{f})$ into positive and negative
values at time $t_{f}$, the variance in $x$ for the distribution
determined by the simulations is reduced below that of the Gaussian
$Q_{+}(x,t)$ {[}Eq. (\ref{eq:fbc}){]} which has a variance of $\sigma_{x}$.
This is pronounced when the two Gaussians $Q_{+}(x,t)$ and $Q_{-}(x,t)$
overlap, at smaller values of $x_{1}$.

We see from Figures \ref{fig:cond-var-inferred} that for smaller
$r$, e.g. $r=0$, where we consider a superposition of coherent states,
the result (\ref{eq:varx-1}) holds for sufficiently large $x_{1}$,
as expected. For the cat state where $r=0$ and $x_{1}=2\alpha_{0}$
is large, the variance $[\Delta(\hat{x}|+)]^{2}$ is reduced to $e^{-2r}=1$.
This is explained as follows. The overall variance in $x$ at the
time $t_{0}=0$ is large, due to there being two states comprising
the superposition, but the final amplified outcome of either $x_{1}=2\alpha_{0}$
or $-x_{1}=-2\alpha_{0}$ (Fig. \ref{fig:x-trajectories-cat-superposition-1})
links the trajectory back to only \emph{one} of these states, $|\alpha_{0}\rangle$
or $|-\alpha_{0}\rangle$, which has a variance in $x$ of $1$.

The numerical values $\Delta(\hat{p}|+)$ shown in Figure \ref{fig:cond-var-inferred}
can be compared with the analytical results plotted in Figure \ref{fig:analytical-var-p}
for $(\Delta p)_{+}^{2}$ as obtained in Eq. (\ref{eq:anp}). We find
excellent agreement.\textcolor{red}{{} }

\section{Measurement via coupling to a meter: two-mode cat states}

In this section, we study the Schrodinger cat generated when the system
has become entangled with a separate system, a macroscopic meter.
This allows us to examine the collapse of the wave function (Conclusion
(2) of this paper). We seek to answer the question: What is it that
the meter measures about the system? If we consider a superposition
of two quantum states, then typically, as in the cat paradox, a measurement
$\hat{O}$ may be constructed to determine ``which of the two states
the system is in''.

We consider the entangled meter-system described by the state (\ref{eq:ent-two-coherent-1}).
We simulate direct measurements on\emph{ }both the system $A$ and
the meter $B$, and examine the solutions for the amplitudes $x_{A}(t_{f})$
and $x_{B}(t_{f})$. Here, $x_{A}(t)$ and $x_{B}(t)$ are the values
of the amplitudes at the time $t$ in the simulation, and $t_{f}$
is the time after the amplification, when the measurement has been
completed. In the Q model of reality, $x_{A}(t_{f})$ and $x_{B}(t_{f})$
give the outcomes of the measurements $\hat{x}_{A}$ and $\hat{x}_{B}$,
defined for two fields by Eq. (\ref{eq:x}). This enables us to demonstrate
the correlation between the sign of the two measurement outcomes $x_{A}(t_{f})$
and $x_{B}(t_{f})$. In this way, we identify what it is about $A$
that we are inferring, from the outcome $x_{B}(t_{f})$ of $\hat{x}_{B}$
of the meter.

For definiteness, we consider that the system $S\equiv A$ is initially
in the superposition
\begin{equation}
|\psi_{S}\rangle=\frac{1}{\sqrt{2}}(|x_{1}\rangle+e^{i\varphi}|-x_{1}\rangle)\label{eq:sup-special}
\end{equation}
where $|x_{1}\rangle$ and $|-x_{1}\rangle$ are eigenstates of $\hat{x}_{A}$.
A measurement $\hat{O}$ is made on system $A$ to  infer ``which
of the two states the system $A$ is in'', $|x_{1}\rangle$ or $|-x_{1}\rangle$.
The measurement $\hat{O}$ can be made by coupling the system $A$
to a meter $M$, which we model as field $B$. A prototype for the
state \emph{after} such a coupling is the entangled two-mode state
(\ref{eq:ent-two-coherent-1})
\begin{eqnarray}
|\psi_{ent}\rangle & = & N_{2}\{|\frac{x_{1}}{2},r\rangle|\frac{x_{1B}}{2},r_{2}\rangle+e^{i\varphi}|-\frac{x_{1}}{2},r\rangle|-\frac{x_{1B}}{2},r_{2}\rangle\}\nonumber \\
\label{eq:ent-cat-1}
\end{eqnarray}
where $N_{2}$ is the normalization constant. We approximate the eigenstates
of $\hat{x}_{A}$ by the squeezed states (\ref{eq:sq}) with $r$
large. The state becomes in the limit of large $r$
\begin{equation}
|\psi_{ent}\rangle=N_{2}\{|x_{1}\rangle|\frac{x_{1B}}{2},r_{2}\rangle+e^{i\varphi}|-x_{1}\rangle|-\frac{x_{1B}}{2},r_{2}\rangle\}\label{eq:ent-meter-x}
\end{equation}
We will later consider the case where for the meter $r_{2}=0$ and
$x_{1B}=2\beta_{0}$. We take $x_{1}$, $x_{1B}$ and $\beta_{0}$
to be real, and $|\beta_{0}\rangle$ and $|-\beta_{0}\rangle$ to
be coherent states for the meter mode $B$. It is understood that
for an effective measurement, $\beta_{0}$ would become large. The
measurement of the quadrature phase amplitude $\hat{x}_{B}=\hat{b}+\hat{b}^{\dagger}$
of mode $B$ would indicate ``whether the system is in the state
$|\beta_{0}\rangle$ or $|-\beta_{0}\rangle$'', and hence also be
a measurement to indicate the state of the system $A$, ``whether
$|x_{1}\rangle$ or $|-x_{1}\rangle$''. 

When $r=r_{2}=0$, the state (\ref{eq:ent-cat-1}) is a two-mode entangled
cat state
\begin{equation}
|\psi_{Cat}\rangle=N_{2}\{|\alpha_{0}\rangle|\beta_{0}\rangle+e^{i\varphi}|-\alpha_{0}\rangle|-\beta_{0}\rangle\}\label{eq:ent-cat}
\end{equation}
\textcolor{green}{}where $\alpha_{0}=x_{1}/2$, and $|\pm\alpha_{0}\rangle$
are coherent states for the system mode $A$ and $|\pm\beta_{0}\rangle$
are coherent states for mode $B$ \citep{wang-two-mode-cat}. Here,
\[
N_{2}=\frac{1}{\sqrt{2(1+[\cos\varphi]e^{-2|\alpha_{0}|^{2}-2|\beta_{0}|^{2}})}}
\]
which becomes $1/\sqrt{2}$ for $\varphi=\pi/2$. The measurement
on $B$ is then intended to infer whether the system $A$ ``is in
state $|\alpha\rangle$ or $|-\alpha\rangle$''.

\subsection{Measurements on the meter and system}

\subsubsection{Hamiltonian and stochastic equations}

We consider a measurement of $\hat{x}$ for both the meter $B$ and
the system $A$. The systems $A$ and $B$ are independently amplified
by interacting with the parametric medium. The interactions are given
by Hamiltonians 
\begin{eqnarray}
H_{amp}^{A} & = & \frac{i\hbar g_{A}}{2}\left[\hat{a}^{\dagger2}-\hat{a}^{2}\right]\nonumber \\
H_{amp}^{B} & = & \frac{i\hbar g_{B}}{2}\left[\hat{b}^{\dagger2}-\hat{b}^{2}\right]\label{eq:two-ham-two}
\end{eqnarray}
as in (\ref{eq:hamb-1}), where $g_{A}$ and $g_{B}$ are real. Here,
$g_{A}>0$ amplifies $\hat{x}_{A}$ and $g_{A}<0$ amplifies $\hat{p}_{A}$.
Similarly, for the meter system $B$, $\hat{x}_{B}$ is amplified
when $g_{B}>0$, and $\hat{p}_{B}$ is amplified when $g_{B}<0$.
It is possible to independently measure either $\hat{x}$ or $\hat{p}$
of the fields $A$ and $B$.

The $Q$ function of the two-mode system is defined as 
\begin{equation}
Q(x_{A},p_{A},x_{B},p_{B},t)=\frac{1}{\pi^{2}}\langle\alpha|\langle\beta|\rho(t)|\beta\rangle|\alpha\rangle\label{eq:q-two-mode}
\end{equation}
where $|\alpha\rangle$ is a coherent state for mode $A$ and $|\beta\rangle$
is a coherent state for mode $B$. Here, $\alpha=(x_{A}+ip_{A})/2$
and $\beta=(x_{B}+ip_{B})/2$. We solve for the dynamics of the amplitudes
$x_{A},$ $p_{A}$, $x_{B}$ and $p_{B}$.

We first consider joint measurements of $\hat{x}_{A}$ and $\hat{x}_{B}$.
Following the procedure given in Refs. \citep{q-measurement,q-retrocausal-model-measurement},
we derive equations for the dynamical evolution of the amplitudes
$x_{A}$, $p_{A}$, $x_{B}$ and $p_{B}$ which are defined by the
$Q$ function. The initial time is $t_{0}=0$. The total time of the
interaction is $t=t_{f}$. The equations for the measurement $\hat{x}_{B}$
on the meter $B$ are
\begin{equation}
\frac{dx_{B}}{dt_{-}}=-g_{B}x_{B}+\sqrt{2g_{B}}\xi_{B1}\label{eq:forwardSDE-2-1}
\end{equation}
where $t_{-}=-t$ with a boundary condition at time $t_{-}=-t_{f}$,
and
\begin{align}
\frac{dp_{B}}{dt} & =-g_{B}p_{B}+\sqrt{2g_{B}}\xi_{B2}\label{eq:backwardSDE-2-1}
\end{align}
with a boundary condition at time $t=t_{0}$. The Gaussian random
noises $\xi_{\mu}\left(t\right)$ satisfy $\left\langle \xi_{B\mu}\left(t\right)\xi_{B\nu}\left(t'\right)\right\rangle =\delta_{\mu\nu}\delta\left(t-t'\right)$.
The equations for the measurement $\hat{x}_{A}$ at $A$ are
\begin{equation}
\frac{dx_{A}}{dt_{-}}=-g_{A}x_{A}+\sqrt{2g_{A}}\xi_{A1}\label{eq:forwardSDE-2-1-1}
\end{equation}
where $t_{-}=-t$ with a boundary condition at time $t_{f}$, and
\begin{align}
\frac{dp_{A}}{dt} & =-g_{A}p_{A}+\sqrt{2g_{A}}\xi_{A2}\label{eq:backwardSDE-2-1-1}
\end{align}
with a boundary condition at time $t_{0}$. The Gaussian random noises
$\xi_{\mu}\left(t\right)$ satisfy $\left\langle \xi_{A\mu}\left(t\right)\xi_{A\nu}\left(t'\right)\right\rangle =\delta_{\mu\nu}\delta\left(t-t'\right)$.

\subsubsection{Boundary conditions and Q functions}

The boundary conditions for the stochastic equations are determined
by the $Q$ functions at times $t=0$ and $t=t_{f}$. The $Q$ function
at the time $t=0$, for the system in state (\ref{eq:ent-cat-1}),
is\textcolor{green}{}\begin{widetext} 
\begin{eqnarray}
Q_{ent}(\lambda,t_{0}) & = & \frac{e^{-p_{A}^{2}/2\sigma_{p_{A}}^{2}-p_{B}^{2}/2\sigma_{p_{B}}^{2}}}{8\pi^{2}\sigma_{p_{A}}\sigma_{p_{B}}\sigma_{x_{A}}\sigma_{x_{B}}f(\varphi)}\Bigl\{ e^{-(x_{A}-x_{1})^{2}/2\sigma_{x_{A}}^{2}-(x_{B}-x_{1B})^{2}/2\sigma_{x_{B}}^{2}}+e^{-(x_{A}+x_{1})^{2}/2\sigma_{x_{B}}^{2}-(x_{B}+x_{1B})^{2}/2\sigma_{x_{B}}^{2}}\nonumber \\
 &  & +2e^{-x_{A}^{2}/2\sigma_{x_{A}}^{2}-x_{B}^{2}/2\sigma_{x_{B}}^{2}-x_{1}^{2}/2\sigma_{x_{A}}^{2}-x_{1B}^{2}/2\sigma_{x_{B}}^{2}}\cos\bigl(\varphi+|x_{1}|p_{A}/\sigma_{x_{A}}^{2}+|x_{1B}|p_{B}/\sigma_{x_{B}}^{2}\bigl)\Bigl\}\nonumber \\
\label{eq:Qent-2-1}
\end{eqnarray}
\end{widetext} where\textcolor{green}{}\textcolor{black}{}
\begin{eqnarray*}
f(\varphi) & = & 1+[\cos\varphi]e^{-x_{1}^{2}\left(1+\sigma_{p_{A}}^{2}/\sigma_{x_{A}}^{2}\right)/2\sigma_{x_{A}}^{2}}\\
 &  & \times e^{-x_{1B}^{2}\left(1+\sigma_{p_{B}}^{2}/\sigma_{x_{B}}^{2}\right)/2\sigma_{x_{B}}^{2}}
\end{eqnarray*}
is a normalization factor, and $f(\varphi)=1$ for $\varphi=\pi/2$.\textcolor{green}{}
Here, $\lambda=(x_{A},p_{A},x_{B},p_{B})$  $\sigma_{x_{A}}^{2}$,
$\sigma_{x_{B}}^{2}$, $\sigma_{p_{A}}^{2}$ and $\sigma_{p_{B}}^{2}$
are the variances of $x_{A}$, $x_{B}$, $p_{A}$ and $p_{B}$ respectively.
Hence $\sigma_{x_{A}}^{2}=1+e^{-2r}$ and $\sigma_{p_{A}}^{2}=1+e^{2r}$
where $r$ is the squeezing parameter defined for the squeezed state
for the mode $A$. Similarly, $\sigma_{x_{B}}^{2}=1+e^{-2r_{2}}$
and $\sigma_{p_{B}}^{2}=1+e^{2r_{2}}$ where $r_{2}$ is the squeezing
parameter for mode $B$. We note there are two Gaussian terms, with
peaks at $x_{A}=x_{1}$, $x_{1B}=x_{2}$ and at $x_{A}=-x_{1}$, $x_{1B}=-x_{2}$
respectively, as well as a third sinusoidal term which arises due
to the entangled nature of the state. The third term vanishes for
the system described as a mixture $\rho_{mix}^{(AB)}$ of the two
states $|\frac{x_{1}}{2},r\rangle|\frac{x_{1B}}{2},r_{2}\rangle$
and $|-\frac{x_{1}}{2},r\rangle|-\frac{x_{1B}}{2},r_{2}\rangle$.
\textcolor{black}{}

The solution for the $Q$ function of the amplified two-mode system
after the local interactions $H_{A}$ and $H_{B}$ for a time $t$
can be readily solved. The solution is given by the function (\ref{eq:Qent-2-1})
where the means ($x_{1}$ and $x_{1B}$ ) and variances ($\sigma_{x_{A}}^{2}$,
$\sigma_{p_{A}}^{2}$, $\sigma_{x_{B}}^{2}$, $\sigma_{p_{B}}^{2}$)
are now given as
\begin{align}
\sigma_{x_{A}}^{2}(t) & =1+e^{2(g_{A}t-r)}\nonumber \\
\sigma_{p_{A}}^{2}(t) & =1+e^{-2(g_{A}t-r)}\nonumber \\
x_{1} & \rightarrow x_{1}e^{g_{A}t}\label{eq:var-ent-amp-2}
\end{align}
and similarly
\begin{align}
\sigma_{x_{B}}^{2}(t) & =1+e^{2(g_{B}t-r_{2})}\nonumber \\
\sigma_{p_{B}}^{2}(t) & =1+e^{-2(g_{B}t-r_{2})}\nonumber \\
x_{1B} & \rightarrow x_{1B}e^{g_{B}t}\label{eq:var-ent-amp-1-1}
\end{align}
 Since $p$ decouples from $x$, the relevant boundary condition for
the trajectories $x_{A}(t)$ and $x_{B}(t)$ is determined by the
joint marginal for $x_{A}$ and $x_{B}$ at time $t_{f}$, which
can be found on integrating over $p_{A}$ and $p_{B}$.\textcolor{green}{}\textcolor{black}{
\[
Q(x_{A},x_{B},t)=\int dp_{A}dp_{B}Q_{ent}(\lambda,t)
\]
We find}\textcolor{green}{}\textcolor{red}{}\textcolor{green}{}\begin{widetext} 

\begin{align}
Q(x_{A},x_{B},t_{f}) & =\frac{1}{4\pi\sigma_{x_{A}}(t)\sigma_{x_{B}}(t)f(\varphi,t_{f})}\Bigl\{ e^{-(x_{A}-Gx_{1})^{2}/2\sigma_{x_{A}}^{2}(t)-(x_{B}-Gx_{1B})^{2}/2\sigma_{x_{B}}^{2}(t)}+e^{-(x_{A}+Gx_{1})^{2}/2\sigma_{x_{A}}^{2}(t)-(x_{B}+Gx_{1B})^{2}/2\sigma_{x_{B}}^{2}(t)}\nonumber \\
 & +2[\cos\varphi]e^{-x_{A}^{2}/2\sigma_{x_{A}}^{2}(t)-x_{B}^{2}/2\sigma_{x_{B}}^{2}(t)}\times e^{-G^{2}x_{1}^{2}\left(1+\sigma_{p_{A}}^{2}(t)/\sigma_{x_{A}}^{2}(t)\right)/2\sigma_{x_{A}}^{2}(t)}e^{-G^{2}x_{1B}^{2}\left(1+\sigma_{p_{B}}^{2}(t)/\sigma_{x_{B}}^{2}(t)\right)/2\sigma_{x_{B}}^{2}(t)}\Bigl\}\label{eq:marg-1}
\end{align}
\end{widetext} where we take $g_{A}=g_{B}=g$, $G=e^{gt_{f}}$. Here,
\begin{align}
f(\varphi,t) & =1+[\cos\varphi]e^{-G^{2}(t)x_{1}^{2}\left(1+\sigma_{p_{A}}^{2}(t)/\sigma_{x_{A}}^{2}(t)\right)/2\sigma_{x_{A}}^{2}(t)}\nonumber \\
 & \times e^{-G^{2}(t)x_{1B}^{2}\left(1+\sigma_{p_{B}}^{2}(t)/\sigma_{x_{B}}^{2}(t)\right)/2\sigma_{x_{B}}^{2}(t)}
\end{align}
 where $G(t)=e^{gt}$. For the cat state, we set $r=r_{2}=0$, and
$x_{1}=2\alpha_{0}$, $x_{1B}=2\beta_{0}$. For the choice of a superposition
of eigenstates ($r\rightarrow\infty$) where $\sigma_{p}^{2}\rightarrow\infty$,
and similarly in the limit of large $\beta_{0}$, where the meter
is macrosc\textcolor{black}{opic, or else when $\cos\varphi=0$, the
integration eliminates the third term }\citep{q-measurement}\textcolor{black}{.}

\subsection{Two convenient meters}

There are two choices of parameters that model a meter. The first
is where the system $B$ is prepared in terms of highly squeezed states
so that $r_{2}\rightarrow\infty$. This implies $\sigma_{p_{B}}^{2}\rightarrow\infty$,
implying that the third term in $Q_{ent}(\lambda,t_{0})$ of Eq. (\ref{eq:Qent-2-1})
vanishes. The two states $|\frac{x_{1B}}{2},r_{2}\rangle$ and $|-\frac{x_{1B}}{2},r_{2}\rangle$
are orthogonal in this limit, and thus enable a projection of the
state of the system for small values of $x_{1B}$.

The second choice of meter is the coherent-state meter, where $r_{2}=0$
and $\varphi=\pi/2$. Where $\beta_{0}\rightarrow\infty$, so that
the meter is macroscopic, we see that the third term in $Q_{ent}(\lambda,t_{0})$
of Eq. (\ref{eq:Qent-2-1}) will decay to become negligible. Conveniently,
however, choosing $\varphi=\pm\pi/2$, we find that the third sinusoidal
term in $Q_{ent}(\lambda,t_{0})$ of Eq. (\ref{eq:Qent-2-1}) will
vanish \emph{exactly}, for all choices of $r$ and $\beta_{0}.$ This
provides a convenient meter, provided also that $x_{2}=2\beta_{0}$
is macroscopic, which ensures the two states $|\beta_{0}\rangle$
and $|-\beta_{0}\rangle$ of the meter are orthogonal. \textcolor{red}{}\textbf{\textcolor{black}{}}In
both cases of meter above, because the third term in $Q_{ent}(\lambda,t_{0})$
of Eq. (\ref{eq:Qent-2-1}) vanishes, the backward trajectories are
as for a mixture $\rho_{mix}^{(AB)}$ of the two states $|\frac{x_{1}}{2},r\rangle|\frac{x_{2}}{2},r_{2}\rangle$
and $|\frac{x_{1}}{2},r\rangle|\frac{x_{2}}{2},r_{2}\rangle$.

\begin{figure}[t]
\begin{centering}
\includegraphics[width=0.5\columnwidth]{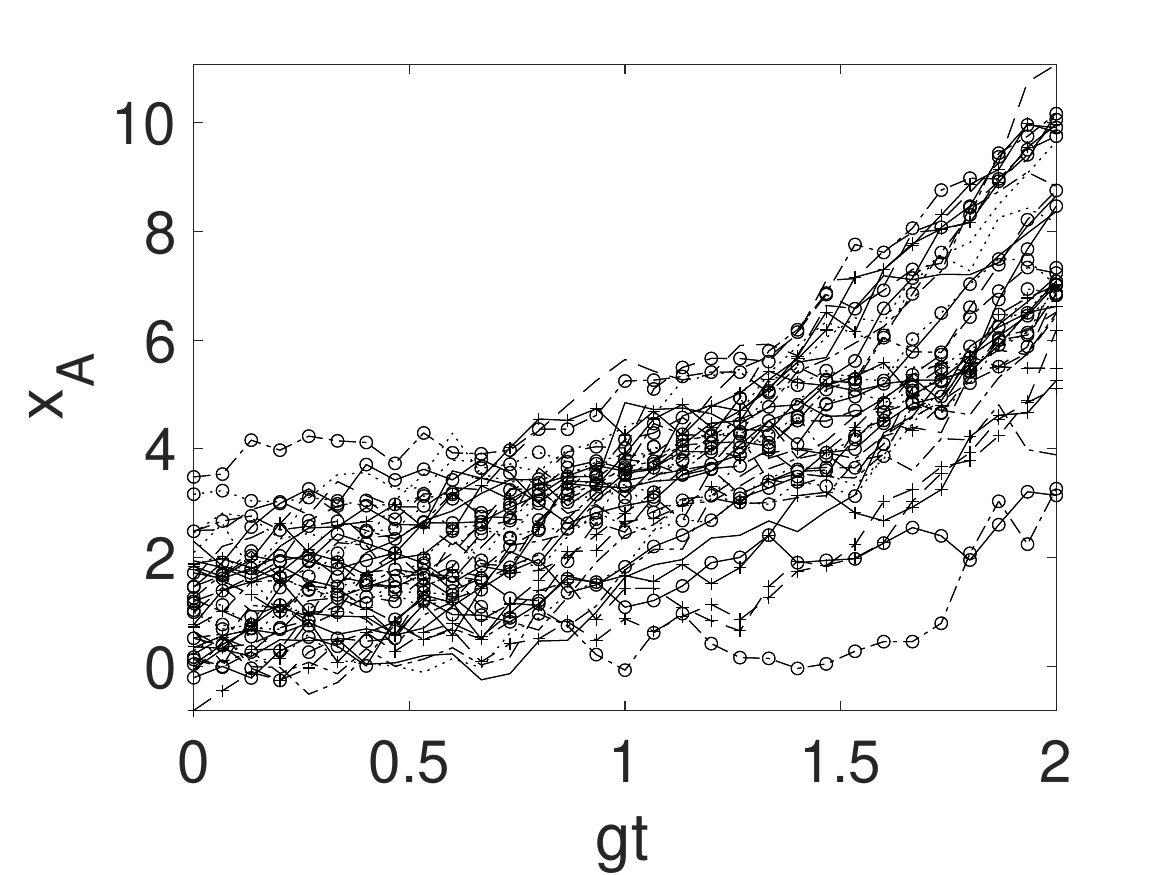}\includegraphics[width=0.5\columnwidth]{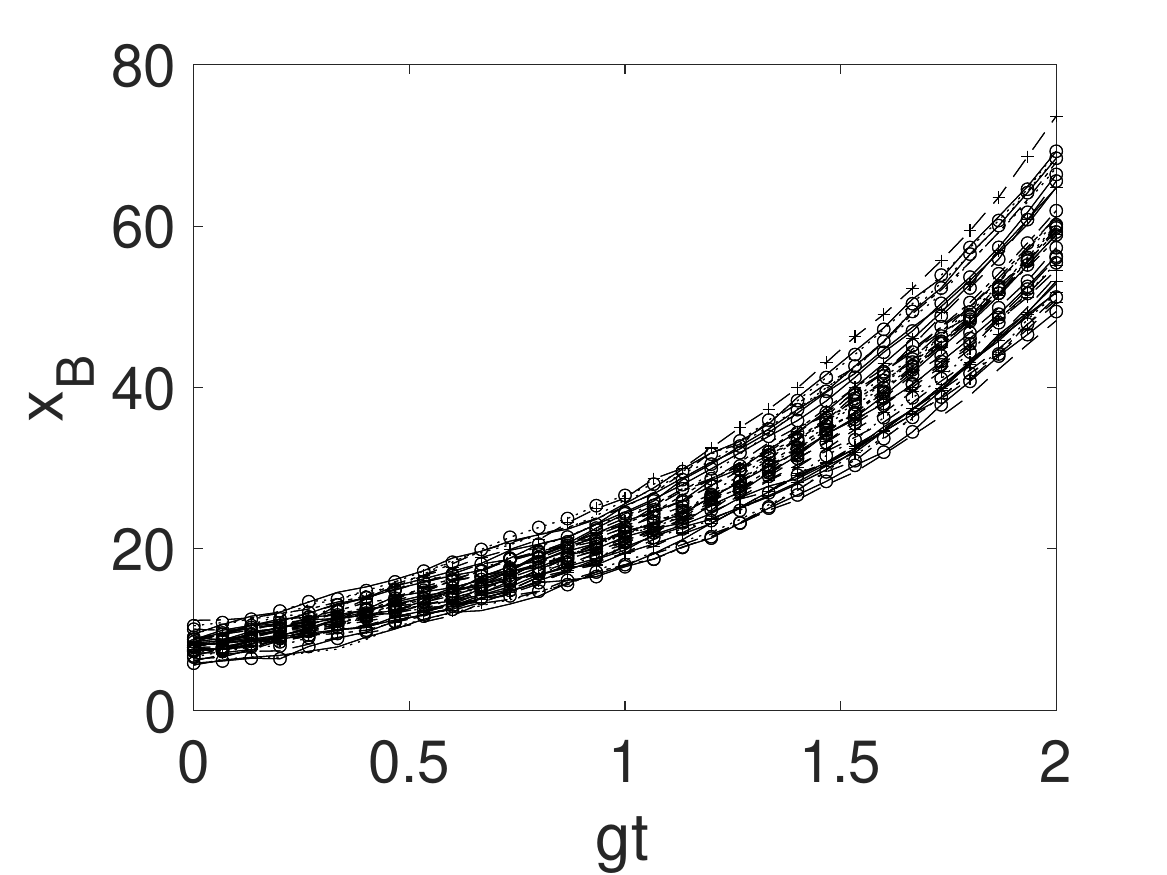}
\par\end{centering}
\begin{centering}
\includegraphics[width=0.5\columnwidth]{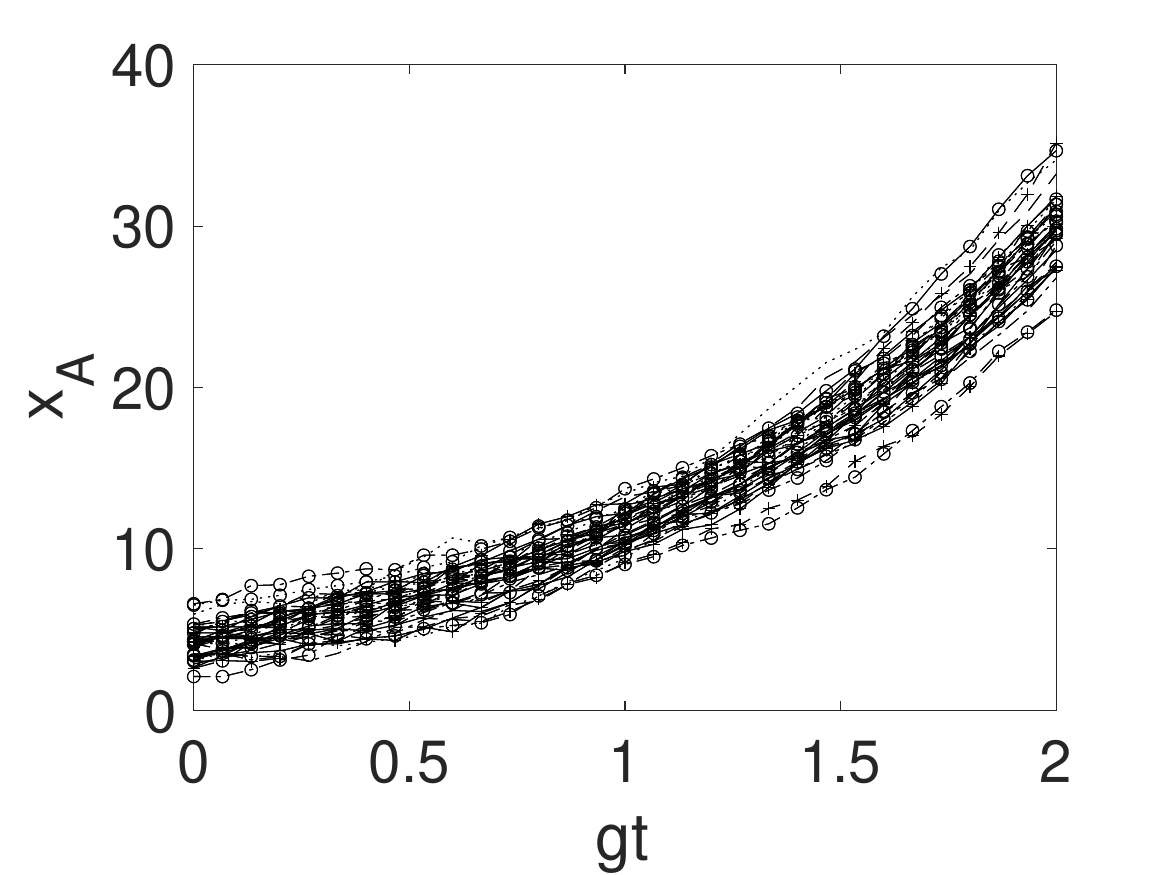}\includegraphics[width=0.5\columnwidth]{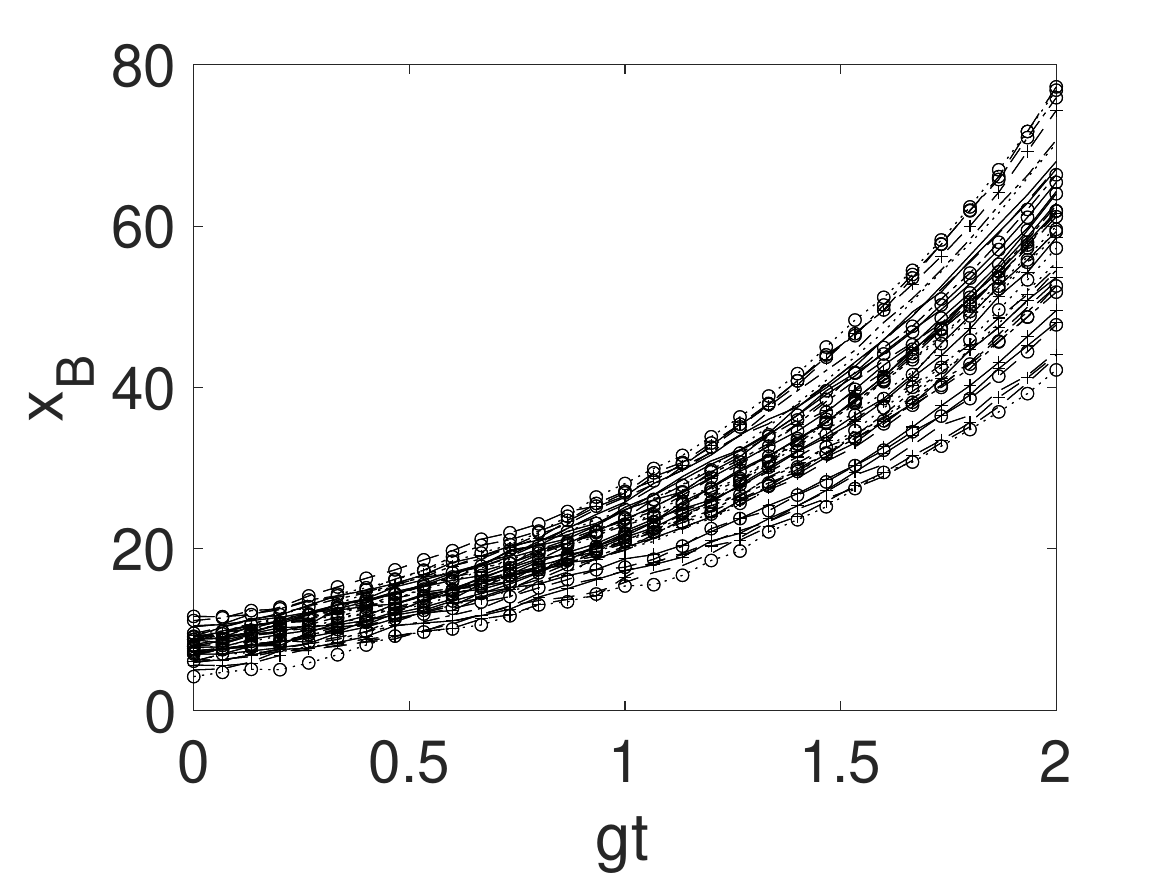}
\par\end{centering}
\caption{Measurement of $\hat{x}_{A}$ and $\hat{x}_{B}$ on the systems prepared
in the entangled state (\ref{eq:Qent-2-1}) with \textcolor{black}{$\varphi=\pi/2$.
Here, $r_{2}=0$, implying a coherent-state meter $B$, where $\beta_{0}=4$.
}\textcolor{red}{}Plots show the trajectories for $x_{A}$ (left
figure) and $x_{B}$ (right figure) conditioned on a positive outcome
for the meter, $x_{B}(t_{f})>0$. We see perfect correlation between
the sign of the final outcomes $x_{A}(t_{f})$ and $x_{B}(t_{f})$,
which indicates an effective meter. We take $x_{1}=1$ (top figures),
and $x_{1}=4$ (lower figures). Here $r=1.5$. \label{fig:ent-meter-2}$10^{5}$
trajectories are plotted.\textcolor{green}{} The same excellent
correlation is observed for solutions conditioned on the negative
sign of $x_{B}(t_{f})$.}
\end{figure}

\begin{figure}
\begin{centering}
\includegraphics[width=0.5\columnwidth]{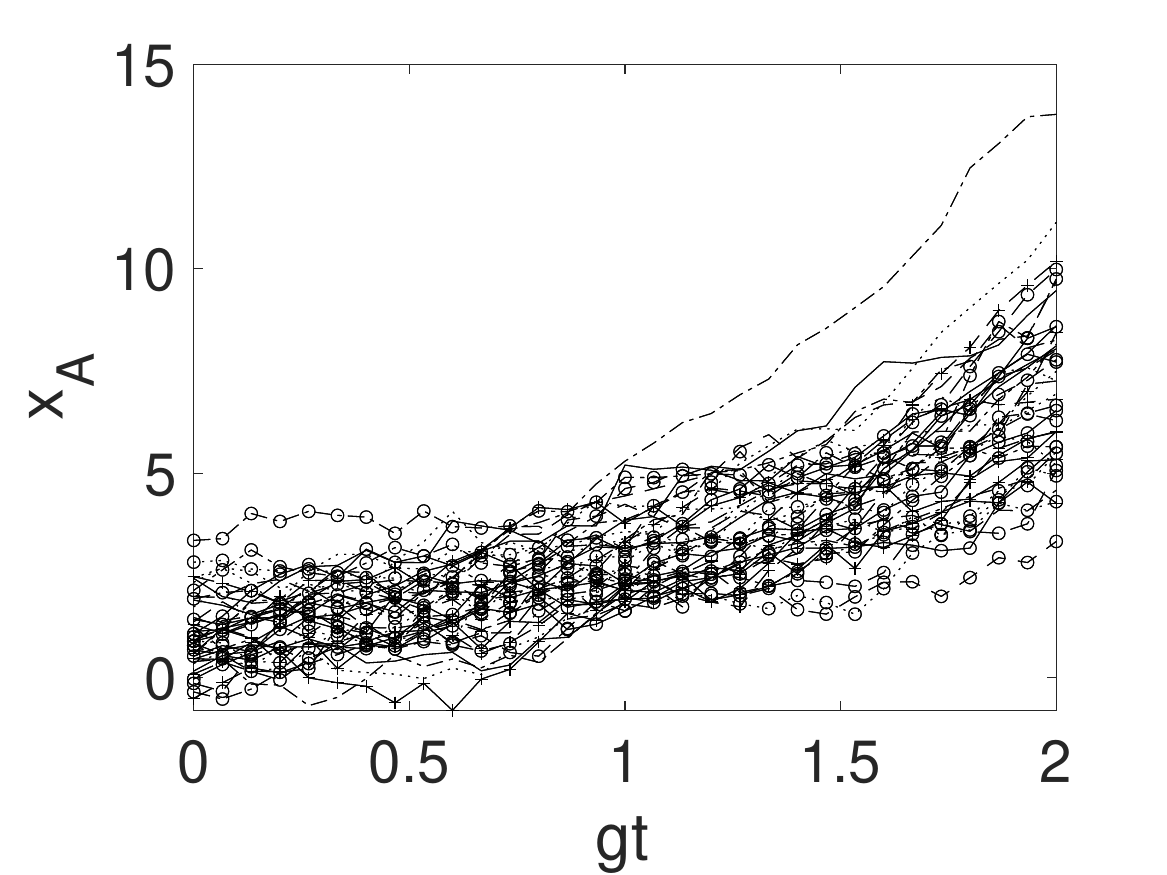}\includegraphics[width=0.5\columnwidth]{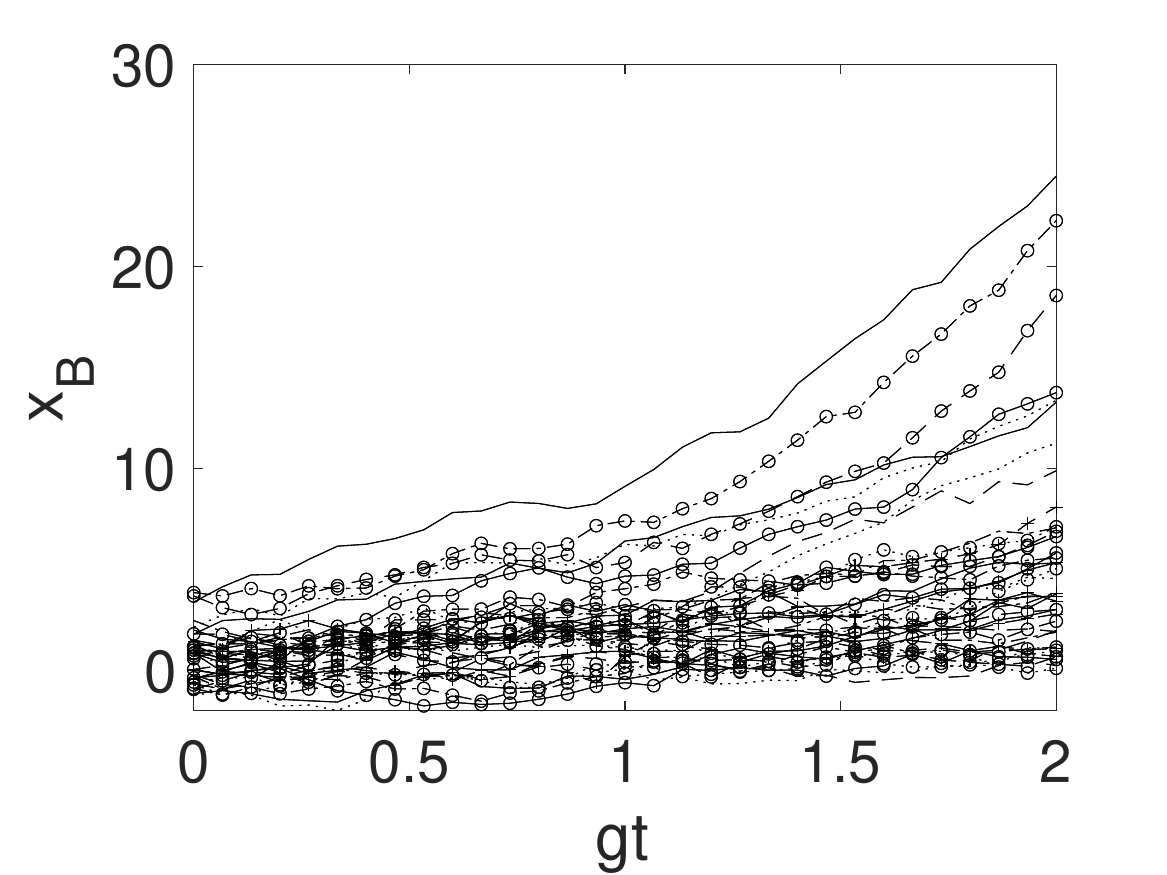}
\par\end{centering}
\begin{centering}
\includegraphics[width=0.5\columnwidth]{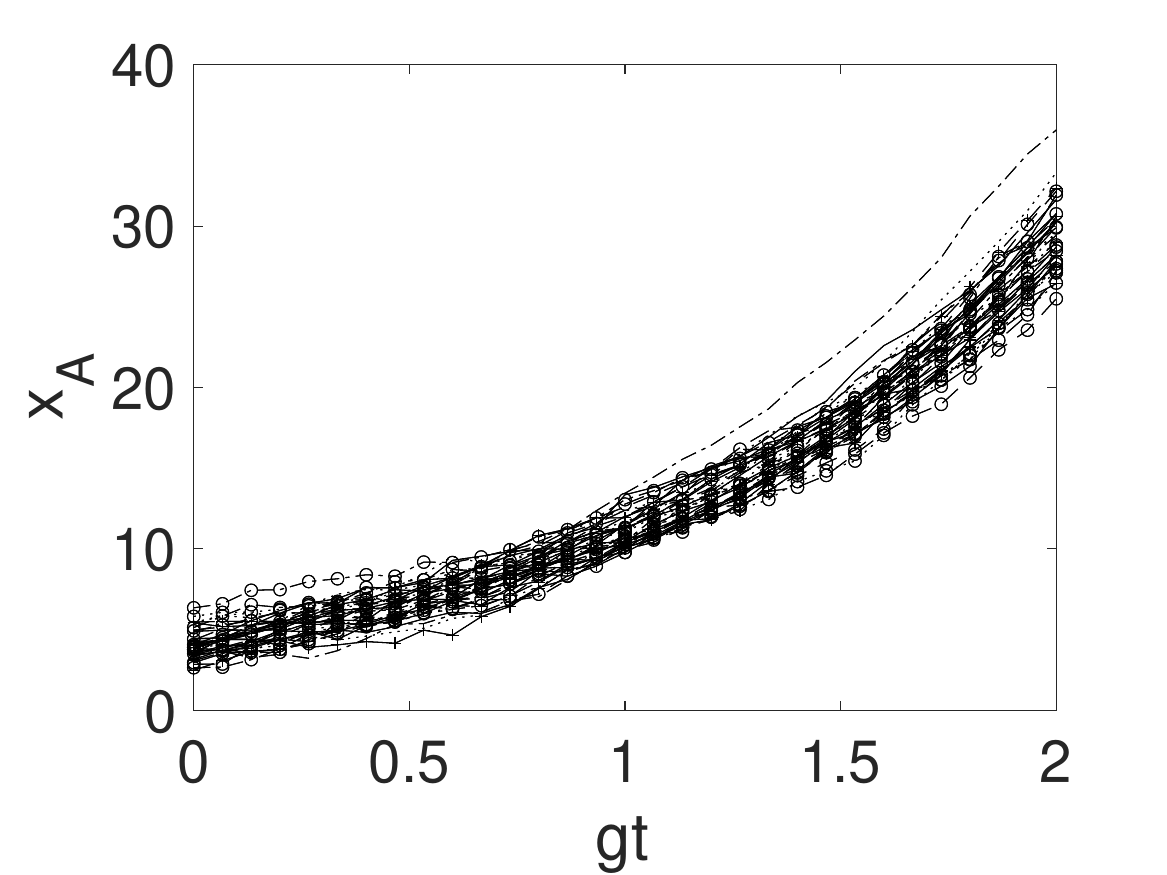}\includegraphics[width=0.5\columnwidth]{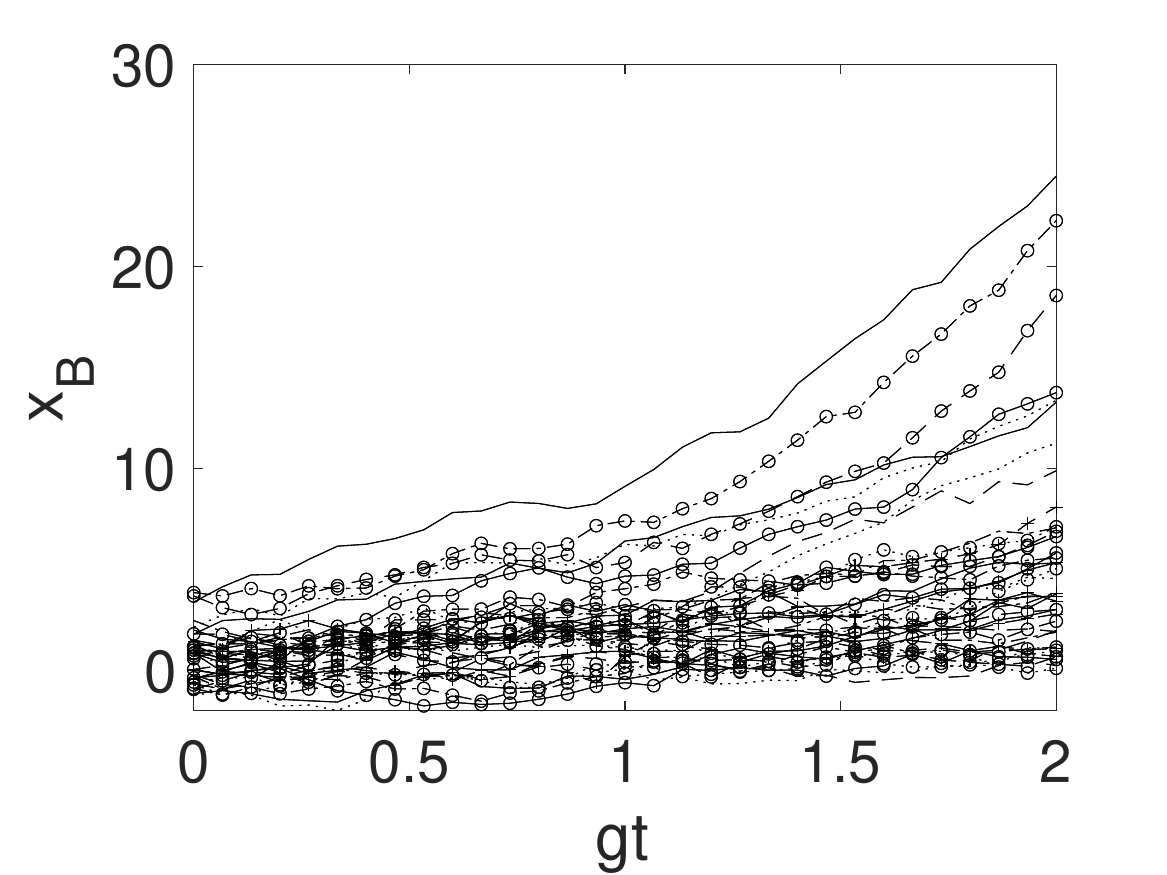}
\par\end{centering}
\begin{centering}
\par\end{centering}
\caption{Measurement of $\hat{x}_{A}$ and $\hat{x}_{B}$ on the systems prepared
in the entangled state (\ref{eq:ent-cat-1}) with \textcolor{black}{$\varphi=\pi/2$.
}Here, $r=1.5$. For the meter $B$, $r_{2}=0$ and $\beta_{0}=0.1$.
Plots show the trajectories for $x_{A}$ (left figure) and $x_{B}$
(right figure) conditioned on a positive outcome for system $B$,
$x_{B}(t_{f})>0$. We see a loss of distinctness between the positive
and negative outcomes of the meter: the smaller value of $\beta_{0}$
implies the system $B$ is an ineffective meter. We take $x_{1}=1$
(top figures), and $x_{1}=4$ (lower figures). \label{fig:ent-meter-2-2}
$10^{5}$ trajectories are plotted.\textcolor{green}{}\textcolor{red}{}}
\end{figure}
\begin{figure}
\begin{centering}
\par\end{centering}
\begin{centering}
\includegraphics[width=0.5\columnwidth]{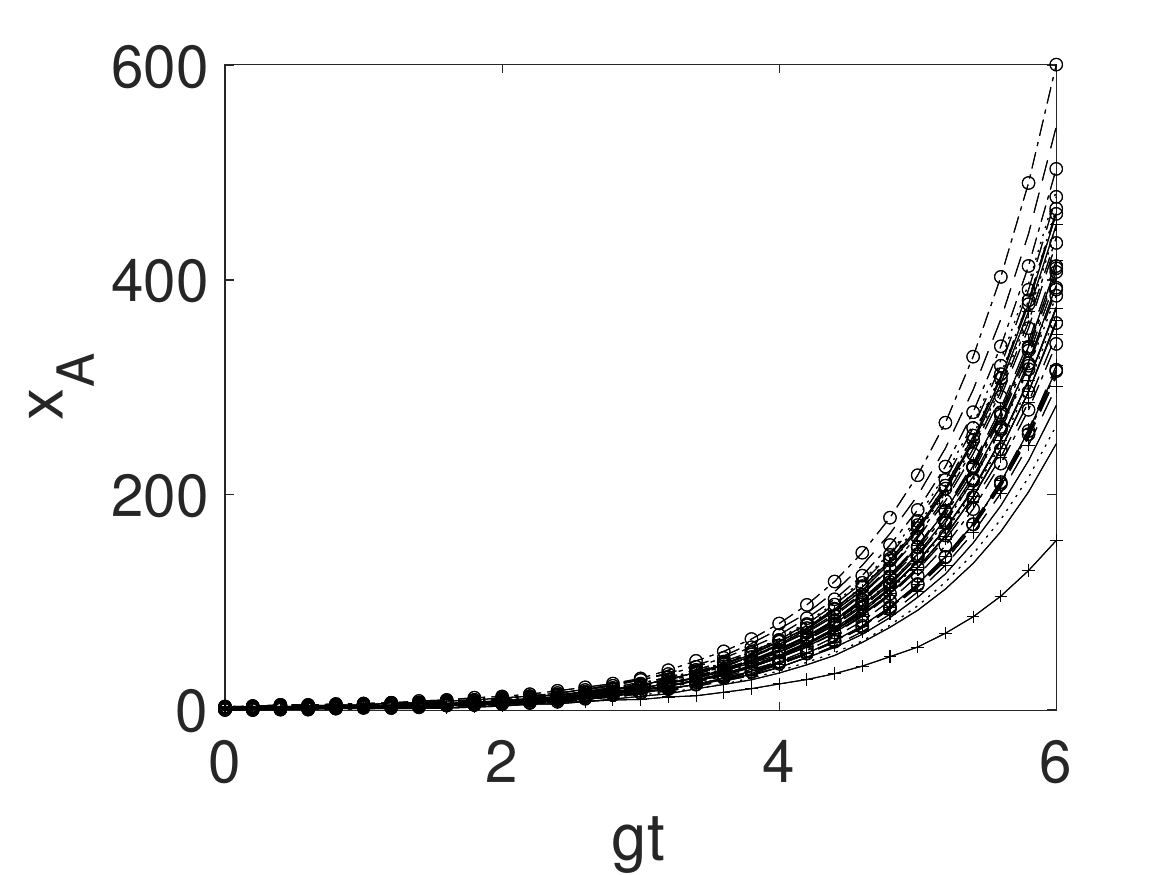}\includegraphics[width=0.5\columnwidth]{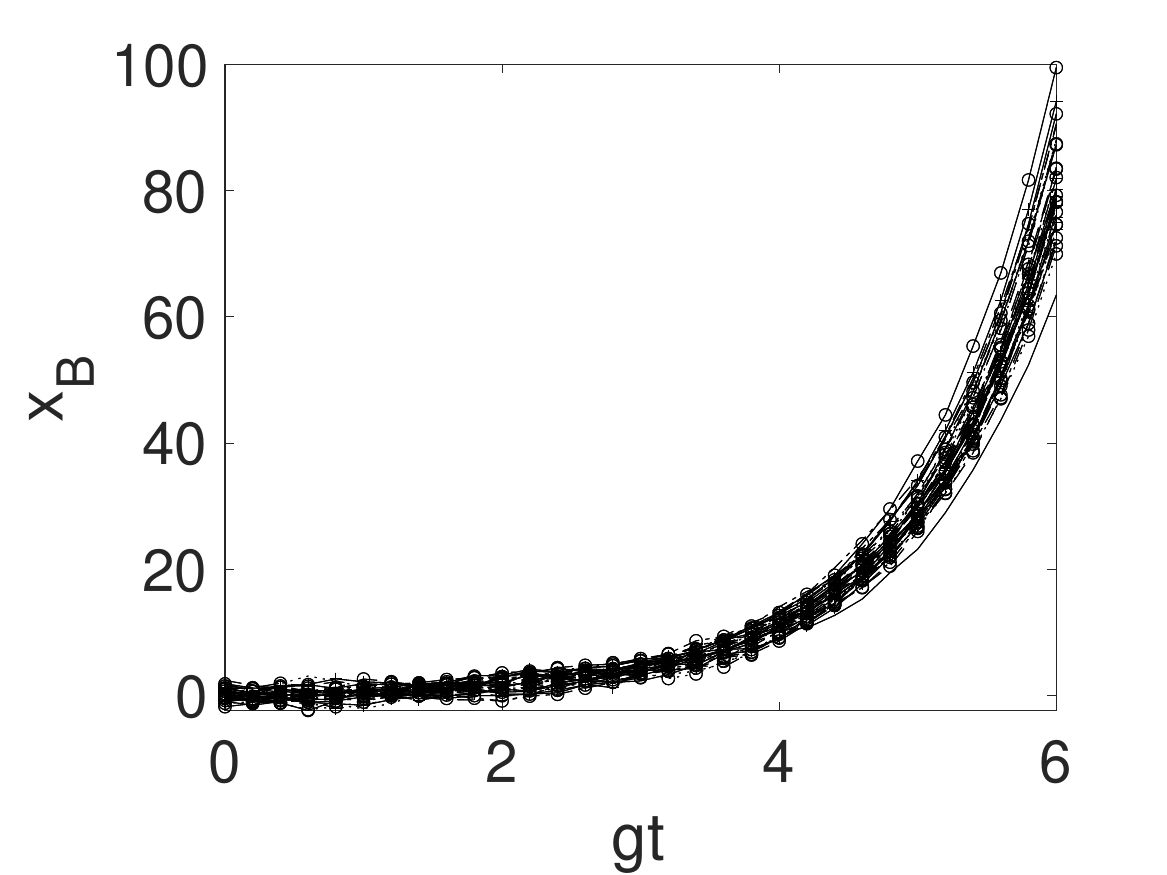}
\par\end{centering}
\begin{centering}
\par\end{centering}
\begin{centering}
\includegraphics[width=0.5\columnwidth]{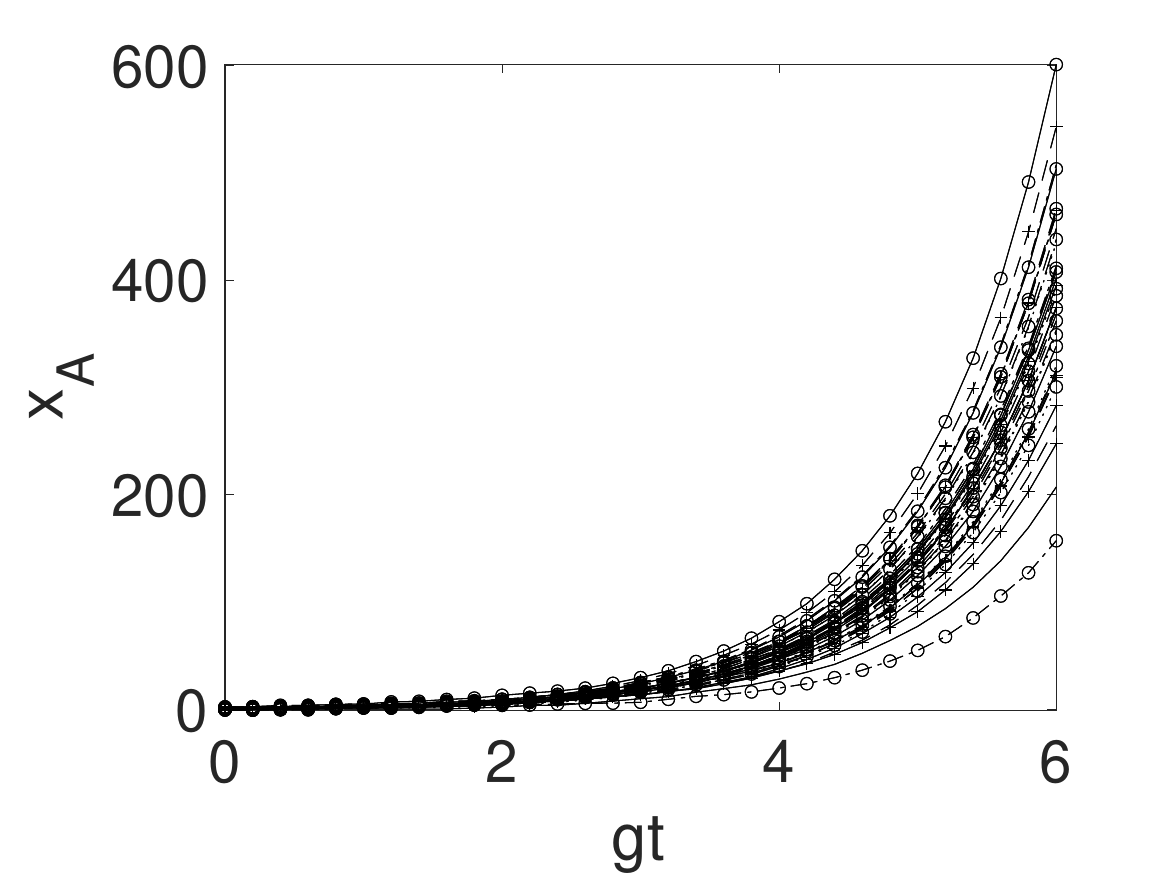}\includegraphics[width=0.5\columnwidth]{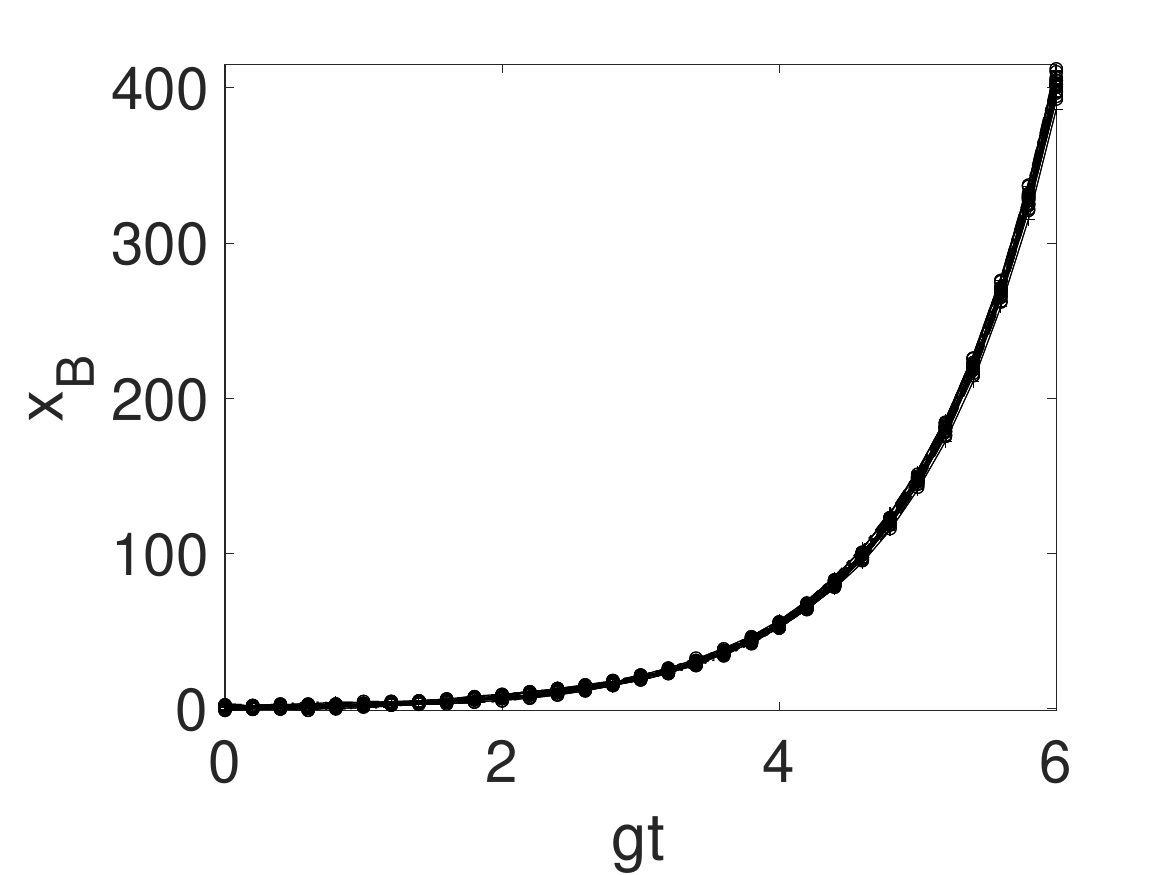}
\par\end{centering}
\begin{centering}
\par\end{centering}
\caption{Measurement of $\hat{x}_{A}$ and $\hat{x}_{B}$ on the meter-system
prepared in the entangled state (\ref{eq:ent-cat-1}) with \textcolor{black}{$\varphi=\pi/2$
and where $r_{2}=4$, so that the meter is highly squeezed. }We show
results for two choices of $x_{1B}$: $x_{1B}=0.2$ (top) and $x_{1B}=1$
(lower). We take $x_{1}=1$.  Here, $r_{1}=1.5$. With higher $r_{2}$,
the meter is effective for smaller values of $x_{1B}$ than possible
with $r_{2}=0$.\textcolor{red}{{} }Plots show the trajectories for
$x_{A}$ (left figure) and $x_{B}$ (right figure) conditioned on
a positive outcome for the meter, $x_{B}(t_{f})>0$. \label{fig:ent-meter-2-1}
$10^{5}$ trajectories are plotted.\textcolor{green}{}}
\end{figure}
Forward-backward stochastic simulations of Eqs. (\ref{eq:forwardSDE-2-1}
- \ref{eq:backwardSDE-2-1-1}) are shown in Figure \ref{fig:ent-meter-2},
for the coherent-state meter where $r_{2}=0$ and $\beta_{0}=4$.
The entangled meter and system are prepared in the state (\ref{eq:ent-cat-1})
with \textcolor{black}{$\varphi=\pi/2$}. Figure \ref{fig:ent-meter-2}
shows results for $r=1.5,$ modeling measurement on system $A$ originally
in a superposition $|x_{1}\rangle+i|-x_{1}\rangle$ of eigenstates
of $\hat{x}_{A}$. There is excellent correlation between the sign
of the outcomes of the measurement $\hat{x}_{A}$ and $\hat{x}_{B}$
for large $\beta_{0}$. We note in Figure \ref{fig:ent-meter-2-2},
where the coherent-state meter has a smaller value of $\beta_{0}$,
that the correlation is lost and the system $B$ is not an effective
meter. However, provided $\beta_{0}$ is sufficiently large (and $gt_{f}\rightarrow\infty$),
the correlation between the sign of the final amplitudes $x_{A}(t_{f})$
and $x_{B}(t_{f})$ is maximum \emph{even for small} $x_{1}$ (Figure
\ref{fig:ent-meter-2}, for $x_{1}=1$) when there is considerable
overlap of the two peaks of the $Q$ function of the superposition
$|x_{1}\rangle+i|-x_{1}\rangle$ for system$A$. This is evident for
$\beta_{0}=4$ and is possible because two eigenstates $|x_{1}\rangle$
and $|x_{2}\rangle$ ($x_{1}\neq x_{2}$), can always be distinguished
provided $gt_{f}$ is sufficiently large (Figure \ref{fig:sup}).

The measurement $\hat{O}$ on the meter-system $B$ prepared in (\ref{eq:ent-meter-x})
is intended to determine ``which state the system $A$ is in, $|x_{1}\rangle$
or $|-x_{1}\rangle$''. Here, $\hat{O}$ is the sign of $\hat{x}_{B}$.
In the Q model of reality, the value of the sign of $x_{A}(t_{f})$
is interpreted as the outcome of the measurement $\hat{O}$. We assume
that the outcome of $\hat{O}$ can also be obtained by direct amplification
of system $A$, to measure $\hat{x}_{A}$ directly. Since different
eigenstates can always be distinguished with sufficient amplification,
we conclude that the sign of $x_{A}(t_{f})$ is an accurate measurement
of $\hat{O}$, consistent with the Q model of reality. Hence, from
Figure \ref{fig:ent-meter-2}, since the sign of $x_{B}(t_{f})$ is
correlated with the sign of $x_{A}(t_{f})$, we interpret that the
detection of $x_{B}(t_{f})$ does indeed determine ``which state
the system $A$ is in'', whether $|x_{1}\rangle$ or $|-x_{1}\rangle$),
regardless of the size of $x_{1}$.

Results for the second convenient meter, where the meter $B$ is prepared
in squeezed states corresponding to a higher value of $r_{2}$, are
shown in Figure \ref{fig:ent-meter-2-1}. We find that the meter is
effective at lower values of mean amplitude $x_{1B}$ than is possible
for the coherent-state meter, where $r_{2}=0$.

\begin{figure}
\begin{centering}
\includegraphics[width=0.5\columnwidth]{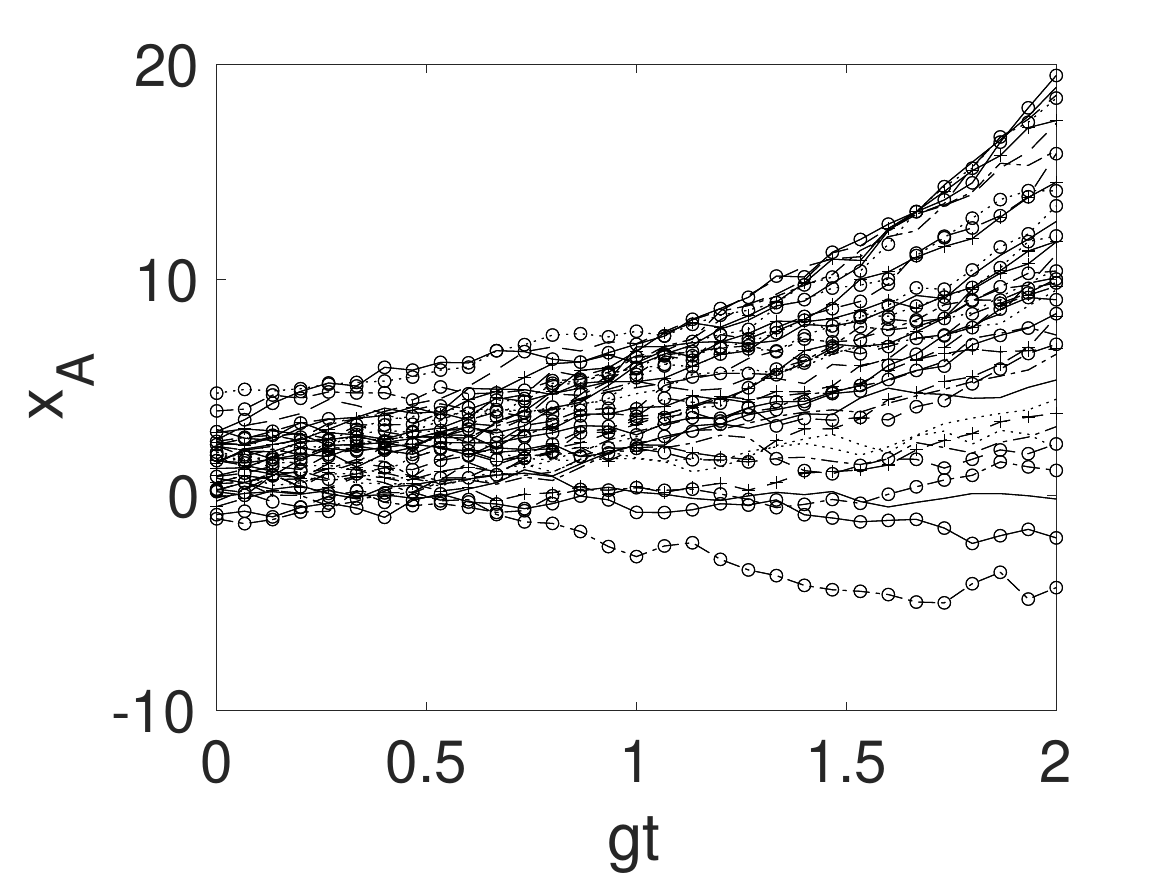}\includegraphics[width=0.5\columnwidth]{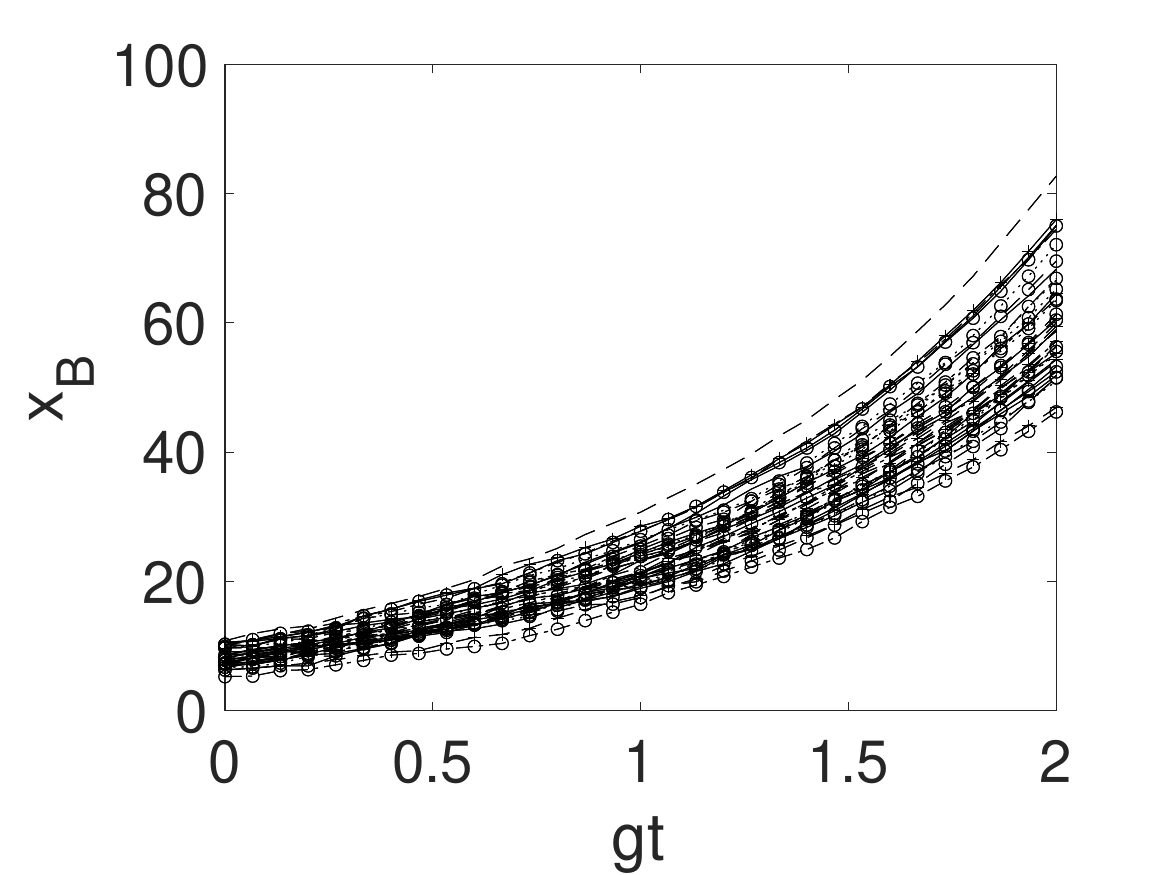}
\par\end{centering}
\begin{centering}
\includegraphics[width=0.5\columnwidth]{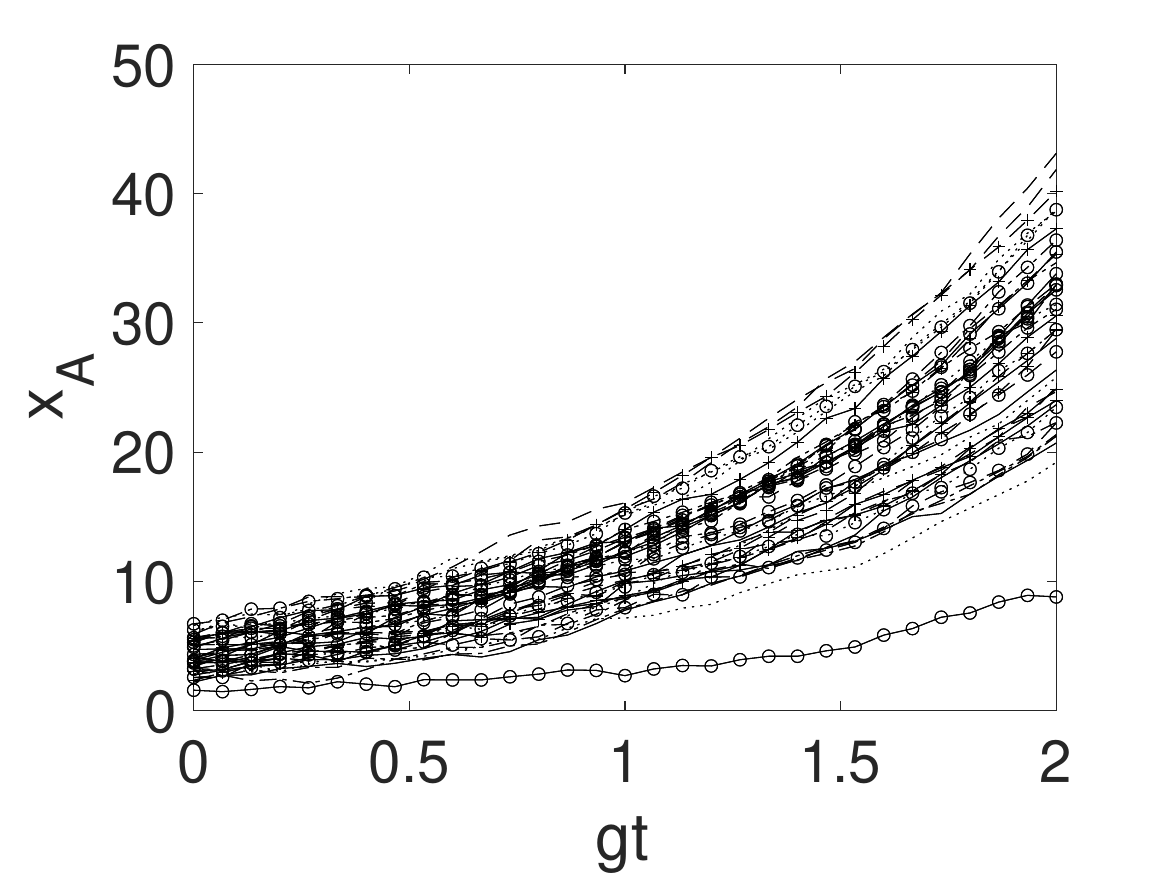}\includegraphics[width=0.5\columnwidth]{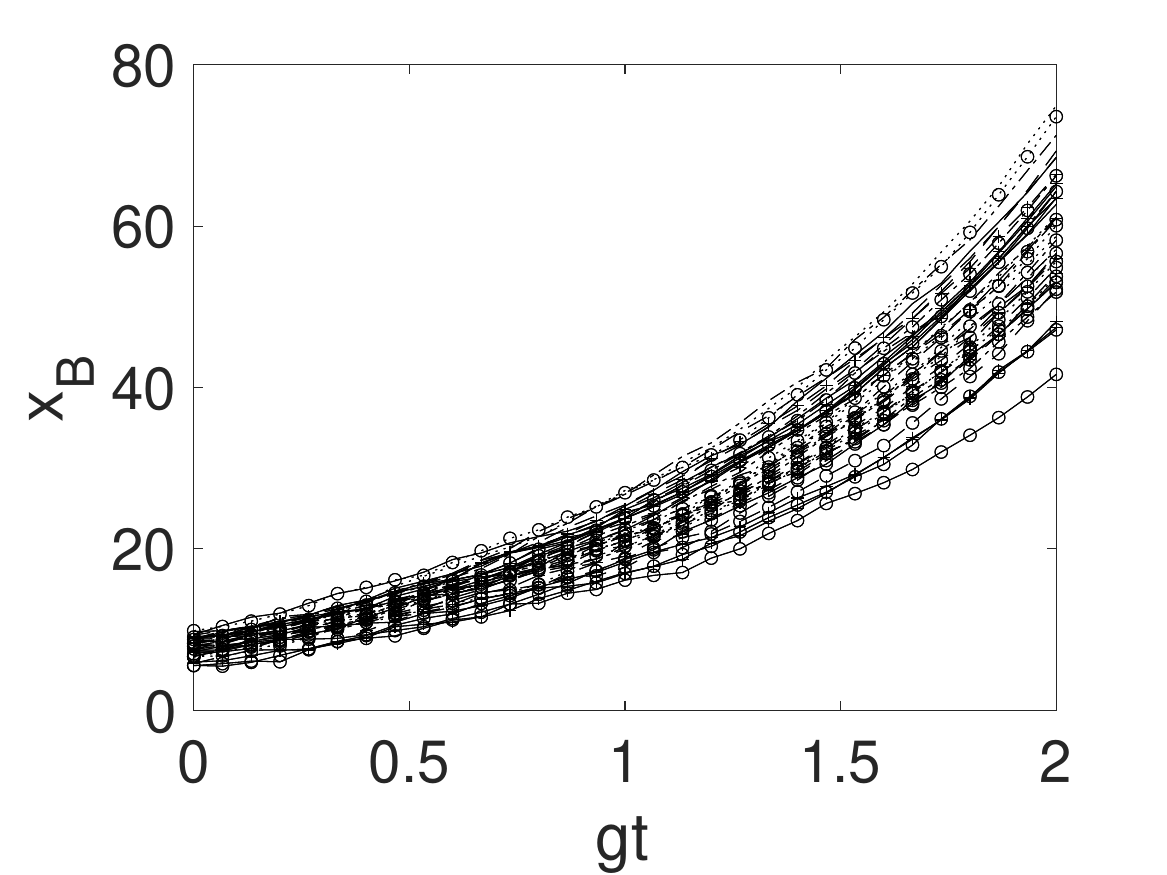}
\par\end{centering}
\caption{As for Figure (\ref{fig:ent-meter-2}), where $\beta_{0}=4$, except
here the system is in the two-mode entangled cat state (\ref{eq:ent-cat}),
where $r=r_{2}=0$. We take $x_{1}=1$ which implies $\alpha_{0}=0.5$
(top) and $x_{1}=4$ which implies $\alpha_{0}=2$ (lower). \label{fig:ent-meter-1}$10^{5}$
trajectories are plotted.\textcolor{green}{}}
\end{figure}

In Figure \ref{fig:ent-meter-1}, we show solutions where the system
$A$ is in a cat state ($r=0$). The measurement $\hat{O}$ on the
meter-system $B$ is intended to determine ``which state the system
$A$ is in, $|\alpha_{0}\rangle$ or $|-\alpha_{0}\rangle$''. There
is a greater level of noise at time $t_{f}$ in the solutions for
$x_{A}(t)$, since the final variance $\sigma_{x_{A}}$ at the time
$t_{f}$ given by (\ref{eq:var-ent-amp-2}) (similar to (\ref{eq:var-gen-cat}))
will show a detected noise level $(\Delta\hat{x})^{2}=1$ associated
with a coherent state. Hence, the signs of $x_{A}(t_{f})$ and $x_{B}(t_{f})$
are perfectly correlated only where $\alpha_{0}$ is sufficiently
large (lower plots). We will find in the next Section however that
the conditional distribution of $x_{A}(t_{0})$ given $x_{B}(t_{f})>0$
will correspond to the distribution of $x_{A}$ for the coherent state
$|\alpha_{0}\rangle$. In this sense, the coherent-state meter regardless
identifies that the system $A$ is in the state $|\alpha_{0}\rangle$
(or $|-\alpha_{0}\rangle$) depending on the sign of $\hat{O}$ (whether
positive or negative).

\section{collapse of the wave function: Inferred state for the system given
an outcome for the meter}

Motivated by this model, we next examine the ``collapse'' of the
system $A$ based on the measurement $\hat{x}_{B}$ of the meter,
system $B$. A question is: \emph{What can be inferred from the
measurement on the meter $B$ about the state at $A$ as it exists
at the initial time $t=0$?} We ask what is inferred for the state
of system $A$,\textcolor{red}{{} }if the outcome at $B$ for $\hat{x}_{B}$
is positive? How or when does the collapse to the eigenstate occur?

We seek to infer the state of the system \emph{prior} to its measurement,
in the context of the Q model, where the amplitudes $x$ and $p$
describe the state of the system at the given time $t_{0}$. This
has a similar but deeper meaning to evaluating the $Q$ function for
system $A$ conditioned on the outcome $x_{B}(t_{f})>0$, which can
be calculated from standard quantum mechanics by projection. For the
system prepared in the two-mode state (\ref{eq:ent-two-coherent-1})
for example, where $x_{1B}\rightarrow\infty$, the state at $A$ conditioned
on the outcome $x_{1B}$ for $\hat{x}_{B}$ is the eigenstate $|x_{1}\rangle$.
With the projection method, it is not clear how the ``collapse''
to the state $|x_{1}\rangle$ occurs, or on what time scales. The
meter and system are spatially separated. A natural question has been
whether there is a sudden nonlocal action-at-a-distance that forces
the system into the state $|x_{1}\rangle$ given the positive outcome
($x_{B}(t_{f})>0$) of the measurement at $B$ \citep{epr-1}? When
is the outcome of the meter $B$ finalized?

\subsection{Postselected state given an outcome for the meter $B$}

Our interest is the evaluation of the state at $A$ at the initial
time $t_{0}=0$, \emph{postselected} on the measurement outcome for
the meter $B$. The system is prepared in the state (\ref{eq:ent-cat-1})
given by the Q function $Q_{ent}(\lambda,t_{0})$ (Eq. (\ref{eq:Qent-2-1}),
where $\lambda=(x_{A},p_{A},x_{B},p_{B})$. We follow the procedure
explained in Section VI for the single-mode system.

We consider the measurement of $\hat{x}_{B}$ on the meter, system
$B$. The boundary condition for the trajectories $x_{B}(t)$ can
be evaluated by integrating the $Q$ function (\ref{eq:Qent-2-1})
over $x_{A}$, $p_{A}$ and $p_{B}$ (i.e. integrating (\ref{eq:marg-1})
over $x_{A}$) to calculate the marginal for $x_{B}$ at time $t$.
We find\textbf{\textcolor{green}{}}\textcolor{green}{}\begin{widetext} \textbf{\textcolor{black}{
\begin{eqnarray}
Q(x_{B},t) & = & \frac{1}{2\sqrt{2\pi}\sigma_{x_{B}}(t)f(\varphi,t)}\times\left\{ e^{-\left(x_{B}-G(t)x_{1B}\right){}^{2}/2\sigma_{x_{B}}^{2}(t)}\right.+e^{-\left(x_{B}+G(t)x_{1B}\right)^{2}/2\sigma_{x_{B}}^{2}(t)}\nonumber \\
 &  & +2\cos\bigl(\varphi\bigl)e^{-x_{B}^{2}/2\sigma_{x_{B}}^{2}(t)}e^{-G^{2}(t)x_{1}^{2}\left(1+\sigma_{p_{A}}^{2}(t)/\sigma_{x_{A}}^{2}(t)\right)/2\sigma_{x_{A}}^{2}(t)}e^{-G^{2}(t)x_{1B}^{2}\left(1+\sigma_{p_{B}}^{2}(t)/\sigma_{x_{B}}^{2}(t)\right)/2\sigma_{x_{B}}^{2}(t)}\biggl\}\label{eq:qmargianlfuture-1-1}
\end{eqnarray}
}}\end{widetext} \textcolor{black}{}Here, the parameters are
as defined for (\ref{eq:marg-1})\textcolor{black}{. Thi}s provides,
by substituting $t=t_{f}$, the boundary condition for the backward
trajectory. We take $x_{1}>0$ and $x_{1B}>0$.

We choose that $\varphi=\pi/2$, which ensures that the third term
in $Q(x_{B},t)$ vanishes. For this case \citep{q-measurement}, the
future boundary condition for the measurement of $\hat{x}_{B}$ is
given {[}similar to Eq. (\ref{eq:fbc-prob-mix}){]} by the 50/50 mixture
of the two Gaussian distributions $Q_{\pm}(x_{B},t_{f})$, defined
by
\begin{equation}
Q_{\pm}(x_{B},t_{f})=\frac{1}{\sqrt{2\pi}\sigma_{x_{B}}(t_{f})}e^{-(x_{B}\mp G(t_{f})x_{1B})^{2}/2\sigma_{x_{B}}^{2}(t_{f})}\label{eq:G-1}
\end{equation}
There are two sets of trajectories for $x_{B}(t_{f})$, where $t_{f}\rightarrow\infty$,
one stemming from $G(t_{f})x_{1B}$ and the other from $-G(t_{f})x_{1B}$.
Where $r_{2}$ is large so that the system is a superposition of eigenstates
of $\hat{x}_{B}$, then these sets corresponding to outcomes for $\hat{x}_{B}$
of $x_{1B}$ and $-x_{1B}$ respectively. We refer to the first set
as the positive branch $B_{+}$, and the second set as the negative
branch $B_{-}$. The treatment also allows for the more general
case of arbitrary $r_{2}$, in particular of the coherent-state meter
($r_{2}=0$). In the limit where $\beta_{0}\rightarrow\infty$, the
entire set of amplitudes $x(t_{f})$ emanating from the positive or
negative Gaussian ($Q_{+}(x_{B},t_{f})$ or $Q_{-}(x_{B},t_{f})$)
will correspond to the outcome for $\hat{x}_{B}$ of $+2\beta_{0}$
or $-2\beta_{0}$, respectively.

In the limit of large $gt_{f}$, the values of $B_{+}$ emanate from
the positive Gaussian $Q_{+}(x_{B},t_{f})$. The set of amplitudes
$x(t)$ propagating from this set of values corresponds to the Gaussian
\begin{equation}
Q_{+}(x_{B})=\frac{1}{\sqrt{2\pi}\sigma_{x_{B}}}e^{-(x_{B}-x_{1B})^{2}/2\sigma_{x_{B}}^{2}}
\end{equation}
at time $t_{0}=0$ (refer Figures \ref{fig:sup} and \ref{fig:sup-2}),
where $\sigma_{x_{B}}^{2}\equiv\sigma_{x_{B}}^{2}(0)$. For each trajectory
beginning from $x(t_{f})$, there is a single value at $x(t_{0})$.
Summed over all trajectories, this defines the distribution $Q(\lambda,t_{0}|B_{+})$
for the set $B_{+}$ at the initial time $t_{0}$, where $\lambda=(x_{A},p_{A},x_{B},p_{B})$.
Hence, the postselected distribution for the combined meter and
system conditioned on the branch $B_{+}$ of the meter is
\begin{eqnarray}
Q(\lambda,t_{0}|B_{+}) & = & Q_{+}(x_{B})Q(\lambda|x_{B})\label{eq:hidden-1}
\end{eqnarray}
Here, the conditional distribution $Q(\lambda|x_{B})$ for the system
at time $t_{0}$ is
\begin{equation}
Q(\lambda|x_{B})=\frac{Q_{ent}(\lambda,t_{0})}{Q(x_{B},t_{0})}\label{eq:cond-4-1}
\end{equation}
where $Q_{ent}(\lambda,t_{0})$ is given by Eq. (\ref{eq:Qent-2-1}).
 From (\ref{eq:qmargianlfuture-1-1}), we find for $t=t_{0}$
and $\varphi=\pi/2$ that\textcolor{green}{}\textcolor{red}{}
\begin{equation}
Q(x_{B},t_{0})=\{Q_{+}(x_{B})+Q_{-}(x_{B})\}/2\label{eq:Q(x)}
\end{equation}
where $Q_{\pm}(x_{B})=Q_{\pm}(x_{B},0)$ as given by (\ref{eq:G-1}).
Hence \begin{widetext} 

\begin{eqnarray}
Q(\lambda|x_{B}) & = & \frac{e^{-p_{A}^{2}/2\sigma_{p_{A}}^{2}-p_{B}^{2}/2\sigma_{p_{B}}^{2}}}{4\sqrt{2}\pi^{3/2}\sigma_{p_{A}}\sigma_{p_{B}}\sigma_{x_{A}}\cosh\left(x_{1B}x_{B}/\sigma_{x_{B}}^{2}\right)}\biggl\{ e^{-(x_{A}-2\alpha_{0})^{2}/2\sigma_{x_{B}}^{2}}e^{|x_{1B}|x_{B}/\sigma_{x_{B}}^{2}}\nonumber \\
 &  & +e^{-(x_{A}+x_{1})^{2}/2\sigma_{x_{B}}^{2}}e^{-|x_{1B}|x_{B}/\sigma_{x_{B}}^{2}}-2e^{-x_{A}^{2}/2\sigma_{x_{A}}^{2}-x_{1}^{2}/2\sigma_{x_{A}}^{2}}\sin\bigl(|x_{1}|p_{A}/\sigma_{x_{A}}^{2}+|x_{1B}|p_{B}/\sigma_{x_{B}}^{2}\bigl)\biggl\}\nonumber \\
\label{eq:cond}
\end{eqnarray}
\end{widetext} \textcolor{red}{}\textcolor{green}{}where $\sigma_{x_{A}}^{2}\equiv\sigma_{x_{A}}^{2}(0)$,
$\sigma_{x_{B}}^{2}\equiv\sigma_{x_{B}}^{2}(0)$, $\sigma_{p_{A}}^{2}\equiv\sigma_{p_{A}}^{2}(0)$,
$\sigma_{p_{B}}^{2}\equiv\sigma_{p_{B}}^{2}(0)$ {[}Eqs. (\ref{eq:var-ent-amp-2})
and (\ref{eq:var-ent-amp-1-1}){]}. The postselected distribution
$Q(\lambda,t_{0}|B_{+})$ readily follows.\textcolor{green}{}

\subsection{Inferred state for system $A$: collapse of the wave function}

We are interested to determine the inferred state $Q_{+,inf}^{(A)}$
for system $A$ at time $t_{0}$, conditioned on a positive outcome
(identified by the positive branch $B_{+}$) of the measurement $\hat{x}_{B}$
on the meter $B$, where $\varphi=\pi/2$ as above. We define 
\begin{eqnarray*}
Q(x_{A},p_{A}|B_{+}) & = & \int dx_{B}dp_{B}Q(\lambda,t_{0}|B_{+})\\
 & = & \int dx_{B}Q_{+}(x_{B})\int dp_{B}Q(\lambda|x_{B})
\end{eqnarray*}
Integrating $Q(\lambda|x_{B})$ {[}Eq. (\ref{eq:cond}){]} over $p_{B}$,
we define $Q(x_{A},p_{A}|x_{B})=\int dp_{B}Q(\lambda|x_{B})$ and
find \textcolor{green}{}\textcolor{green}{}\textcolor{green}{}\textcolor{red}{}\begin{widetext} 

\begin{eqnarray}
Q(x_{A},p_{A}|x_{B}) & = & \frac{e^{-p_{A}^{2}/2\sigma_{p_{A}}^{2}}}{2\pi\sigma_{p_{A}}\sigma_{x_{A}}S(x_{B})}\biggl\{ e^{-(x_{A}-x_{1})^{2}/2\sigma_{x_{A}}^{2}}e^{-(x_{B}-x_{1B})^{2}/2\sigma_{x_{B}}^{2}}+e^{-(x_{A}+x_{1})^{2}/2\sigma_{x_{A}}^{2}}e^{-(x_{B}+x_{1B})^{2}/2\sigma_{x_{B}}^{2}}+\mathcal{I}nt\biggl\}\nonumber \\
\label{eq:acond}
\end{eqnarray}
where $S(x_{B})=e^{-(x_{B}-x_{1B})^{2}/2\sigma_{x_{B}}^{2}}+e^{-(x_{B}+x_{1B})^{2}/2\sigma_{x_{B}}^{2}}$
and\textcolor{green}{} 
\[
\mathcal{I}nt=-2e^{-x_{A}^{2}/2\sigma_{x_{A}}^{2}-x_{1}^{2}/2\sigma_{x_{A}}^{2}}e^{-x_{B}^{2}/2\sigma_{x_{B}}^{2}}e^{-x_{1B}^{2}\left(1+\sigma_{p_{B}}^{2}/\sigma_{x_{B}}^{2}\right)/2\sigma_{x_{B}}^{2}}\sin\bigl(|x_{1}|p_{A}/\sigma_{x_{A}}^{2}\bigl)
\]
\end{widetext} Hence the inferred state for $A$ is given by
\begin{eqnarray}
Q(x_{A},p_{A}|B_{+}) & = & \int dx_{B}Q_{+}(x_{B})Q(x_{A},p_{A}|x_{B})\nonumber \\
\label{eq:int-sA}
\end{eqnarray}
When system $B$ is a meter, this is the inferred distribution $Q_{+,inf}^{(A)}$
for system $A$ based on the measurement by the meter: 
\[
Q_{+,inf}^{(A)}=Q(x_{A},p_{A}|B_{+})
\]
Before calculating this distribution, we investigate the limit of
a macroscopic coherent-state meter.

\subsubsection{Analytical limit of a macroscopic meter}

We gain insight by examining the limit of a macroscopic coherent-state
meter, where we consider $r_{2}=0,$ $x_{1B}=\beta_{0}$ and $\beta_{0}$
large. The $x_{B}\equiv x_{B}(t_{0})$ are justified to be (mainly)
positive, based on the plots of the trajectories for $x_{B}$ that
emanate from $x_{B}(t_{f})>0$. Then we see that the fringe term $\mathcal{I}nt$
in (\ref{eq:acond}) is heavily damped and the conditional distribution
becomes\textcolor{green}{}
\begin{equation}
Q(x_{A},p_{A}|x_{B})\rightarrow\frac{e^{-p_{A}^{2}/2\sigma_{p_{A}}^{2}}e^{-\left(x_{A}-x_{1}\right)^{2}/2\sigma_{x_{A}}^{2}}}{2\pi\sigma_{p_{A}}\sigma_{x_{A}}}\label{eq:Qreduced}
\end{equation}
The distribution $Q_{+,inf}^{(A)}$ can be evaluated by averaging
over all positive $x_{B}(t_{f})$ {[}as in Eq. (\ref{eq:int-sA}){]}
but in the limit corresponding to a measurement where $\beta_{0}$
is large, we see that \textcolor{green}{}
\begin{equation}
Q_{+,inf}^{(A)}=\frac{e^{-p_{A}^{2}/2\sigma_{p_{A}}^{2}}e^{-\left(x_{A}-x_{1}\right)^{2}/2\sigma_{x_{A}}^{2}}}{2\pi\sigma_{p_{A}}\sigma_{x_{A}}}\label{eq:reduced-cat}
\end{equation}
The inferred state of system $A$ corresponds to the Q function of
$|x_{1}/2,r\rangle$, in agreement with the state projected from (\ref{eq:ent-cat}),
using standard quantum mechanics. This is verified in Section VIII.B.3
, where the postselected state is evaluated from the numerical simulation.

For $r=0$, where system $A$ is a cat state, we see from (\ref{eq:Qreduced})
that \textcolor{green}{}
\begin{eqnarray}
Q(x_{A},p_{A}|x_{B}) & \rightarrow & \frac{e^{-p_{A}^{2}/4}e^{-\left(x_{A}-2\alpha_{0}\right)^{2}/4}}{4\pi}\label{eq:limit-meter-cat-collapse}
\end{eqnarray}
which is the Q function of a coherent state $|\alpha_{0}\rangle$,
a result that is independent of $x_{B}$. This is based on the amplification
due to $\beta_{0}\rightarrow\infty$ and holds regardless of the size
of $\alpha_{0}$. The distribution $Q_{+,inf}^{(A)}$ in the limit
corresponding to a measurement where $\beta_{0}$ is large is hence
\begin{equation}
Q_{+,inf}^{(A)}=\frac{e^{-p_{A}^{2}/4}e^{-\left(x_{A}-2\alpha_{0}\right)^{2}/4}}{4\pi}\label{eq:reduced-cat-2}
\end{equation}
The inferred state of system $A$ is $|\alpha_{0}\rangle$, in agreement
with the state projected from (\ref{eq:ent-cat}), using standard
quantum mechanics. This will be verified in Section VIII.B.3, where
the postselected state is evaluated from the stochastic solutions.

\begin{figure}
\begin{centering}
\includegraphics[width=0.8\columnwidth]{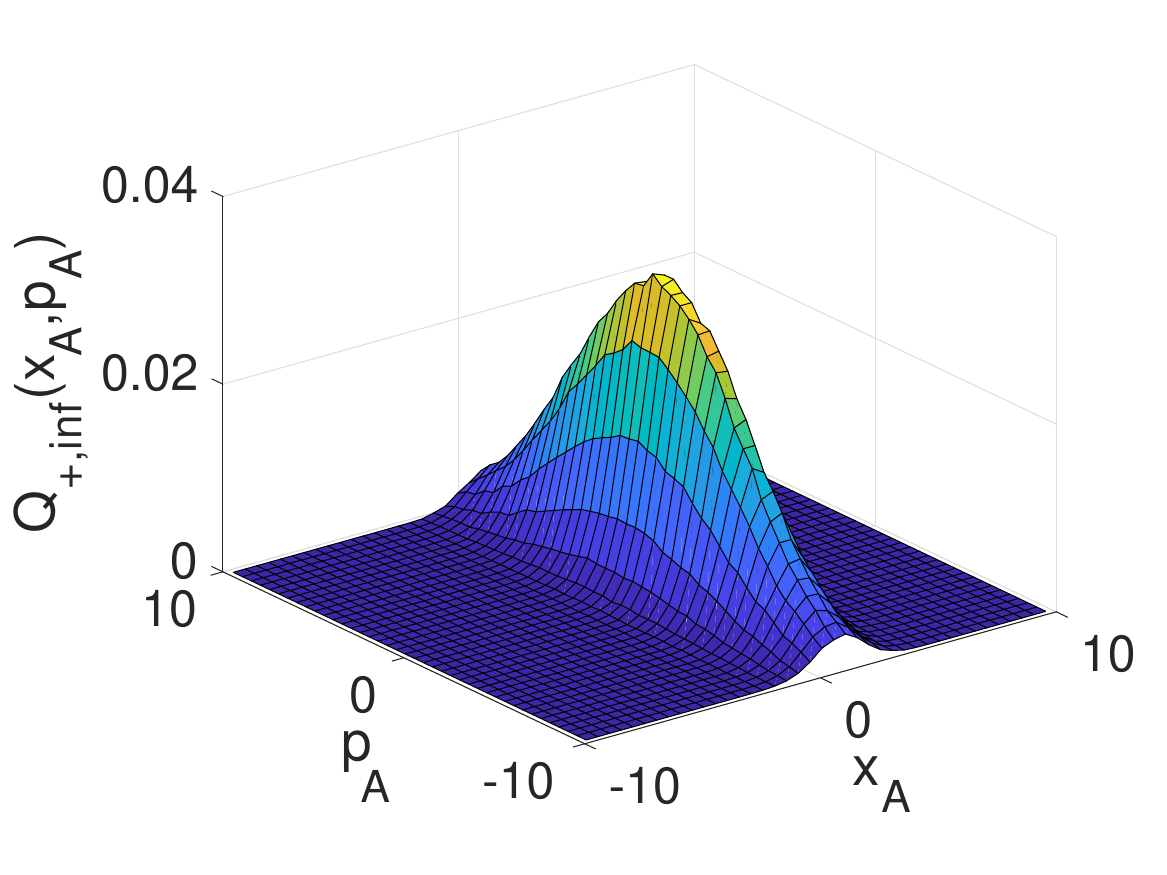}
\par\end{centering}
\begin{centering}
\includegraphics[width=0.8\columnwidth]{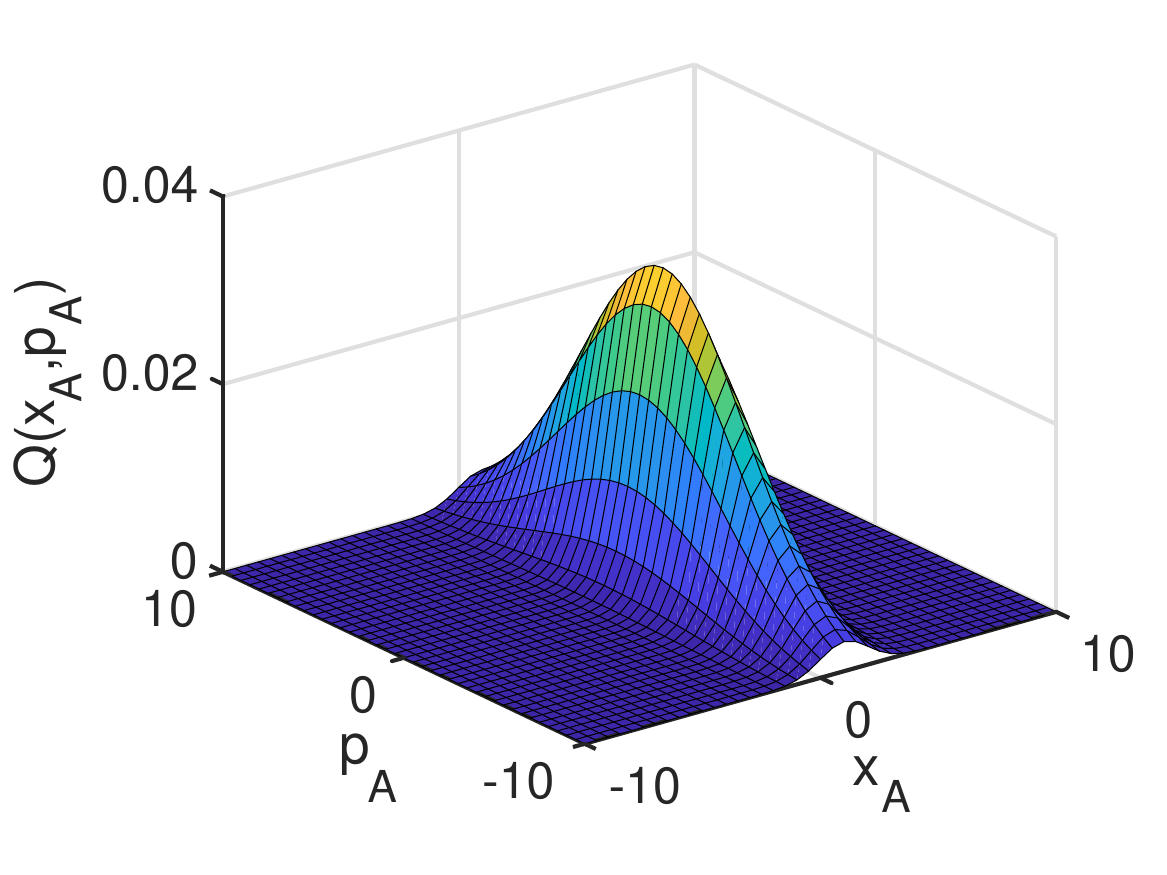}
\par\end{centering}
\begin{centering}
\par\end{centering}
\begin{centering}
\includegraphics[width=0.8\columnwidth]{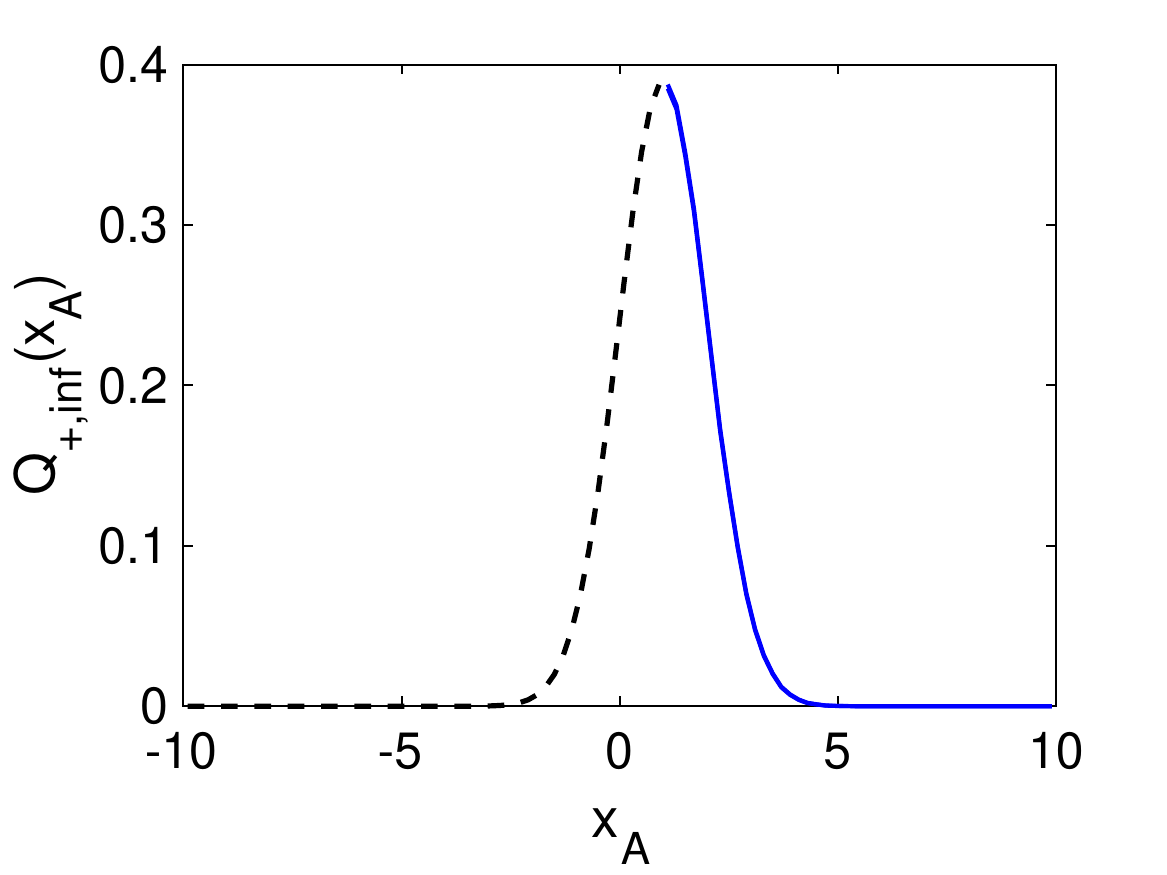}
\par\end{centering}
\begin{centering}
\par\end{centering}
\caption{Collapse of the wave function: Measurement of $\hat{x}_{A}$ of system
$A$ by a coherent-state meter $B$. The entangled meter-system is
in the state (\ref{eq:ent-cat-1}) where $\varphi=\pi/2$ and $r_{2}=0$.
For system $A$, $r=1.5$\textcolor{red}{{} }and $x_{1}=1$. Top: Plot
of the inferred state $Q_{+,inf}(x_{A},p_{A})$ for $A$ conditioned
on a positive outcome $+$ of $\hat{x}_{B}$, where for the meter
$B$, $\beta_{0}=2$. Plot shows $1.2\times10^{6}$ trajectories.
$gt_{f}=2$. \textcolor{red}{}\textcolor{green}{}Centre: Plot of
the Q function $Q(x_{A},p_{A})$ of the squeezed state $|x_{1}/2,r\rangle$.
Lower: Thick blue curve shows the marginal distribution $Q_{+,inf}(x_{A})$
calculated from the forward-backward stochastic solutions. The black
dashed line shows the distribution calculated from $Q(x_{A},p_{A})$.
Both solutions are symmetrical about the $x_{A}=0$.\textcolor{magenta}{{}
\label{fig:ent-meter-infa-eigenstate-1-1}}\textcolor{green}{}\textcolor{red}{}\textcolor{green}{}\textcolor{red}{}}
\end{figure}

\begin{figure}
\begin{centering}
\par\end{centering}
\begin{centering}
\includegraphics[width=0.8\columnwidth]{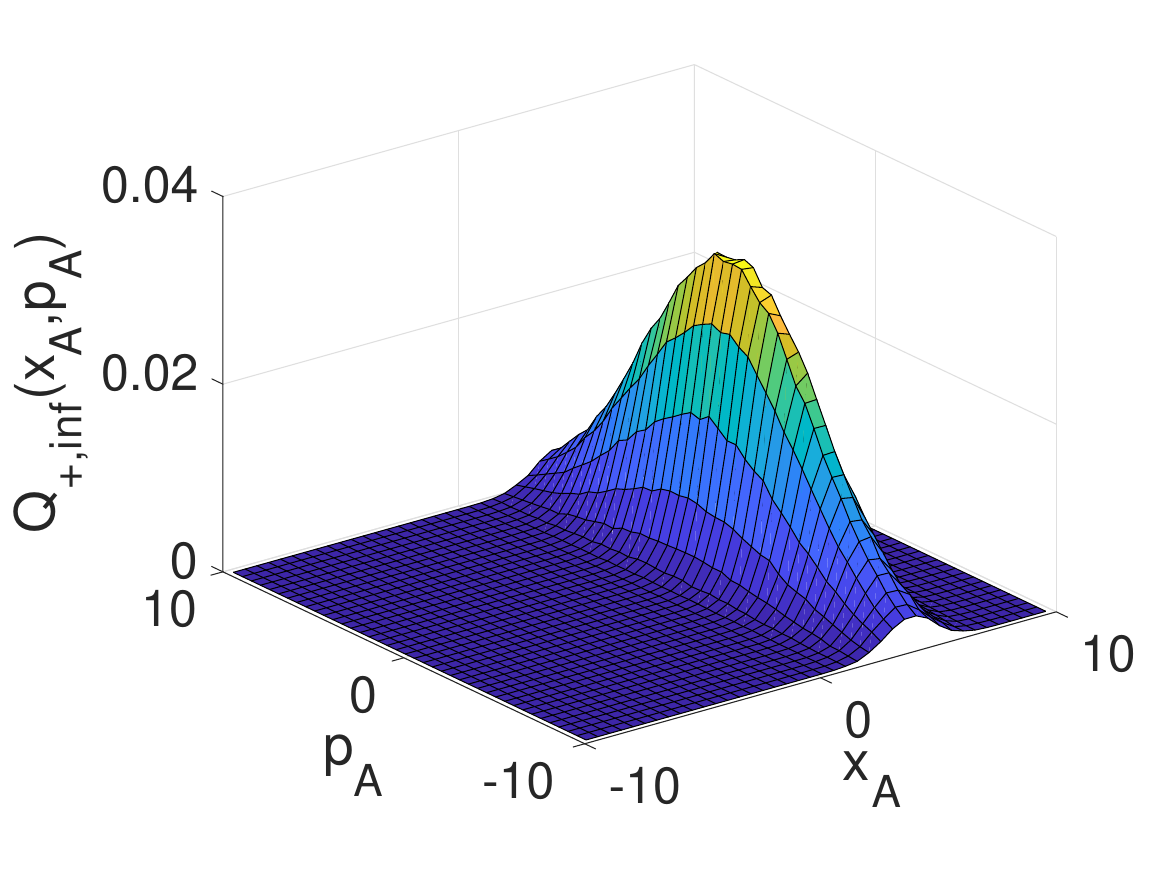}
\par\end{centering}
\begin{centering}
\includegraphics[width=0.8\columnwidth]{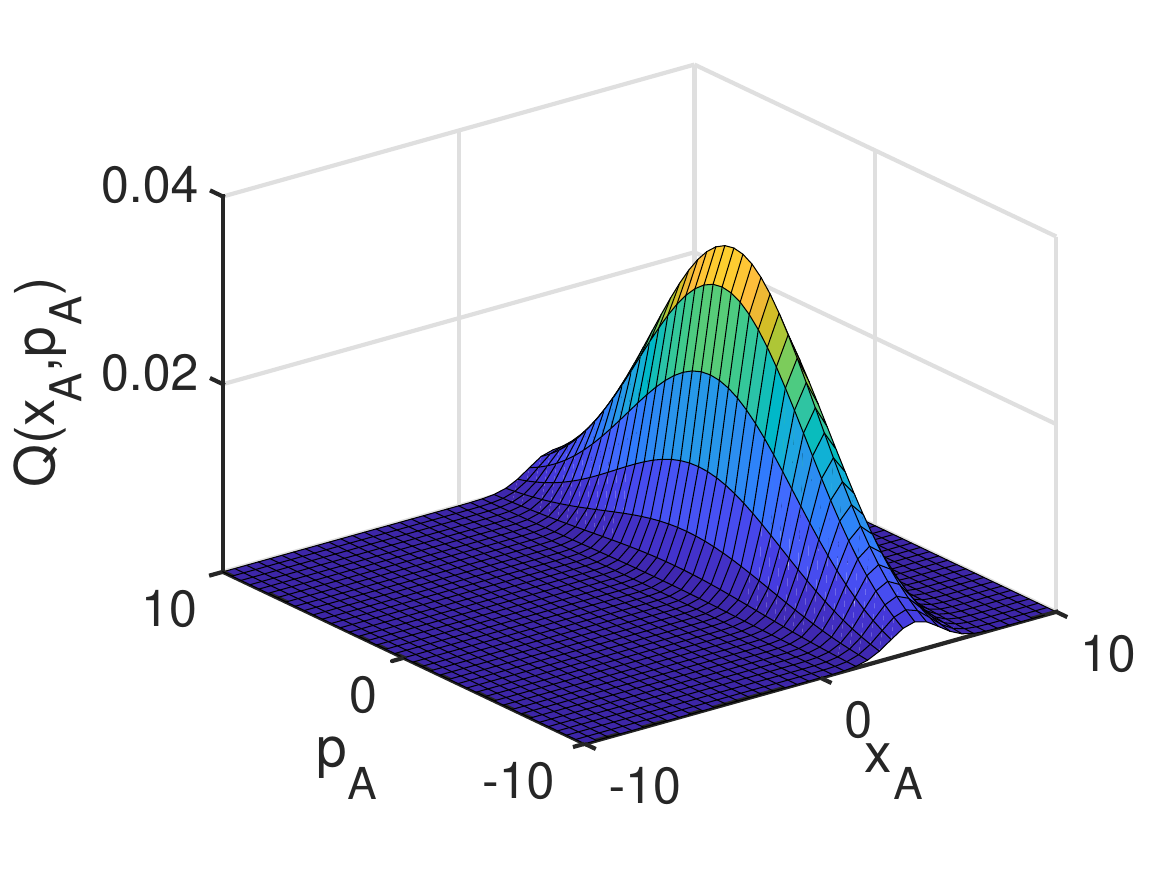}
\par\end{centering}
\begin{centering}
\par\end{centering}
\begin{centering}
\includegraphics[width=0.8\columnwidth]{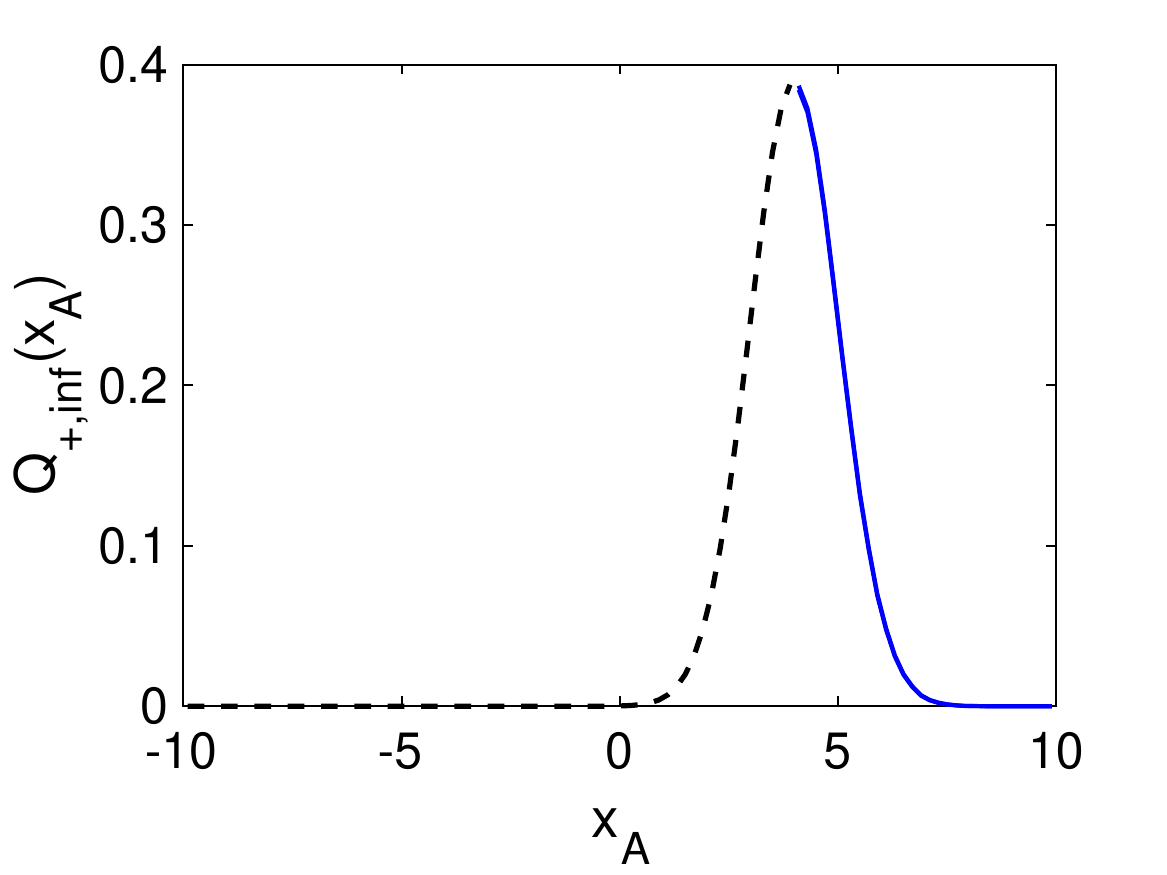}
\par\end{centering}
\caption{Collapse of the wave function: Measurement of $\hat{x}_{A}$ of system
$A$ by a coherent-state meter $B$. As for Figure \ref{fig:ent-meter-infa-eigenstate-1-1},
for the meter, $r_{2}=0$ and $\beta_{0}=2$. Here for the system
$A$, $r=1.5$\textcolor{red}{{} }with $x_{1}=4$. Top: Plot of the
inferred state $Q_{+,inf}(x_{A},p_{A})$ for $A$ conditioned on the
positive outcome $+$ of $\hat{x}_{B}$ at the meter $B$. $gt_{f}=2$.\textcolor{red}{}\textcolor{green}{{}
}Plot shows $1.2\times10^{6}$ trajectories.\textcolor{green}{{} }Centre:
Plot of the Q function $Q(x_{A},p_{A})$ of the squeezed state $|x_{1}/2,r\rangle$.
Lower: Blue solid curve shows the marginal distribution $Q_{+,inf}(x_{A})$.
The black dashed line shows the distribution calculated from $Q(x_{A},p_{A})$.
Both curves are symmetrical about $x_{A}=0$. \textcolor{red}{} \textcolor{magenta}{\label{fig:ent-meter-infa-eigenstate}}\textcolor{green}{
}\textcolor{red}{}}
\end{figure}

\subsubsection{Analytical calculation of the inferred state of system $A$}

We wish to establish the distribution $Q(x_{A},p_{A}|B_{+})$ inferred
for system $A$ at time $t_{0}$ conditioned on the positive outcome
(identified by the positive branch $B_{+}$) of the measurement $\hat{x}_{B}$
of the meter $B$. Continuing from (\ref{eq:int-sA}), we evaluate

\begin{eqnarray*}
Q(x_{A},p_{A}|B_{+}) & = & \int\int Q(\lambda,t_{0}|B_{+})dx_{B}dp_{B}\\
 & = & T_{1}+T_{2}+T_{3}
\end{eqnarray*}
\textcolor{green}{}where we will take the limit $x_{1B}\rightarrow\infty$,
where the amplitude of the meter $B$ is very large. In calculating
the inferred state for $A$, there is integration over three terms
$T_{1}$, $T_{2}$ and $T_{3}$ that appear, in order, in $Q(\bm{\lambda},t_{0}|B_{+})$.
For the first two Gaussian terms $T_{1}$ and $T_{2}$, the integrals
simplify as follows, when $x_{1B}$ is large. 
\begin{align*}
\int dx_{B}\frac{e^{-(x_{B}-x_{1B})^{2}/2\sigma_{x_{B}}^{2}}}{1+e^{-2x_{B}x_{1B}/\sigma_{x_{B}}^{2}}} & \rightarrow\sqrt{2\pi}\sigma_{x_{B}}
\end{align*}
\begin{align*}
\int dx_{B}\frac{e^{-(x_{B}-x_{1B})^{2}/2\sigma_{x_{B}}^{2}}}{1+e^{2x_{B}x_{1B}/\sigma_{x_{B}}^{2}}} & \rightarrow0
\end{align*}
 For the third term $T_{3}$, we use the exact results: 

\begin{align*}
\int dx_{B}\frac{e^{-(x_{B}-2\beta_{0})^{2}/2\sigma_{x_{B}}^{2}}}{\cosh\left(2x_{B}\beta_{0}/\sigma_{x_{B}}^{2}\right)} & =e^{-4\beta_{0}^{2}/2\sigma_{x_{B}}^{2}}\sqrt{2\pi}\sigma_{x_{B}}
\end{align*}
and
\begin{eqnarray*}
 &  & \int dp_{B}e^{-p_{B}^{2}/2\sigma_{p_{B}}^{2}}\sin\bigl(x_{1}p_{A}/\sigma_{x_{A}}^{2}+x_{1B}p_{B}/\sigma_{x_{B}}^{2}\bigl)\\
 & = & \sqrt{2\pi}\sigma_{p_{B}}\sin\left(x_{1}p_{A}/\sigma_{x_{A}}^{2}\right)e^{-\sigma_{p_{B}}^{2}x_{1B}^{2}/2\sigma_{x_{B}}^{4}}
\end{eqnarray*}
\textcolor{green}{}\textcolor{red}{}\textcolor{black}{Hence, we
write} for the inferred state of system $A$ conditioned on an outcome
$x_{1B}$ of the meter $B$ \begin{widetext} 

\begin{eqnarray}
Q(x_{A},p_{A}|B_{+}) & \rightarrow & \frac{e^{-p_{A}^{2}/2\sigma_{x_{A}}^{2}}}{2\pi\sigma_{x_{A}}\sigma_{p_{A}}}\Biggl\{ e^{-(x_{A}-x_{1})^{2}/2\sigma_{x_{A}}^{2}}-e^{-\left(x_{A}^{2}+x_{1}^{2}\right)/2\sigma_{x_{A}}^{2}}e^{-x_{1B}^{2}\left(1+\sigma_{p_{B}}^{2}/\sigma_{x_{B}}^{2}\right)/2\sigma_{x_{B}}^{2}}\sin\left(\frac{p_{A}x_{1}}{\sigma_{x_{A}}^{2}}\right)\Biggl\}\nonumber \\
\nonumber \\
\label{eq:QA-limit}
\end{eqnarray}
\end{widetext} \textcolor{black}{We note that for consistency with
the limit taken of a large meter-amplitude $x_{1B}$, a correction
term }in (\ref{eq:QA-limit}) resulting from $T_{2}$ exists, which
is a function of $x_{A}$ but not $p_{A}$.

For the convenient coherent-state meter, $\varphi=\pi/2$, $r_{2}=0$
and $x_{1B}=2\beta_{0}$ with $\beta_{0}$ large. \textcolor{black}{Where
the system is a two-mode cat state (so that $x_{1}=2\alpha_{0}$,
$\sigma_{x_{A}}^{2}=\sigma_{p_{A}}^{2}=2$, $\sigma_{x_{B}}^{2}=\sigma_{p_{B}}^{2}=2$)
with $\varphi=\pi/2$, we find}

\textcolor{black}{
\begin{align*}
Q(x_{A},p_{A}|B_{+}) & =\frac{e^{-p_{A}^{2}/4}}{4\pi}\Bigl\{ e^{-(x_{A}-x_{1})^{2}/4}\\
 & -e^{-\left(x_{A}^{2}+4\alpha_{0}^{2}\right)/4}e^{-2\beta_{0}^{2}}\sin\left(p_{A}\alpha_{0}\right)\Bigl\}
\end{align*}
}Taking $\beta_{0}$ large, the $Q(x_{A},p_{A}|B_{+})$ is the inferred
state for system $A$ based on the measurement by the meter: $Q_{+,inf}^{(A)}=Q(x_{A},p_{A}|B_{+})$.
The above form shows that for the coherent-state meter when $x_{1B}=2\beta_{0}$
is large, 
\begin{align}
Q(x_{A},p_{A}|B_{+}) & \rightarrow\frac{e^{-p_{A}^{2}/2\sigma_{p_{A}}^{2}}e^{-(x_{A}-x_{1})^{2}/2\sigma_{x_{A}}^{2}}}{2\pi\sigma_{p_{A}}\sigma_{x_{A}}}
\end{align}
\textcolor{green}{}\textcolor{red}{}in agreement with the result
from Sec. VI.1.

\subsubsection{Numerical calculation of the inferred state for system $A$}

The simulation allows an accurate calculation of the inferred function
$Q_{+,inf}^{(A)}$ for the system $A$, based on a positive outcome
$+$ at $B$. In this case, we define the inferred function directly
from the trajectories of the amplitudes $x(t_{f})>0$ where $gt_{f}\rightarrow\infty$.
There is a set of backward trajectories emanating from the set $x_{B}(t_{f})>0$,
for a given $gt_{f}$. For each such trajectory, there is a single
$x_{B}(0)$ at the time $t_{0}=0$. Over the whole set of amplitudes,
this defines a distribution $P(x_{B}(0))$ for $x_{B}(0)$ at the
time $t_{0}=0$. Using the expression Eq. (\ref{eq:cond}) for the
conditional distribution $Q(\lambda|x_{B})$ ($\lambda=(x_{A},p_{A},x_{B},p_{B})$)
we evaluate the postselected distribution $Q(\lambda,t_{0}|+)$, given
by
\[
Q(\lambda,t_{0}|+)=\sum_{x_{B}(0)}P(x_{B}(0))Q(\lambda|x_{B})
\]
 Hence,
\[
Q(x_{A},p_{A}|+)=\int dx_{B}dp_{B}Q(\lambda,t_{0}|+)
\]
The inferred distribution for system $A$ as evaluated numerically
is 
\[
Q_{+,inf}^{(A)}\equiv\lim_{gt_{f}\rightarrow\infty}Q(x_{A},p_{A}|+)
\]
The expression differs from that derived analytically when the meter
is not perfect. Similarly, $Q(\lambda,t_{0}|+)$ can be different
to $Q(\lambda,t_{0}|B_{+})$. This is due to the states $|x_{1B}/2,r_{2}\rangle$
and $|-x_{1B}/2,r_{2}\rangle$ of the meter not being orthogonal,
so the detection of the positive amplitude $x(t_{f})>0$ does not
always imply when the state $|x_{1B}/2,r_{2}\rangle$ of the meter.
Orthogonality is achieved for the coherent-state meter when $\beta_{0}\rightarrow\infty$.

\begin{figure}
\begin{centering}
\includegraphics[width=0.8\columnwidth]{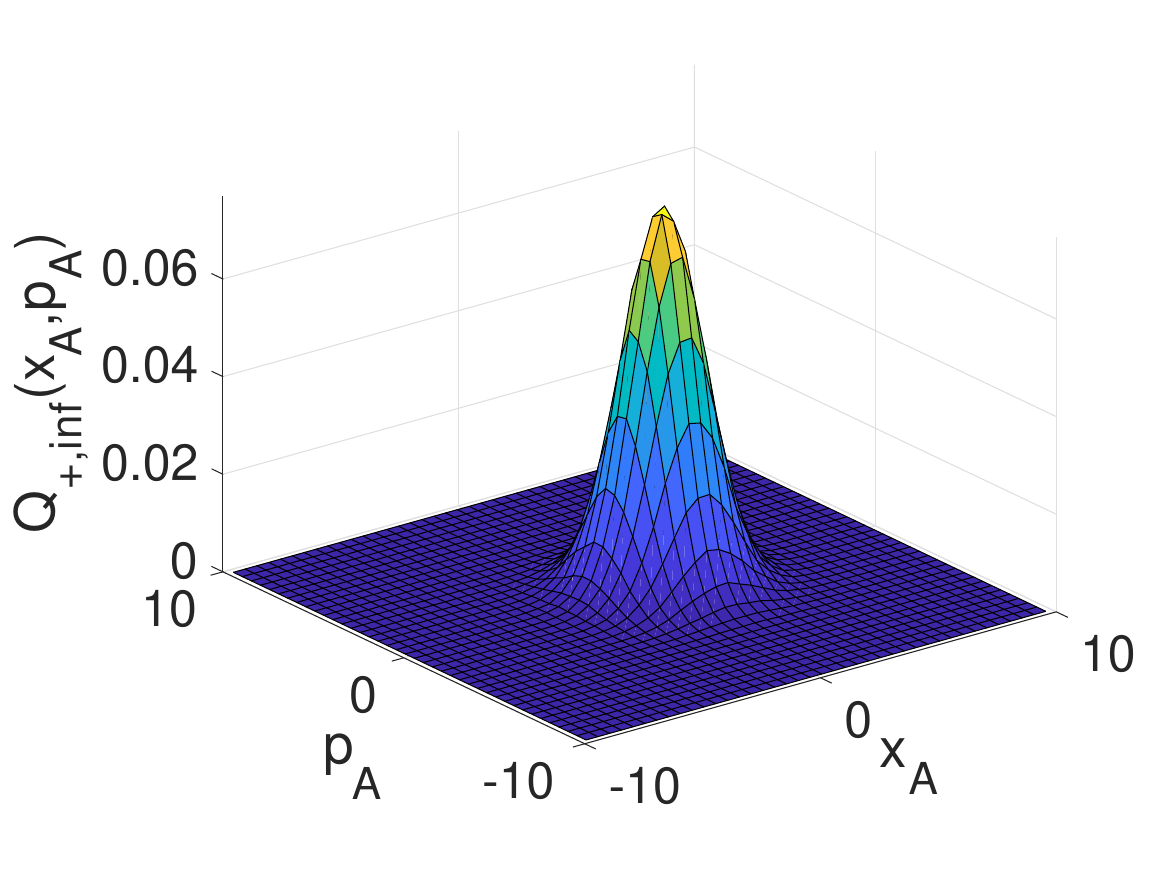}
\par\end{centering}
\begin{centering}
\includegraphics[width=0.8\columnwidth]{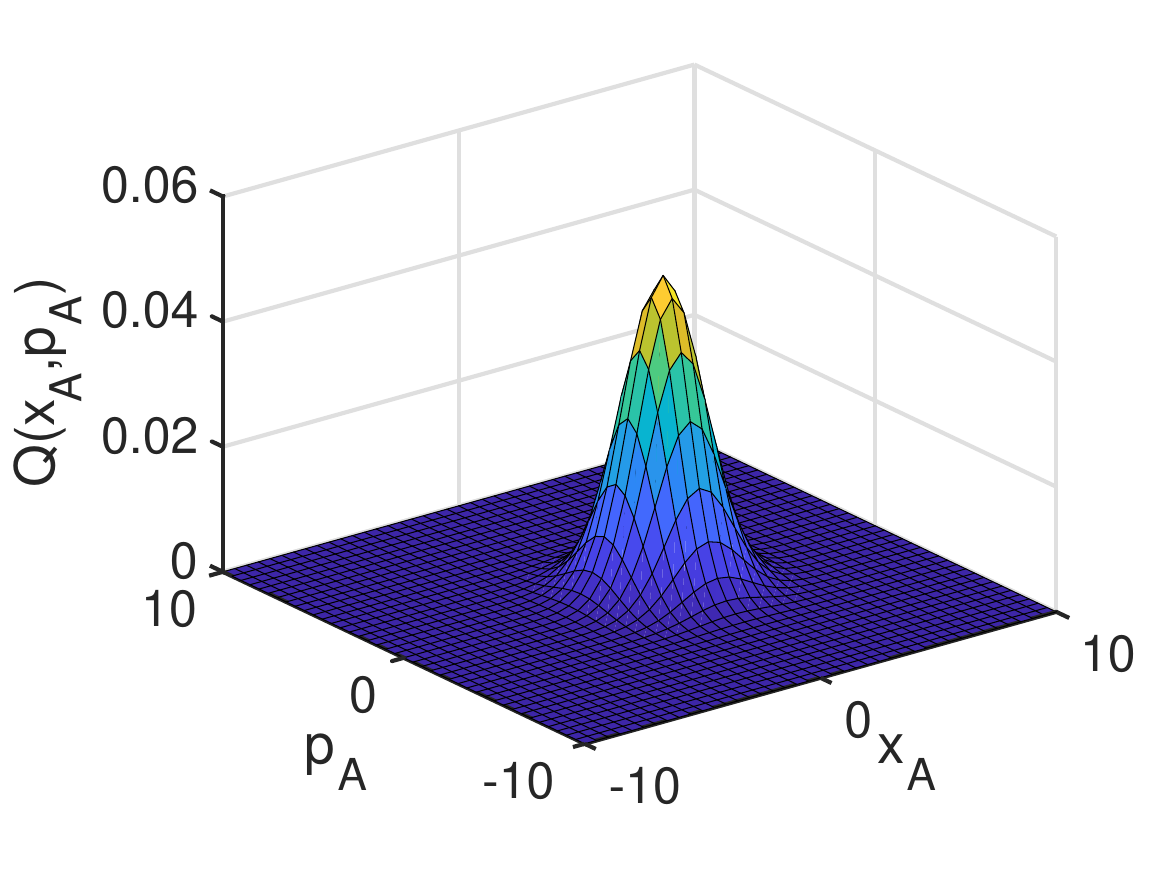}\\
\par\end{centering}
\begin{centering}
\includegraphics[width=0.8\columnwidth]{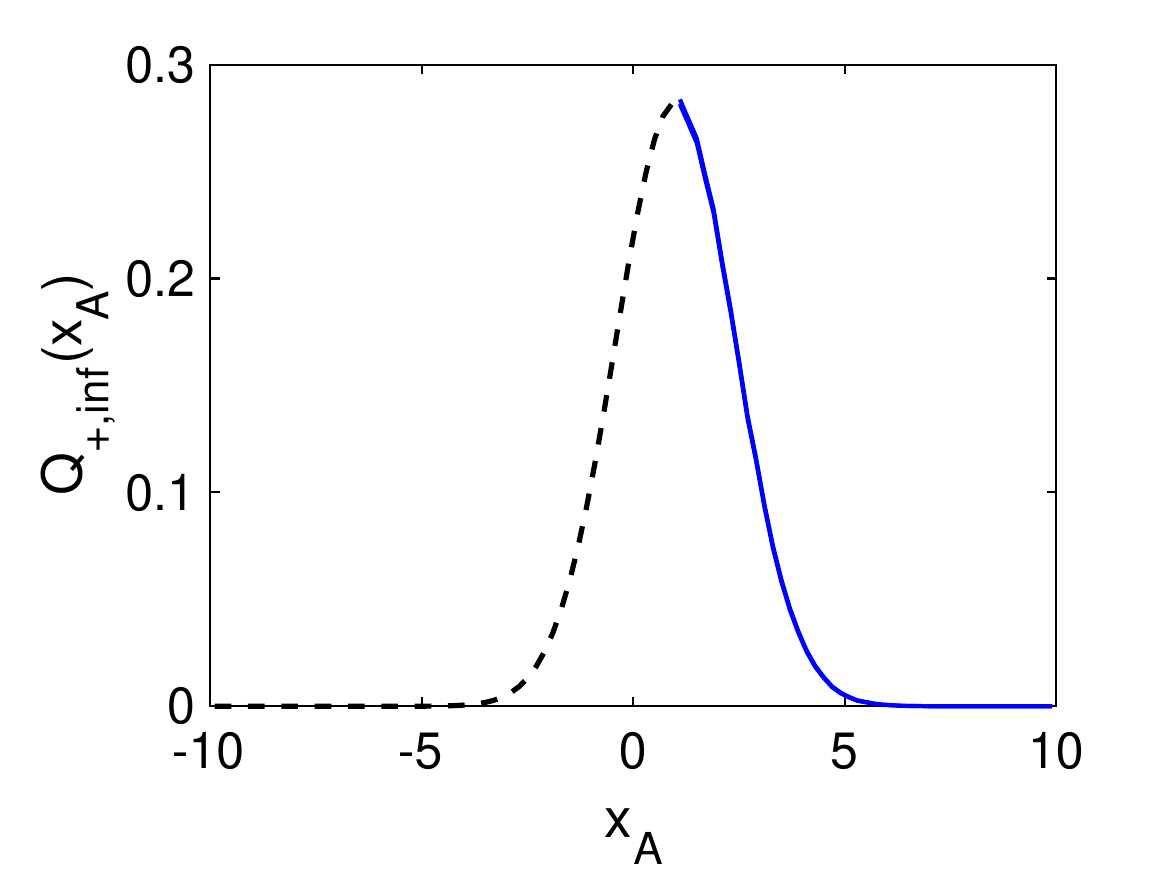}
\par\end{centering}
\caption{Collapse of the wave function: As for Figure \ref{fig:ent-meter-infa-eigenstate-1-1},
but here system $A$ is in a cat state, so that for system $A$, $r=0$
and $x_{1}=1$. $gt_{f}=2$.\textcolor{red}{} \textcolor{magenta}{\label{fig:ent-meter-cat}}\textcolor{green}{}\textcolor{red}{}}
\end{figure}

\begin{figure}
\begin{centering}
\includegraphics[width=0.8\columnwidth]{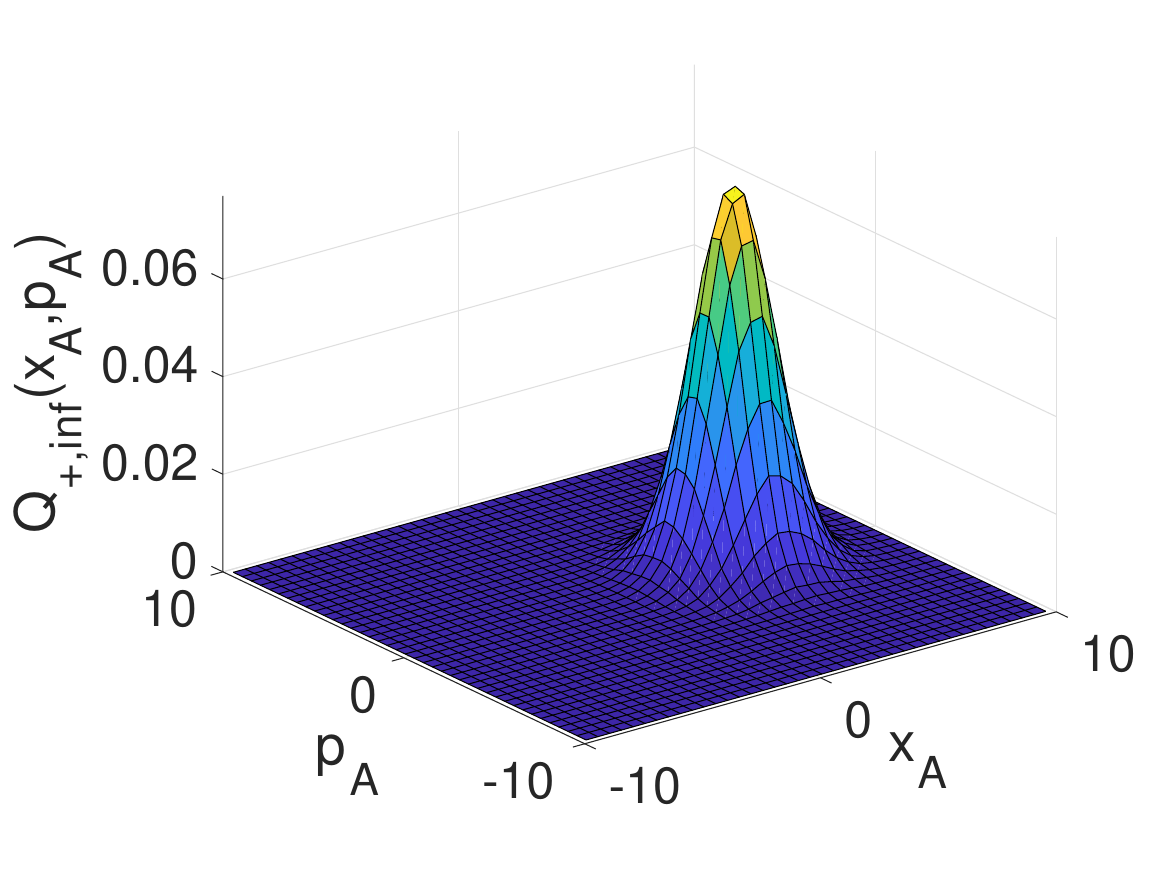}
\par\end{centering}
\begin{centering}
\includegraphics[width=0.8\columnwidth]{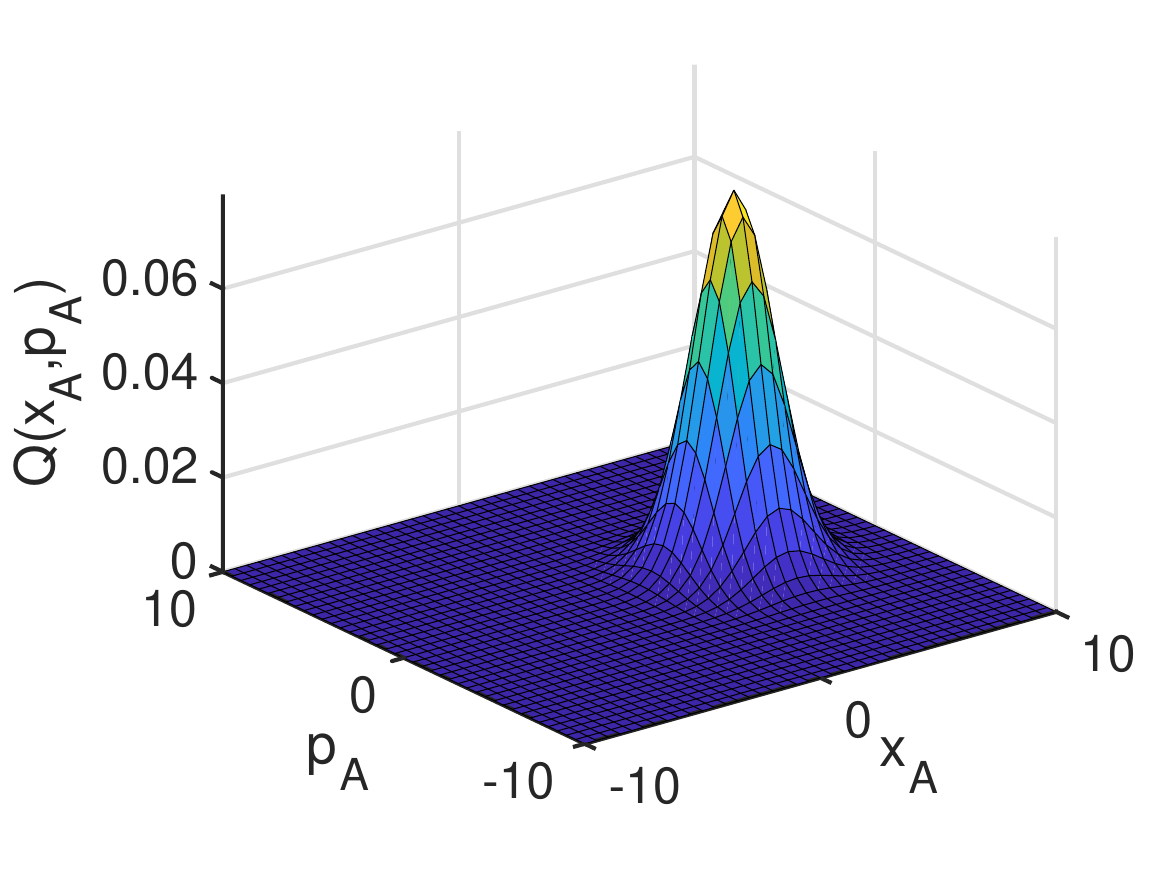}
\par\end{centering}
\begin{centering}
\includegraphics[width=0.8\columnwidth]{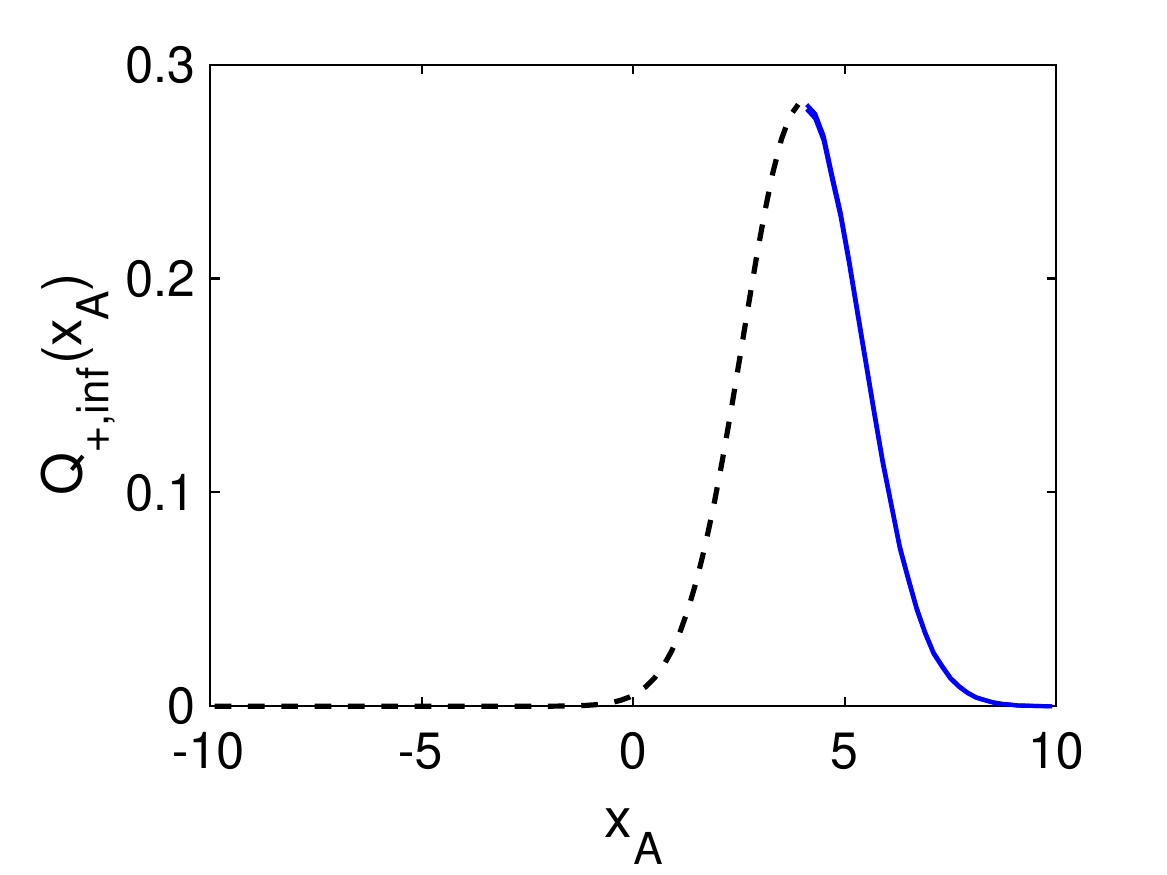}
\par\end{centering}
\caption{Collapse of the wave function for the cat state: As for Figure \ref{fig:ent-meter-infa-eigenstate-1-1},
but here system $A$ is in a cat state, so that for system $A$, $r=0$
and $x_{1}=1$. $gt_{f}=2$.\textcolor{magenta}{\label{fig:ent-meter-cat-2}}\textcolor{green}{}\textcolor{red}{}}
\end{figure}

Results are plotted in Figures \ref{fig:ent-meter-infa-eigenstate-1-1}
- \ref{fig:ent-meter-cat-2} for the system prepared in the convenient
coherent-state meter\textcolor{red}{{} } 
\begin{equation}
|\psi_{ent}\rangle=\frac{1}{\sqrt{2}}\{|x_{1}/2,r\rangle|\beta_{0}\rangle+i|-x_{1}/2,r\rangle|-\beta_{0}\rangle\label{eq:cat-meter}
\end{equation}
where $|\beta_{0}\rangle$ are coherent states of the meter. We choose
$\beta_{0}$ and $gt$ sufficiently large, and consider system $A$
to be in a superposition of eigenstates, approximated with $r=1.5$
(Figures \ref{fig:ent-meter-infa-eigenstate-1-1} and \ref{fig:ent-meter-infa-eigenstate}).
As the meter becomes macroscopic, with $\beta_{0}$ large, the function
$Q_{+,inf}^{(A)}$ becomes that of the $Q$ function for the squeezed
state $|\frac{x_{1}}{2},r\rangle$. This is true even for small $x_{1}$,
when the Q function peaks associated with each eigenstates of system
$A$ overlap. We conclude that the state of system $A$ at time $t_{0}$
conditioned on the outcome $\beta_{0}$ for $\hat{x}_{B}$ corresponds
to the ``collapsed'' or ``projected'' state $|\frac{x_{1}}{2},r\rangle$,
consistent with the measurement postulates. Figures \ref{fig:ent-meter-cat}
and \ref{fig:ent-meter-cat-2} show similar results for the two-mode
cat state ($r=0$). The inferred state $Q_{+,inf}^{(A)}$ for system
$A$ when the meter is macroscopic corresponds to a coherent state
$|\alpha_{0}\rangle$.

\subsubsection{Interpretation of wave-function collapse}

In the Q model of reality, each amplitude $x$ and $p$ is a possible
realization of a state for the system, at the time $t_{0}$. Hence,
the model allows an interpretation about when and how the measured
system $A$ ``collapses'' to the state $|x_{1}/2,r\rangle$ (Conclusion
(2) of this paper).

We see that the collapse is a two-stage process, which does not suddenly
happen with the final measurement at the meter $B$, when $x_{B}(t_{f})$
is detected. Rather, the system $A$ is in one or other of two states
(with a distribution of amplitudes $x_{A}(t_{0})$) which is correlated
with a definite outcome, positive or negative, for measurement of
$\hat{x}_{B}$ of the meter $\hat{x}_{B}$. This is because the meter
is \emph{macroscopic}, at the time $t_{0}$, and the interference
term in the Q function (\ref{eq:Qent-2-1}) has become negligible
with large $\beta_{0}$. This we see for the special case of the two-mode
cat state (Eq. (\ref{eq:cat-meter}) where $r=0$) by examining the
two-mode $Q$ function (\ref{eq:Qent-2-1}) where we set ($r=r_{2}=0$,
and $x_{1}=2\alpha_{0}$, $x_{1B}=2\beta_{0}$), which for $\beta_{0}\rightarrow\infty$
becomes two two-mode Gaussians with correlated means $(2\alpha_{0},2\beta_{0})$
and $(-2\alpha_{0},-2\beta_{0})$. The system has ``collapsed''
to the final state $|\alpha_{0}\rangle$ or $|-\alpha_{0}\rangle$
in a limiting sense for $\beta_{0}$ large, by the time $t_{0}$.
The essential aspect of the collapse was created by the prior interaction
$H_{int}$ (refer Section II.B) which coupled the system $A$ to the
macroscopic meter system $B$. In other words, the measurement process
is based on amplification, and the interference terms do not amplify
but decay through this process.

The ``collapse'' brought about by the interaction $H_{int}$ is
not complete, however, since the interaction $H_{int}$ is unitary
and can hence be reversed. This is possible because the fringe terms
in (\ref{eq:Qent-2-1}) although small do not completely vanish, for
large $\beta_{0}$. However, if the system $A$ is decoupled from
$B$, then reversibility is not possible. The decoupling amounts to
a loss of information for each system. The $Q$ function for system
$A$ in this case is found by integrating over $x_{B}$ and $p_{B}$,
conditioned on the outcome for the meter. The integration over $p_{B}$
removes the interference term in the $Q$ function (\ref{eq:Qent-2-1}):
There is a\emph{ loss of information about the complementary de-amplified
variable $p_{B}$ of the meter} that results from the measurement
process, which is based on amplification of $x_{B}$. The resulting
$Q$ function conditioned on a positive outcome ($x_{B}(t_{f})>0$)
for the meter is that of the projected state $|x_{1}/2,r\rangle$
for system: Averaging over both outcomes of the meter, the overall
system is in a statistical mixture of the two states $|x_{1}/2,r\rangle|\beta_{0}\rangle$
and $|-x_{1}/2,r\rangle|-\beta_{0}\rangle$. At this stage, the
system $A$ is precisely in one or other states $|x_{1}/2,r\rangle$
or $|-x_{1}/2,r\rangle$ and the collapse is completed, being irreversible
in the decoupled limit.

\subsection{Hidden Variables}

Generalizing the analysis of Section VI to two modes, we next
examine the distribution $Q(\lambda,t_{0}|B_{+})$  (and $Q(\lambda,t_{0}|+)$,
as derived numerically) for the amplitudes at time $t_{0}$ conditioned
on an outcome for the measurement on the meter $B$. We show that
this distribution does not correspond to the Q function of a quantum
state. The variables $\lambda=(x_{A},p_{A},x_{B},p_{B})$ are hence
``hidden variables'' in the reality model associated with the measurement.

\textcolor{green}{}We consider the superposition (\ref{eq:ent-cat-1}),
given by the $Q$-function (\ref{eq:Qent-2-1}), with $\varphi=\pi/2$.
A measurement of $\hat{x}_{B}$ is made on the meter. The postselected
state $Q(\lambda,t_{0}|B_{+})$ for the combined meter-system conditioned
on the positive branch $B_{+}$ of the meter was calculated in the
last section.  In the numerical simulation, $Q(\lambda,t_{0}|B_{+})$
is given by $Q(\lambda,t_{0}|+)$, the two being equal for the perfect
meter.

We wish to examine the distribution $Q(x_{B},p_{B}|B_{+})$ at the
time $t_{0}$ for $B$ alone, conditioned on a positive outcome for
$\hat{x}_{B}$ of the meter. We find by integrating over $x_{A}$
and $p_{A}$:
\begin{eqnarray}
Q(x_{B},p_{B}|B_{+}) & = & \int dx_{A}dp_{A}Q(\lambda,t_{0}|B_{+})\nonumber \\
 & = & Q_{+}(x_{B})\int dx_{A}\int dp_{A}Q(\lambda|x_{B})\nonumber \\
\label{eq:hid5}
\end{eqnarray}
We evaluate\textcolor{red}{{} }\textcolor{green}{}\textcolor{red}{}$Q(p_{B}|x_{B})=\int dx_{A}\int dp_{A}Q(\lambda|x_{B})$.
\textcolor{green}{}\textcolor{red}{}\textcolor{green}{}\textcolor{red}{}\textcolor{black}{We
find}\textcolor{black}{{} }\textcolor{red}{}\begin{widetext} 
\begin{eqnarray}
Q(x_{B},p_{B}|B_{+}) & = & \frac{e^{-(x_{B}-|x_{1B}|)^{2}/2\sigma_{x_{B}}^{2}}}{\sqrt{2\pi}\sigma_{x_{B}}}\frac{e^{-p_{B}^{2}/2\sigma_{p_{B}}^{2}}}{\sqrt{2\pi}\sigma_{p_{B}}}\Bigr(1-e^{-x_{1}^{2}(1+\sigma_{p_{A}}^{2}/\sigma_{x_{A}}^{2})}\frac{\sin\left(p_{B}|x_{1B}|/\sigma_{x_{B}}^{2}\right)}{\cosh\left(x_{B}|x_{1B}|/\sigma_{x_{B}}^{2}\right)}\Bigl)\label{eq:qb+}
\end{eqnarray}

\end{widetext} \textcolor{red}{}\textcolor{green}{}We next evaluate
$Q(p_{B}|B_{+})$. Integrating $Q(x_{B},p_{B}|B_{+})$ over $x_{B}$
we find\textcolor{green}{}
\begin{align}
Q(p_{B}|B_{+}) & =\frac{e^{-p_{B}^{2}/2\sigma_{p_{B}}^{2}}}{\sqrt{2\pi}\sigma_{p_{B}}}\Bigl\{1-e^{-x_{1B}^{2}/2\sigma_{x_{B}}^{2}}\nonumber \\
 & \times e^{-x_{1}^{2}\left(1+\sigma_{p_{A}}^{2}/\sigma_{x_{A}}^{2}\right)/2\sigma_{x_{A}}^{2}}\sin\bigl(|x_{1B}|p_{B}/\sigma_{x_{B}}^{2}\bigl)\Bigl\}
\end{align}
where we use Eq. (\ref{eq:int}). \textcolor{black}{}We note the
distribution $Q(p_{B}|B_{+})$ conditioned on a negative outcome of
$\hat{x}_{B}$ is the same as $Q(p_{B}|B_{+})$. The function is independent
of the sign of the outcome i.e. which branch of the superposition.\textcolor{green}{}

\begin{figure}
\begin{centering}
\includegraphics[width=0.7\columnwidth]{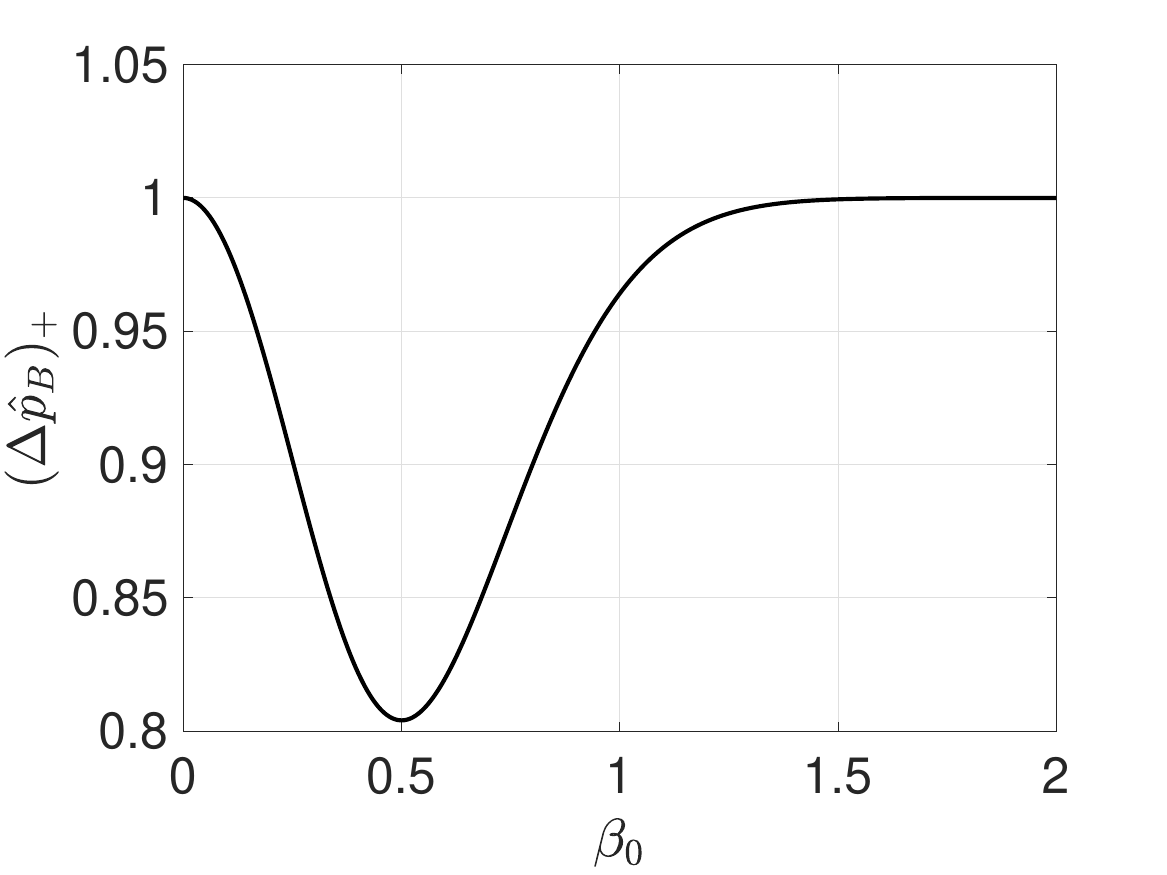} \includegraphics[width=0.7\columnwidth]{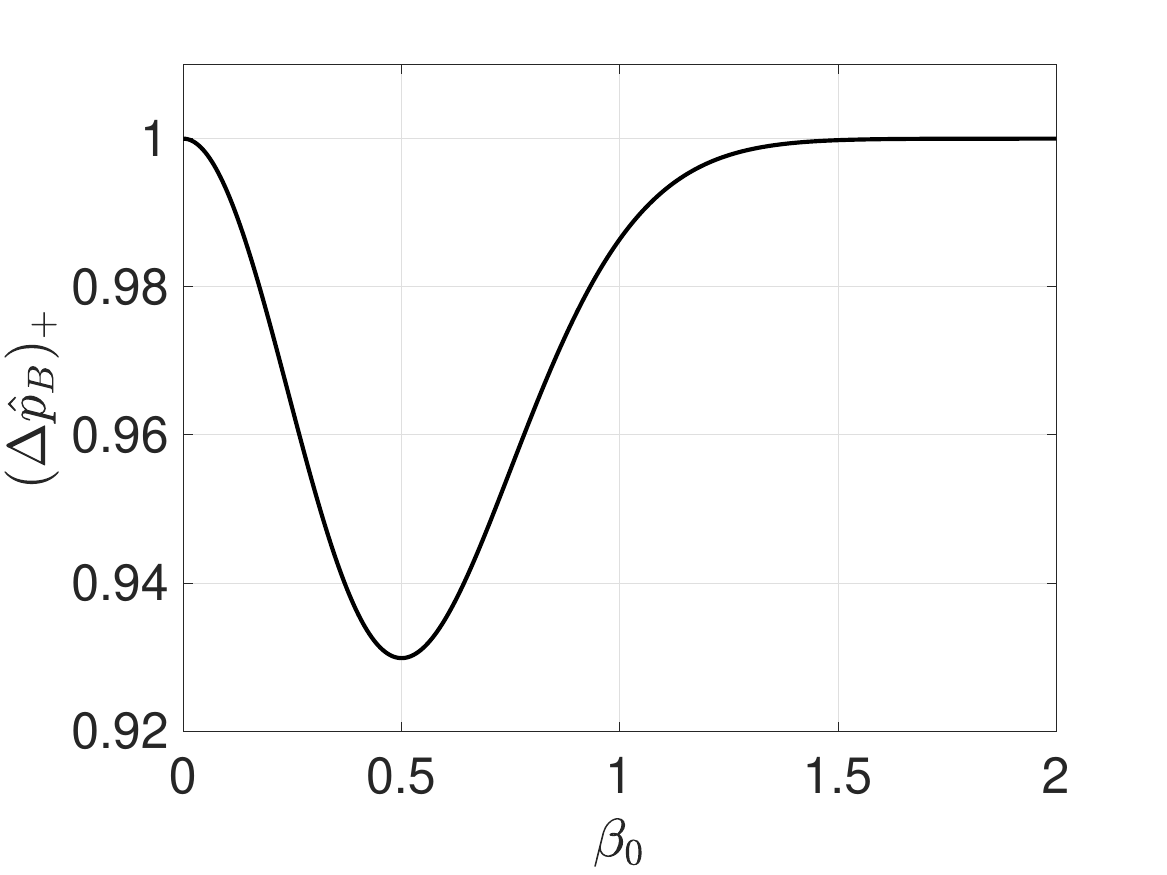}
\par\end{centering}
\centering{}\includegraphics[width=0.7\columnwidth]{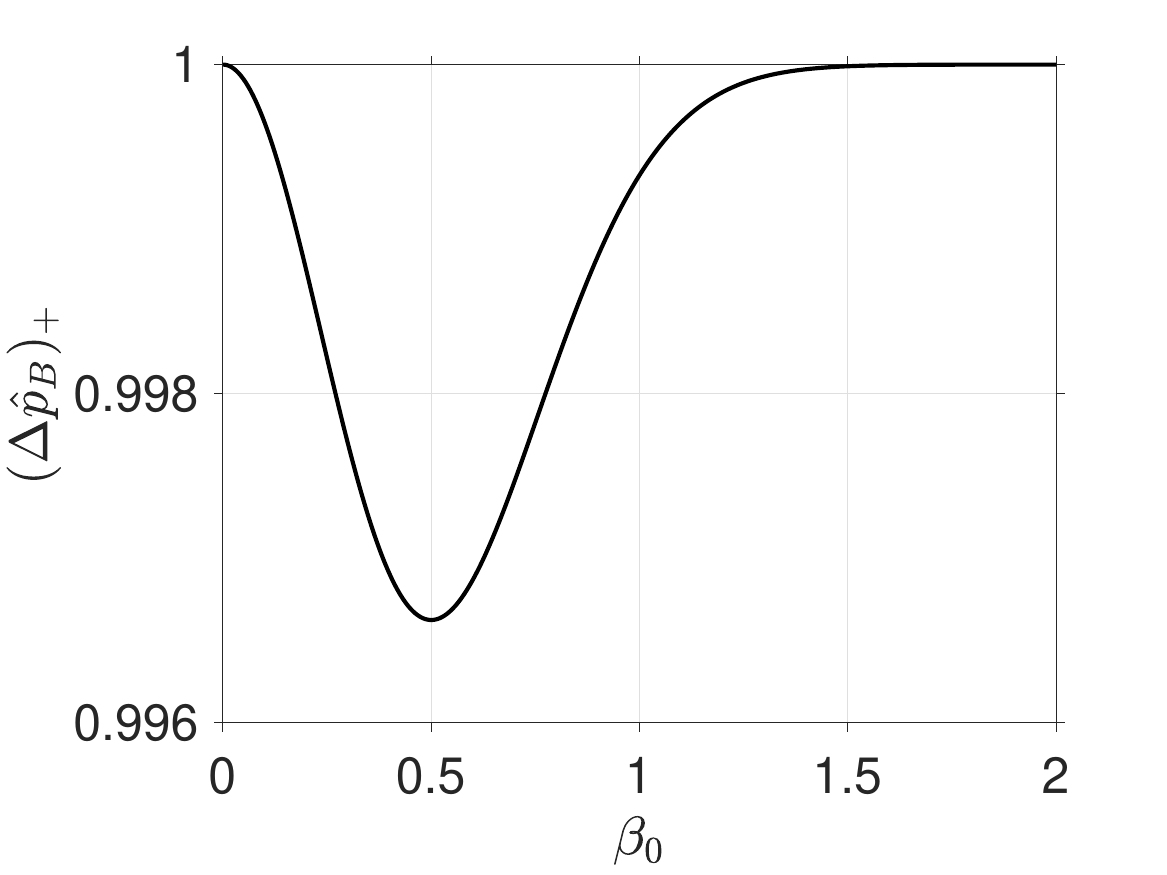}\caption{The analytical functions (Eq. (\ref{eq:tmcvarp})) obtained for $(\Delta\hat{p}_{B})_{+}^{2}$
vs $\beta_{0}$ for fixed $\alpha_{0}$ and $\varphi=\pi/2$, where
we examine the two-mode cat state.\textcolor{red}{{} }Here $\alpha_{0}=0.1$
(top), $\alpha_{0}=0.5$ (centre) and $\alpha_{0}=1$ (lower). \label{fig:analytic-var}
\textcolor{red}{}}
\end{figure}

\begin{figure}
\begin{centering}
\includegraphics[width=0.7\columnwidth]{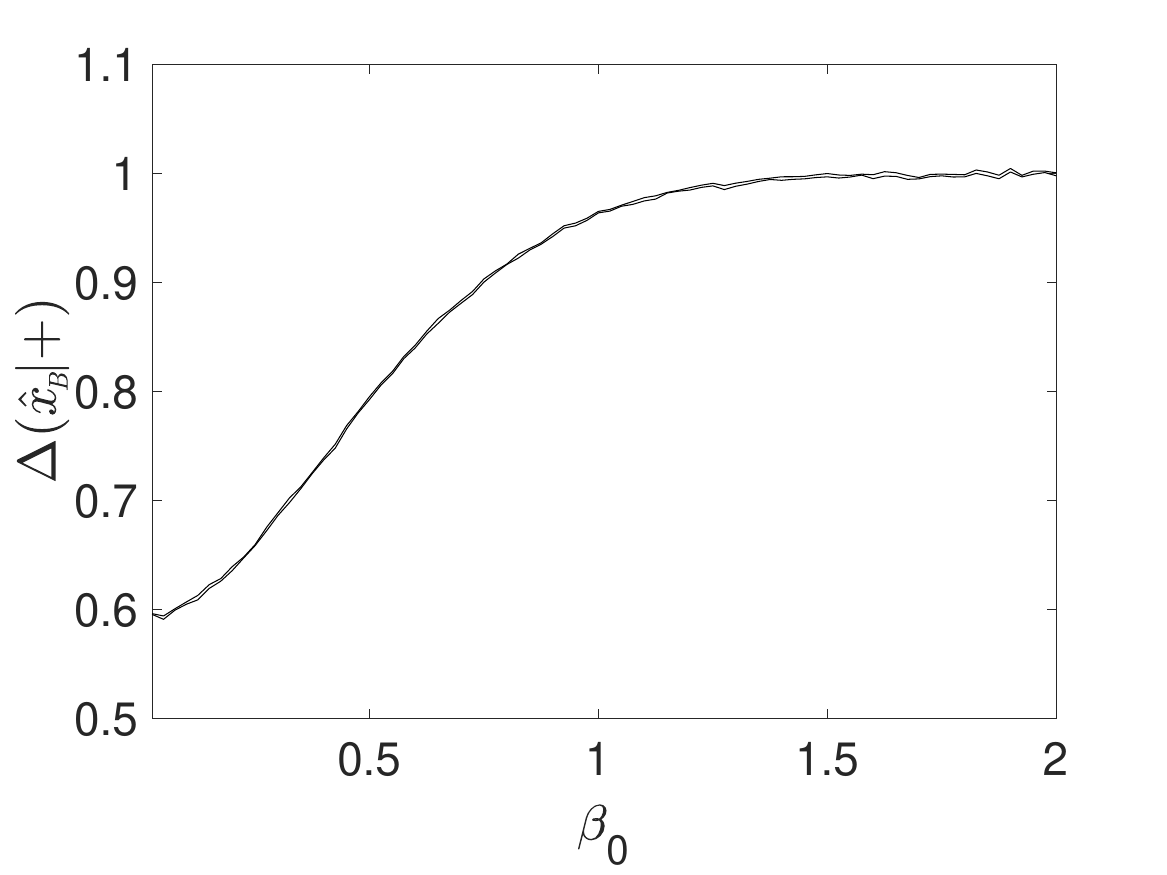}
\par\end{centering}
\begin{centering}
\includegraphics[width=0.7\columnwidth]{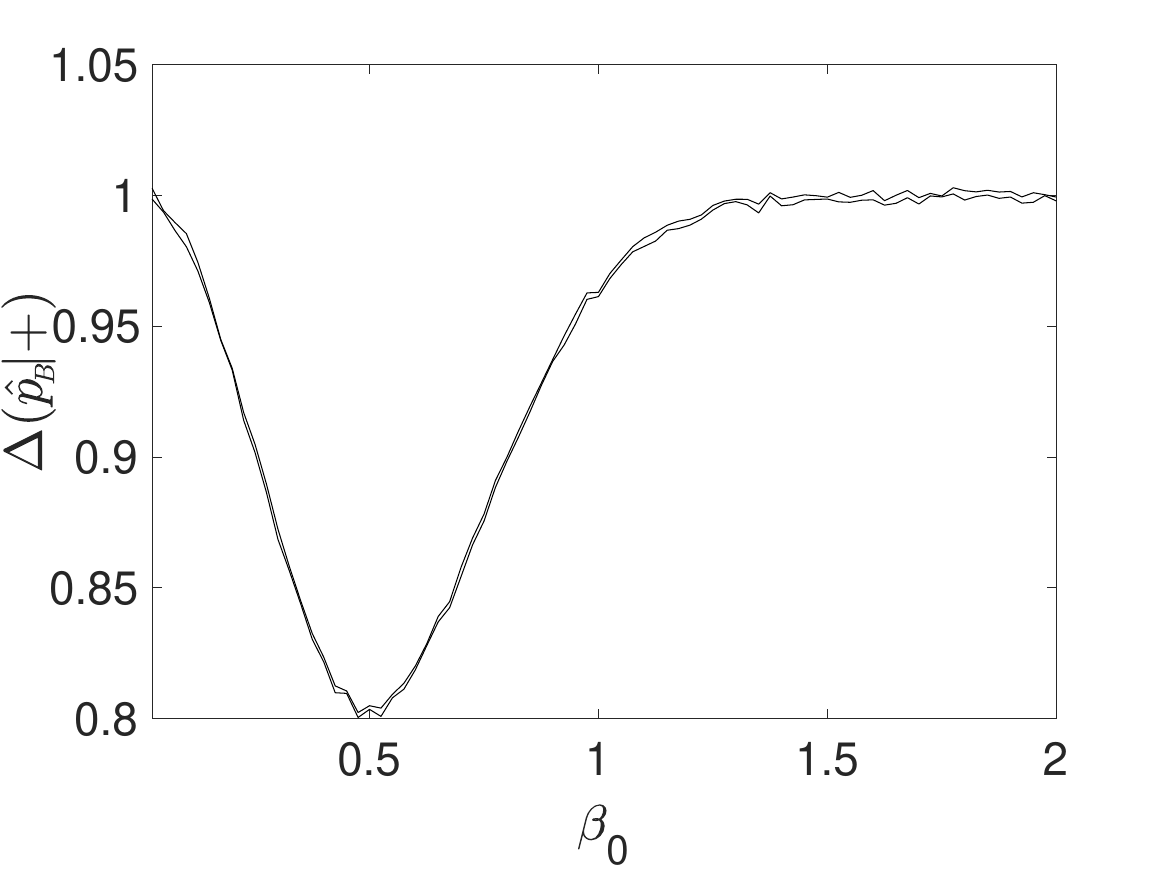}
\par\end{centering}
\begin{centering}
\par\end{centering}
\begin{centering}
\includegraphics[width=0.7\columnwidth]{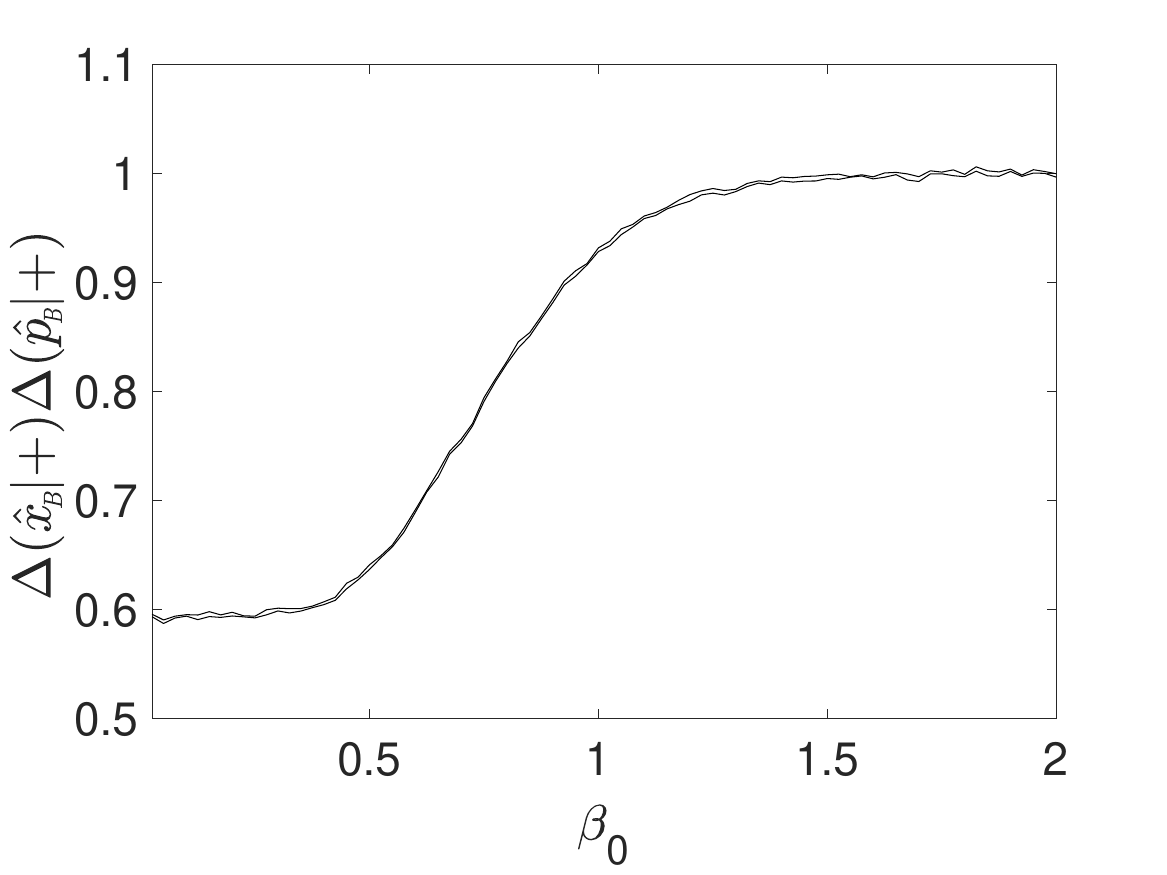}
\par\end{centering}

\caption{Plot of the uncertainty in the inferred state of the meter $B$ given
the outcome $x_{B}(t_{f})$ is positive, as determined from the numerical
calculations. Here $\alpha_{0}=0.1$, $r=0$, \textcolor{black}{and
$\varphi=\pi/2$} and $gt=2$. The two parallel lines indicate the
upper and lower error bounds from sampling errors, with $1.2\times10^{6}$
trajectories.\label{fig:up}\textcolor{green}{}\textcolor{red}{}\textcolor{green}{}}
\end{figure}

We also calculate the distribution $Q(x_{B}|B_{+})$ of $x_{B}$,
given the outcome $x_{1B}$. Where\textcolor{black}{{} $\varphi=\pi/2$,
\begin{equation}
Q(x_{B}|B_{+})=\frac{1}{\sqrt{2\pi}\sigma_{x_{B}}}e^{-(x_{B}-|x_{1B}|)^{2}/2\sigma_{x_{B}}^{2}}\label{eq:qx-1}
\end{equation}
}The variance $(\Delta x_{B})_{+}^{2}\equiv\langle x_{B}^{2}\rangle-\langle x_{B}\rangle^{2}$
of the distribution $Q(x_{B},p_{B}|B_{+})$ is hence $(\Delta x_{B})_{+}^{2}=\sigma_{x_{B}}^{2}=1+e^{-2r}$.
The measured variance is {[}Eq. (\ref{eq:varp}){]}
\begin{equation}
(\Delta\hat{x}_{B})_{+}^{2}=(\Delta x_{B})_{+}^{2}-1=e^{-2r}\label{eq:measure-x}
\end{equation}
Next, we take the special case where system $A$ is the cat state,
with $\sigma_{x_{A}}^{2}=2$ and $x_{1}=2\alpha_{0}$, but $\varphi=\pi/2$.
Then \textcolor{red}{}\textcolor{green}{}\textcolor{red}{}
\begin{align}
Q(p_{B}|B_{+}) & =\frac{e^{-p_{B}^{2}/2\sigma_{p_{B}}^{2}}}{\sqrt{2\pi}\sigma_{p_{B}}}\Bigl\{1-e^{-2\alpha_{0}^{2}}\nonumber \\
 & \times e^{-x_{1B}^{2}/2\sigma_{x_{B}}^{2}}\sin\bigl(|x_{1B}|p_{B}\bigl)\Bigl\}\label{eq:qp-2}
\end{align}
The variance $(\Delta p_{B})_{+}^{2}\equiv\langle p_{B}^{2}\rangle-\langle p_{B}\rangle^{2}$
of the distribution $Q(x_{B},p_{B}|B_{+})$ is hence calculated. We
find\textcolor{red}{}\textcolor{green}{} $\langle p_{B}^{2}\rangle_{+}=\sigma_{p_{B}}^{2}$
and
\begin{eqnarray}
\langle p_{B}\rangle_{+} & = & -\frac{|x_{1B}|\sigma_{p_{B}}^{2}}{\sigma_{x_{B}}^{2}}e^{-2\alpha_{0}^{2}}e^{-\frac{x_{1B}^{2}}{2\sigma_{x_{B}}^{2}}\left(1+\frac{\sigma_{p_{B}}^{2}}{\sigma_{xB}^{2}}\right)}\label{eq:meanp}
\end{eqnarray}
Inserting\textcolor{red}{{} } $\sigma_{x_{B}}^{2}=1+e^{-2r}$ and
$\sigma_{p_{B}}^{2}=1+e^{2r}$ into the resulting expression, we find\begin{widetext} 
\begin{align}
(\Delta p_{B})_{+}^{2} & =\left(1+e^{2r_{2}}\right)\left(1-\frac{x_{1B}^{2}\left(1+e^{2r_{2}}\right)}{\left(1+e^{-2r_{2}}\right)^{2}}e^{-4\alpha_{0}^{2}}e^{-\frac{x_{1B}^{2}}{\left(1+e^{-2r_{2}}\right)}\left(1+\frac{\left(1+e^{2r_{2}}\right)}{\left(1+e^{-2r_{2}}\right)}\right)}\right)
\end{align}
\end{widetext} For $r_{2}$ large, \textcolor{black}{$(\Delta p_{B})_{+}^{2}\rightarrow1+e^{2r_{2}}$.
For the special case of the coherent-state meter, we find }\textcolor{green}{}$(\Delta p_{B})_{+}^{2}=2-4\beta_{0}^{2}e^{-4\alpha_{0}^{2}}e^{-4\beta_{0}^{2}}$\textcolor{red}{}\textcolor{green}{}\textcolor{black}{}.
Hence, the measured variance is
\begin{equation}
(\Delta\hat{p}_{B})_{+}^{2}=1-4\beta_{0}^{2}e^{-4\alpha_{0}^{2}}e^{-4\beta_{0}^{2}}\label{eq:tmcvarp}
\end{equation}
as plotted in Figure \ref{fig:analytic-var}. The product $(\Delta\hat{x}_{B})_{+}(\Delta\hat{p}_{B})_{+}$
is reduced below 1 for finite $r$, which would violate the Heisenberg
uncertainty relation.\textcolor{red}{}\textcolor{black}{}

We demonstrate the hidden nature of the amplitudes directly from the
simulation, by extending the procedure taken for the single-mode case,
in Section VI.C. The postselected distribution $Q(x_{A},p_{A}|+)$
has been calculated numerically in Section VIII.B.3. The variances
$[\Delta(x_{B}|+)]^{2}$ and $[\Delta(p_{B}|+)]^{2}$ associated with
this distribution can also be numerically evaluated. The associated
measured variances $[\Delta(\hat{x}_{B}|+)]^{2}=[\Delta(x_{B}|+)]^{2}-1$
and $[\Delta(\hat{p_{B}}|+)]^{2}=[\Delta(\hat{p}_{B}|+)]^{2}-1$ are
calculated and shown in Figure \ref{fig:up}, for various choices
of parameters. In the numerical estimates, the trajectories for the
set $B_{+}$ are defined by the values $x(t_{f})>0$, and are hence
cut-off at the transition to negative values of $x(t_{f})$. The result
(\ref{eq:measure-x}) for $(\Delta x_{B})_{+}^{2}$ is valid only
in the limit of infinite $\beta_{0}$, because the Gaussian functions
$Q_{\pm}(x_{B},t)$ (Eq. (\ref{eq:G-1})) are only entirely positive
or entirely negative in the limit of $\beta_{0}\rightarrow\infty$,
or else, for eigenstates where $r\rightarrow\infty$, in the limit
of $gt_{f}\rightarrow\infty$. Hence, the variance $[\Delta(\hat{x}_{B}|+)]^{2}$
evaluated numerically is less than that given by (\ref{eq:measure-x})
for the perfect meter: $[\Delta(\hat{x}_{B}|+)]e^{2r}<1$.

In summary, the variances associated with the postselected distribution
for the meter, conditioned on an outcome for the meter, are incompatible
with the Heisenberg uncertainty relation. This demonstrates that hidden-variable
nature of the amplitudes and distribution associated with the the
Q based model of reality.\textcolor{red}{}\textcolor{green}{}\textcolor{green}{}\textcolor{red}{}\textcolor{green}{}

\section{Conclusion}

This paper presents solutions of forward-backward stochastic differential
equations that form the basis for an interpretation of measurement
in quantum mechanics. The solutions are for phase-space amplitudes
$x$ and $p$ of the Q function $Q(x,p)$ that uniquely represents
a single-mode quantum field. In a Q-based model of reality, the field
is described at a time $t$ by amplitudes $x(t)$ and $p(t)$. Measurement
of a quadrature field amplitude $\hat{x}$ corresponds to an amplification
of $x(t)$ to a macroscopic level, where it is directly detectable.
We study two methods of measurement of $\hat{x}$: one where the system
$A$ being measured is directly amplified, and the second where the
system $A$ is measured by coupling to a second system, a meter $B$.

A summary of how the Q-based model for measurement contributes to
a resolution of the measurement problem is given in the Introduction,
where three main conclusions {[}Conclusions (1), (2) and (3){]} are
presented. The conclusions elucidate aspects of the Schrodinger-cat
paradox. In this paper, we show how the solutions for $x(t)$ and
$p(t)$ are compatible with the predictions of quantum mechanics,
by explicitly demonstrating that the probability density of the amplitudes
$x(t)$ in the large amplification limit agrees with Born's rule.

Conclusion (1) of this paper concludes compatibility with macroscopic
realism e.g. a system in a superposition of two macroscopically distinct
coherent states $|\alpha_{0}\rangle$ and $|-\alpha_{0}\rangle$ ($\alpha_{0}$
is real), that can be distinguished by a measurement $\hat{x}$, has
a predetermined value for the outcome of $\hat{x}$. This conclusion
is consistent with known violations of Leggett-Garg inequalities \citep{legggarg-1},
which falsify macro-realism. The consistency is possible, because
the Leggett-Garg inequalities are derived from two assumptions: macroscopic
realism (MR) and noninvasive measurability (NIM). The conclusions
of this paper imply that in the Q-based model of reality, it is NIM
that fails.

Conclusion (2) elucidates how the collapse of the wave function occurs.
The measurement on the meter collapses the system that is being measured
(by the meter) into the eigenstate $|x_{j}\rangle$ given by the outcome
$x_{j}$ of the measurement. This is seen as a two-stage process,
due first to amplification process, which results in the amplitudes
$x(t_{f})$ (of the meter) that are detected separating into \emph{distinct
branches} associated with the distinct eigenvalues $x_{j}$. This
quantifies the real property that predetermines the outcome of the
measurement, at time $t_{f}$. The second stage of the collapse is
the inference about the state of the system being measured, conditioned
on the detected value $x(t_{f})$ of the meter. There is a loss of
information about $p$ of the meter: a distribution is inferred for
the system only.

Conclusions (2) and (3) show how the assumption of MR implies an
incompleteness of quantum mechanics: This gives an illustration of
the essential feature of Schrodinger's argument, as emphasized in
letters between Einstein and Schrodinger \citep{Einstein-letters-sch}.
The postelected state of the system at time $t$, as defined by the
amplitudes $x(t)$ and $p(t)$, corresponding to a given outcome (as
in ``dead'' or ``alive'') does not correspond to a quantum state
for the system. The amplitudes $x(t)$ and $p(t)$ are hence justifiably
referred to as hidden variables. There is no contradiction with Bell's
theorem however. The conclusions are consistent with the violations
of Bell inequalities, in particular those for macroscopic systems,
where the qubits correspond to the two states $|\alpha_{0}\rangle$
and $|-\alpha_{0}\rangle$ \citep{manushan-bell-cat-lg-1}. As pointed
out in Ref. \citep{ghz-cat-1}, this is because in the present paper,
macroscopic realism (MR) is defined  in its weakest form: MR applies
to the system at the time $t$ after all measurement settings have
been fixed. The derivation of Bell inequalities assumes hidden variables
that are defined for the system prior to the choice of settings, and
makes the additional assumption of locality.

The model of this paper is experimentally verifiable. The amplification
of quantum noise via parametric amplification is realized in many
experiments that detect squeezing of light (e.g. Refs. \citep{wu-squeezing-exp,schnabel-1-1-1,schnabel-2-1-1}).
The cat states have also been generated in may experiments (e.g. Refs.
\citep{vlastakis-cat,cat-states-haroche-wine,kirchmair-cat,wang-two-mode-cat,cat-states-review-1-1,brune-cat,monroe-cat,grangier-cat}).
The Q functions are also measurable \citep{tomo}.
\begin{acknowledgments}
This publication was made possible through the support of Grant 62843
from the John Templeton Foundation. 
\end{acknowledgments}

\end{document}